\documentclass[12pt]{article}
\usepackage{amsmath,amsfonts,amsthm,mathrsfs}

\newcommand{\R}{{\mathbb R}} \newcommand{\N}{{\mathbb N}}
\newcommand{\K}{{\mathbb K}} \newcommand{\Z}{{\mathbb Z}}
  \def\C{{\mathbb C}}
\newcommand{\Prm}{{\mathbb P}}
\newcommand{\wt}{\widetilde }
\renewcommand{\epsilon}{\varepsilon } 
 
\renewcommand{\rho}{\varrho } 
\renewcommand{\phi}{\varphi }

\newcommand{\EE}{{\mathbb E}}
\newcommand{\q}{{\rm q }}
\newcommand{\ran}{{\rm ran }}

\newcommand{\de}{{\rm det }}
\newcommand{\ca}{{\rm card}}

\newcommand{\set}{{\rm set }}

\def\ls{\left<}
\def\rs{\right>}
\def\lg{\left|}
\def\rg{\right|}

\def\ls{\left<}
\def\rs{\right>}

\newtheorem{theorem}{Theorem}[section]
\newtheorem{lemma}[theorem]{Lemma}
\newtheorem{corollary}[theorem]{Corollary}
\newtheorem{proposition}[theorem]{Proposition}

\newtheorem{remark}[theorem]{Remark}

\begin {document}
 \title{
Quantum Approximation and Integration in Sobolev Spaces -- Low Smoothness}

\author {Stefan Heinrich\\
Department of Computer Science\\
University of Kaiserslautern\\
D-67653 Kaiserslautern, Germany 
\\\\
{\it Dedicated to Henryk Wo\'zniakowski}\\ {\it on the occasion of his 80th birthday}}
  
\date{}
\maketitle
\begin{abstract}
We study the complexity of 
approximation and integration in  Sobolev spaces  in the quantum setting of information-based complexity theory (IBC). Previous results have been obtained under the assumption that the embedding condition holds. Here we drop this assumption and use a quantum version of a previously developed stochastic discretization technique.
\end{abstract}

\section{Introduction}
\label{sec:1} 
We continue to study the complexity of 
approximation and integration of functions from the Sobolev space $W_p^r(D)$, $D=[0,1]^d$ in the quantum setting of IBC. In this framework the complexity of integration was first studied for H\"older classes in \cite{Nov01}, followed by \cite{Hei01b}, where Sobolev spaces were considered. The quantum complexity of approximation of functions from $W_p^r(D)$ in the norm of $L_q(D)$ was first investigated in \cite{Hei04b}. The results of \cite{Hei01b,Hei04b} have been obtained under the assumption that $W_p^r(D)$ is continuously embedded in $C(D)$ (embedding condition). In later papers, see   \cite{Goc23,HY05, YH08,SY08,Ye10,YH010,Ye12,Ye12a,LY26,NSW04,KU26}, the quantum complexity of integration and approximation in various function classes was studied,  again under the assumption of corresponding embedding conditions. Integration in a quantum setting with randomized queries was considered in \cite{Woz06}.

In this paper we investigate the case that the embedding condition of $W_p^r(D)$ into $C(D)$ does not hold. 
It is shown that without further restriction on the input set $\mathcal{B}_{W_p^r(D)}$ (the unit ball of $W_p^r(D)$),
the quantum setting gives no rate whatsoever. On the other hand, if we restrict to uniformly equicontinuous subsets $\mathcal{E}$ of $\mathcal{B}_{W_p^r(D)}$, then we obtain optimal rates independently of $\mathcal{E}$ in the following sense: we give matching (up to logarithmic factors) upper and lower estimates of
$$
\sup_{\mathcal{E}\subset \mathcal{B}_{W_p^r(D)}} e_n^\q(J,\mathcal{E},L_q(D)), \quad \text{resp.} \quad \sup_{\mathcal{E}\subset \mathcal{B}_{W_p^r(D)}} e_n^\q(I,\mathcal{E},\K),
$$
where $J:W_p^r(D)\to L_q(D)$ is the embedding operator and $I:W_p^r(D)\to \K$ the integration operator, the supremum being taken  over all uniformly equicontinuous subsets $\mathcal{E}$ of $\mathcal{B}_{W_p^r(D)}$.
 This is similar to Theorem 3 in \cite{Hei01}, where quantum integration in $L_p([0,1]^d)$ was considered. 
In \cite{Hei01b} we developed a deterministic discretization technique which we also used in \cite{Hei04b}. Here we employ a quantum version of the  stochastic discretization technique, recently developed for studying parametric integration in the absence of the embedding condition \cite{Hei23b}. 

The paper is organized as follows. In Section \ref{sec:2} we give a brief introduction into the quantum setting of IBC. Furthermore, some general auxiliary results are recalled and new ones are shown. Section \ref{sec:3} introduces parametric integration, with the special cases of approximation and integration. After a short review of the stochastic discretization technique for parametric integration of \cite{Hei23b}, we introduce and study  a quantum version of it. The analysis of full parametric integration is convenient since it allows a unified treatment of the two special cases considered later. The main results are contained in Sections \ref{sec:4} (approximation) and \ref{sec:5} (integration).  The final Section \ref{sec:6} summarizes the results in tables and provides comparisons to the classical settings. Our results extend (and recover) previous ones from \cite{Hei01b} and \cite{Hei04b}.

\section{Preliminaries}\label{sec:2}
We denote ${\mathbb{N}}=\{1,2,\dots\}$,  ${\mathbb{N}}_0={\mathbb{N}}\cup\{0\}$. The symbol ${\mathbb{K}}$ stands for the scalar field ${\mathbb R}$ or ${\mathbb C}$.
We often use the same symbol
$c, c_0,c_1,\dots$ for possibly different constants, even if they appear in a sequence
of relations. However, some constants are supposed to have the same meaning throughout the paper --  these are denoted (or renamed later) by symbols $c(0),c(1),\dots$.
The unit ball of a normed space $G$ is denoted by $\mathcal{B}_G$,  the space of continuous linear functionals on $G$ by $G^*$, and  $\log$ always means $\log_2$.

To emphasize the essential parts 
of the estimates we use the following notation. For functions 
$a,b:\N\to [0,\infty)$, we write 
$a(n)\asymp_{\log} b(n)$ if there are constants 
$c_1,c_2>0$, $n_0\in \N$, $\alpha_1,\alpha_2\in \R$ such that 
\begin{equation}
\label{AB3}
c_1(\log(n+1))^{\alpha_1}b(n)\le a(n)\le c_2(\log(n+1))^{\alpha_2}b(n)
\end{equation}
for all 
$n\in \N$ with $n \ge n_0$. We write $a(n)\asymp b(n)$ iff \eqref{AB3} holds with $\alpha_1=\alpha_2=0$.

In the following we give a short introduction into the quantum setting of information-based complexity theory, more detailed explanations can be found e.g. in \cite{Hei01} and \cite{Hei05}.
A numerical problem is given by a tuple  
\begin{equation}
\label{AP3}
\mathcal{P}=(F,G,S,K,\Lambda),
\end{equation}
where
$F$ is a non-empty set, $G$ a normed space over $\K$, where $\K$ 
stands for the set of real or complex numbers, $S$ a mapping from $F$ to $G$, $K$ a non-empty
set and $\Lambda$ a non-empty set of mappings from $F$ to $K$.   We seek to approximate $S(f)$ for $f\in F$ 
by means of quantum computation. 
Usually $F$ is a set in a function space, $S$ is the solution
operator, which maps the input $f\in F$ to the exact solution $S(f)$.
$\Lambda$ is usually a set of linear functionals, supplying information $\lambda(f)$ about $f$ through which the algorithm 
can access the input $f$. $K$ is mostly $\K$, $G$ is a space 
containing both the solutions and the approximations, usually also a space of functions,  and the error is measured in the norm of $G$. 

In the sequel it will be convenient to consider $f\in F$ also as a function on $\Lambda$ with values in $K$
by setting $f(\lambda):=\lambda(f)$. Let $\mathcal{F}(\Lambda,K)$ denote the set of all functions from $\Lambda$ to $K$. 

A problem $\mathcal{P}=(F,G,S,K,\Lambda)$ is called linear, if  $K=\K$, $F$ is an absolutely convex subset of a linear space $F_1$ over $\K$, $S$ is the restriction to $F$ of a linear operator $S_1$ from $F_1$ to $G$, and each $\lambda\in \Lambda$ is the restriction to $F$ of a an element of $F_1^\#$, the algebraic dual of $F_1$, that is, the space of linear functionals on $F_1$.

For some of the definitions and statements below tuples of the form 
\begin{equation}
\label{AP4}
\mathcal{P}'=(F,G,K,\Lambda),\quad \text{or}\quad \mathcal{P}''=(F,K,\Lambda)
\end{equation}
with $F,G,K,\Lambda$ as in \eqref{AP3} suffice.

Let $H_1$ be the 
two-dimensional complex Hilbert space $\C^2$, with its unit vector
basis $\{e_0,e_1\}$, let
$$
 H_m=\underbrace{H_1\otimes\dots\otimes H_1}_{m}, 
$$
equipped with the tensor
Hilbert space structure. 
Denote
$\Z[M,N) := \{M,\dots,N-1\}$
for $M, N\in\N_0$ with $M\le N$, and correspondingly $\Z[M,N]$.
Let $\mathcal{C}_m = \{\lg i\rs:\, i\in\Z[0,2^m)\}$ be the canonical basis of
$H_m$, where  $\lg i \rs$ stands for 
$e_{j_0}\otimes\dots\otimes e_{j_{m-1}}$ with $i=\sum_{k=0}^{m-1}j_k2^{m-1-k}$.
 Let $\mathcal{U}(H_m)$ denote the set of unitary operators on $H_m$.

A quantum query  on $\mathcal{P}''$ is given by a tuple
$$
Q=(m,m',m'',Z,\tau,\beta),
$$
where $m,m',m''\in \N, m'+m''\le m, Z\subseteq \Z[0,2^{m'})$ is a nonempty 
subset, and $\tau:Z\to \Lambda$ and 
$\beta:K\to\Z[0,2^{m''})$
are arbitrary mappings. Let $m(Q):=m$ denote the number of qubits $Q$ is acting on. 

Given a query $Q$, we define for each $f\in F$ the unitary operator 
$Q_f\in\mathcal{U}(H_m)$ by setting for  
$\lg i\rs\lg x\rs\lg y\rs\in \mathcal{C}_m
=\mathcal{C}_{m'}\otimes\mathcal{C}_{m''}\otimes\mathcal{C}_{m-m'-m''}$:
\begin{equation}
\label{Z6}
Q_f\lg i\rs\lg x\rs\lg y\rs=
\left\{\begin{array}{ll}
\lg i\rs\lg x\oplus\beta(f(\tau(i)))\rs\lg y\rs &\quad \mbox {if} \quad i\in Z\\
\lg i\rs\lg x\rs\lg y\rs & \quad\mbox{otherwise,} 
 \end{array}
\right. 
\end{equation}
where $\oplus$ means addition modulo $2^{m''}$. 

A quantum algorithm on $\mathcal{P}''$  with no measurement is a tuple $A=(Q,(X_j)_{j=0}^n)$
where $Q$ is a quantum query on $\mathcal{P}''$, $n\in\N_0$  and
$X_{j}\in \mathcal{U}(H_m)\,(j=0,\dots,n)$, with $m=m(Q)$.
Given $f\in F$,
we define $A_f\in \mathcal{U}(H_m)$ as
\begin{equation}
\label{Z7}
A_f = X_n Q_f X_{n-1}\dots X_1 Q_f X_0.
\end{equation}
We denote by $n_\q(A):=n$ the number of queries and by $m(A)=m=m(Q)$ the 
number of qubits of $A$. Let $(A_f(x,y))_{x,y\in \Z[0,2^m)}$ be the matrix of the 
transformation $A_f$ in the canonical basis $\mathcal{C}_{m}$, that is,

\begin{equation}
\label{Z8}
A_f\lg y\rs=\sum_{x\in\Z[0,2^m)}A_f(x,y)\lg x\rs.
\end{equation}

A quantum algorithm  on ${\mathcal{P}}'$ with $k$ measurements  is a tuple 
$A=((A_l)_{l=0}^{k-1},(h_l)_{l=0}^{k-1},\varphi)$,
\begin{equation*}
A=((A_l)_{l=0}^{k-1},(h_l)_{l=0}^{k-1},\phi),\quad A_l=(Q_l,
(X_{l j})_{j=0}^{n_l}), \quad Q_l=(m_{l},m_{l}',m_l'',Z_l,\tau_l,\beta_l),
\end{equation*}
where $k\in\N$, $A_l=(Q_l,
(X_{l j})_{j=0}^{n_l})$ with $Q_l=(m_{l},m_{l}',m_l'',Z_l,\tau_l,\beta_l)\;(l=0,\dots,k-1)$ are quantum algorithms
on $\mathcal{P}'$ with no measurements, and
$$
h_0\in\Z[0,2^{m_0}),\quad h_l:\prod_{i=0}^{l-1}\Z[0,2^{m_i}) \to \Z[0,2^{m_l})\quad 
(1\le l \le k-1),\quad \varphi:\prod_{l=0}^{k-1}\Z[0,2^{m_l}) \to G.
$$
Given $f\in F$, the algorithm acts as follows. The starting state for $A_0$ is $\lg h_0\rs$, to which the sequence of unitary transformations of $A_0$ is applied,
\begin{equation*}
A_{0,f}\lg h_0\rs = X_{0,n_0} Q_{0,f} X_{0,n_0-1}\dots X_{0,1} Q_{0,f} X_{0,0}\lg h_0\rs.
\end{equation*}
Then the state 
$$
A_{0,f}\lg h_0\rs=\sum_{x\in\Z[0,2^{m_0})}A_{0,f}(x,h_0)\lg x\rs,
$$
see \eqref{Z8}, is measured, giving a classical state $\lg x_0\rs\in \mathcal{C}_{m_0}$ with probability $|A_{0,f}(x_0,h_0)|^2$. Next the starting state of $A_1$ is determined as $\lg h_1(x_0)\rs$ and after applying the unitary transforms of $A_1$ we reach 
$$
A_{1,f}\lg h_1(x_0)\rs=\sum_{x\in\Z[0,2^{m_1})}A_{1,f}(x,h_1(x_0))\lg x\rs,
$$
which is measured and gives  $\lg x_1\rs\in \mathcal{C}_{m_1}$ with probability $|A_{1,f}(x_1,h_1(x_0))|^2$. This goes on until finally $A_{k-1,f}$ is applied to the starting state $\lg h_{k-1}(x_0,\dots,x_{k-2})\rs$, leading to 
$$
A_{k-1,f}\lg h_{k-1}(x_0,\dots,x_{k-2})\rs=\sum_{x\in\Z[0,2^{m_{k-1}})}A_{k-1,f}(x,h_{k-1}(x_0,\dots,x_{k-2}))\lg x\rs,
$$
which after measurement gives $\lg x_{k-1}\rs\in \mathcal{C}_{m_{k-1}}$ with probability $$
|A_{k-1,f}(x_{k-1},h_{k-1}(x_0,\dots,x_{k-2}))|^2
.$$ 
From the measurement results, the output of $A$ at input $f\in F$ is determined to be 
$$
\phi(x_0,\dots,x_{k-1})\in G.
$$
More formally, the output of $A$ at input $f\in F$ will be a probability measure $A(f)$ on $G$, 
defined as follows: First put for $x_l\in\Z[0,2^{m_l})$
\begin{eqnarray}   
p_{A,f}(x_0,\dots, x_{k-1})&=&
|A_{0,f}(x_0,h_0)|^2 |A_{1,f}(x_1,h_1(x_0))|^2\dots\nonumber\\
&&\dots |A_{k-1,f}(x_{k-1},h_{k-1}(x_0,\dots,x_{k-2}))|^2.\label{B8}
\end{eqnarray}
Then define $A(f)$ by setting for any subset $C\subseteq G$
\begin{equation}
A(f)(C)=\sum_{\phi(x_0,\dots,x_{k-1})\in C}p_{A,f}(x_0,\dots, x_{k-1}).\label{B9}
\end{equation}
Let 
$n_\q(A):=\sum_{l=0}^{k-1} n_\q(A_l)$
denote the number of queries used by $A$.
For more details and background see \cite{Hei01}.
Below we  use the term `quantum algorithm', 
meaning a quantum algorithm with measurement(s). If $S$ is given, as well, we also speak of a quantum algorithm for $\mathcal{P}$. On the other hand, we also use the term quantum algorithm from $F$ to $G$, if $K$ and $\Lambda$ are well-understood.  

\begin{remark} \label{rem:1} A quantum algorithm $A$ on $(F,G,K,\Lambda)$ can also be considered as one on \newline
$(\mathcal{F}(\Lambda,K),G,K,\Lambda)$
(identifying $\Lambda$ with the set of Dirak functionals $\{\delta_{\lambda}:\,\lambda\in\Lambda\}$ on $\mathcal{F}(\Lambda,K)$). Furthermore, if  $F_0\subset F$, then $A$ is also a quantum algorithm on $(F_0,G,K,\Lambda)$ (considering each $\lambda\in\Lambda$ as a mapping on $F_0$). For example, if $X$ is a linear space, $F\subset X$, and $\Lambda\subset X^\#$ is a subset of the space of linear functionals on $X$, then a quantum algorithm on $(X,G,K,\Lambda)$ defines also a quantum algorithm on $(F,G,K,\Lambda)$. In particular, if $S$ is also defined on the whole space $X$ (here we do not distinguish between the operator and its restriction to $F$), we may define a quantum algorithm on the whole space $X$ while having in mind the problem $(F,S,G,K,\Lambda)$. On the other hand, the r\^ole of $F$ itself is of course crucial for the definition of error given next.  
\end{remark}

For $\theta\ge 0$ and a quantum algorithm $A$ for $\mathcal{P}$ we 
define the probabilistic error at $f\in F$ as follows. Let $\zeta$ be a random variable with
distribution $A(f)$. Then
\begin{equation}
e(S,A,f,G,\theta)=\inf\left\{\varepsilon\ge0\,\,|\,\,\Prm\{\|S(f)-\zeta\|>\varepsilon\}\le\theta
\right\}
\label{Z5}
\end{equation}
and the infimum is always attained. Let
$$
e(S,A,F,G,\theta)=\sup_{f\in F} e(S,A,f,G,\theta) 
$$
(this quantity can take the value $+\infty$). Furthermore, we set
\begin{eqnarray*}
e_n^\q(S,F,G,\theta)
&=&\inf\{e(S,A,F,G,\theta)\,\,|\,\,A\,\,
\mbox{is any quantum algorithm for $\mathcal{P}$ with}\,\, n_\q(A)\le n\}.
\end{eqnarray*}
 We denote
$$
e(S,A,f,G)=e(S,A,f,G,1/4)
$$
and similarly,
$$
e(S,A,F,G)=e(S,A,F,G,1/4),\quad e_n^\q(S,F,G)=e_n^\q(S,F,G,1/4).
$$
The quantity $e_n^\q(S,F,G)$ is the 
$n$-th minimal query error, 
that is, the smallest error which can be reached using at most $n$ queries.
Note that it essentially suffices to study
$e_n^\q(S,F,G)$ instead of $e_n^\q(S,F,G,\theta)$, 
since with $\mathcal{O}(\nu)$ repetitions, the error probability 
can be reduced to $2^{-\nu}$, see Lemma \ref{lem:2f} below.

Let us recall from \cite{Hei01} the definition of the composition of quantum algorithms $A_\iota$ on tuples of type \eqref{AP4}
$$
\mathcal{P}'_\iota=(F,G_\iota,K,\Lambda)\quad(\iota=0,\dots,M-1),
$$  
where $M\in\N$ and $G, G_\iota\; (\iota=0,\dots,M-1)$ are normed spaces over $\K$. 
Let
$$
\psi:G_0\times\dots\times G_{M-1} \to G
$$
be any mapping. We define the composed 
algorithm $A=\psi (A_0,\dots,A_{M-1})$ on $\mathcal{P}'=(F,G,K,\Lambda)$
as follows: Let 
$$
A_\iota=((A_{\iota,l})_{l=0}^{k_\iota-1},(h_{\iota,l})_{l=0}^{k_\iota-1},\varphi_\iota),
$$
put $k=\sum_{\iota=0}^{M-1} k_\iota$, let the set
$$
\Upsilon=\{ (\iota,l)\,\,|\,\, \iota=0,\dots,M-1,\, l=0,\dots,k_\iota-1\}
$$
be equipped with the lexicographical order, and put 
$m_{\iota,l}=m(A_{\iota,l})$ for $(\iota,l)\in \Upsilon$. 
Define
$$
h'_{\iota,l}:\prod_{i=0}^{\iota-1}\prod_{j=0}^{k_i-1}\Z[0,2^{m_{i,j}})\times
\prod_{j=0}^{l-1}\Z[0,2^{m_{\iota,j}}) \to \Z[0,2^{m_{\iota,l}}),
$$
as
$$
h'_{\iota,l}(x_{0,0},\dots,x_{\iota-1,k_{\iota-1}-1},x_{\iota,0},\dots,x_{\iota,l-1})
=h_{\iota,l}(x_{\iota,0},\dots,x_{\iota,l-1}).
$$
Furthermore, let 
$$
\phi=\psi\circ(\phi_0\times\dots,\times\phi_{M-1}),
$$
thus 
\begin{equation*}
\phi:\prod_{\iota=0}^{M-1}\prod_{j=0}^{k_\iota-1}\Z[0,2^{m_{\iota,j}})
\overset{\phi_0\times\dots,\times\phi_{M-1}}{\longrightarrow} \prod_{\iota=0}^{M-1}G_\iota\overset{\psi}{\longrightarrow}G.
\end{equation*}
Then we set
$$
\psi (A_0,\dots,A_{M-1})=((A_\upsilon)_{\upsilon\in\Upsilon},
(h'_\upsilon)_{\upsilon\in\Upsilon},\phi).
$$

Let $\mu_\iota\in \mathscr{M}_0(G_\iota)$ $(\iota=0,\dots,M-1)$, where  $\mathscr{M}_0(G_\iota)$ denotes the set of all probability measures on $G_\iota$ whose
support is a finite set. We define 
$\psi(\mu_0,\dots,\mu_{M-1})\in \mathscr{M}_0(G)$ to be the measure induced by 
$\mu_0\times\dots\times\mu_{M-1}$
via $\psi$ on $G$, that is, for any subset $C\subseteq G$,
$$
\psi(\mu_0,\dots,\mu_{M-1})(C)=(\mu_0\times\dots\times\mu_{M-1})(\psi^{-1}(C)).
$$

The following is Lemma 2 of \cite{Hei01}. 
\begin{lemma}
\label{lem:2a}
For each $f\in F$,
$$ 
\psi (A_0,\dots,A_{M-1})(f)=\psi (A_0(f),\dots, A_{M-1}(f)),
$$
or stated equivalently, if  $(\zeta_i)_{i=0}^{M-1}$ are independent random 
variables with distribution $A_i(f)$ respectively, then  
$$
\psi (A_0,\dots,A_{M-1})(f)={\rm dist}(\psi(\zeta_0,\dots,\zeta_{M-1})).
$$
Moreover, 
$$
n_\q(\psi (A_0,\dots, A_{M-1}))=\sum_{i=0}^{M-1}n_\q(A_i).
$$
\end{lemma}
Given a quantum algorithm $A$ on $(F,G,K,\Lambda)$ and a mapping $\psi$ from
$G$ to $\wt{G}$, the algorithm $\psi(A)$ is defined by replacing 
$\phi$ in the definition of $A$  by $\psi\circ\phi$ -- this is the special case $M=1$
of  the composition above.   Recall that $\psi$ 
is said to be Lipschitz, if there is a constant $c\ge 0$ such that
$$
\|\psi(x)-\psi(y)\|_{\wt{G}}\le c\,\|x-y\|_G \quad (x,y\in G).
$$ 
The Lipschitz constant $\|\psi\|_{\rm Lip}$ is the smallest constant $c$ such that
the relation above holds. The  next result is Lemma 1 of \cite{Hei04a}.
\begin{lemma}
\label{lem:7}
Let  $A$ be a quantum algorithm on $\mathcal{P}=(F,G,S,K,\Lambda)$. Let $\psi$
be a Lipschitz mapping from
$G$ to $\wt{G}$. Then for each $f\in F$ and $\theta\ge 0$,
$$
e(\psi\circ S,\psi(A),f,\wt{G},\theta)\le \|\psi\|_{\rm Lip} e(S,A,f,G,\theta).
$$
Consequently, for each $n\in\N$,
$$
e_n^\q(\psi\circ S,F,\wt{G},\theta)\le \|\psi\|_{\rm Lip} e_n^\q(S,F,G,\theta).
$$
\end{lemma}

For the following we refer to Lemma 2 of \cite{Hei04a}.
\begin{lemma}
\label{lem:2b}
Let $\mathcal{P}'=(F,G,K,\Lambda)$  be as in \eqref{AP3}, $k\in\N_0$, 
let $S_l:F\to G$ $(l=0,\dots,k)$ be any mappings and  $A_l$  quantum algorithms for $\mathcal{P}'$. Define $S:F\to G$ by
$S(f)=\sum_{l=0}^k S_l(f)\;(f\in F)$ and let $A=\sum_{l=0}^k A_l$ be the composed algorithm according to the definition above. 
Then for $\theta_0,\dots,\theta_k\ge 0$
$$
e(S,A,F,G,\sum_{l=0}^k \theta_l)\le\sum_{l=0}^k e(S_l,A_l,F,G,\theta_l).
$$
\end{lemma}
Let  $\nu\in \N$ and let 
$\mathfrak{m}:\R^\nu\to \R$ be the mapping given by the median, that is, 
$\mathfrak{m}(a_0,\dots,a_{\nu-1})$ is the value 
of the $\lceil (\nu+1)/2\rceil$-th element of the
non-decreasing rearrangement of $(a_i)$. For the following construction
we consider $G$ as a space over $\R$ (each normed space over $\C$ can also be considered as
a normed space over $\R$).  Let 
$\mathcal{T}\subseteq \mathcal{B}_{G^*}$
be a norming set, that is,
\begin{equation*}
\|g\|=\sup_{t\in \mathcal{T}}|t(g)|\quad(g\in G).
\end{equation*}
Then $G$ can be identified with a subspace 
of $\ell_\infty(\mathcal{T})$,  the space of all bounded real-valued functions 
on $\mathcal{T}$, endowed with the supremum norm 
$\|g\|_{\ell_\infty(\mathcal{T})}=\sup_{t\in \mathcal{T}}|g(t)|$, via the embedding map $J:G\to \ell_\infty(\mathcal{T})$ 
defined by $Jg=(t(g))_{t\in \mathcal{T}}$.
This mapping preserves the norm: $\|Jg\|_{\ell_\infty(\mathcal{T})}=\|g\|_G$. Let   $\bar{\mathfrak{m}}:\ell_\infty(\mathcal{T})^\nu\to \ell_\infty(\mathcal{T})$ be the component-wise application of the median, that is,
$$
\bar{\mathfrak{m}}(g_0,\dots,g_{\nu-1})=(\mathfrak{m}(g_0(t),\dots,g_{\nu-1}(t)))_{t\in \mathcal{T}}.
$$
Let $\delta>0$ and let  $\pi_\delta:\ell_\infty(\mathcal{T})\to G$ be such that 
$$
\|x-\pi_\delta(x)\|_{\ell_\infty(\mathcal{T})}
\le (1+\delta)\inf_{g\in G}\|x-g\|_{\ell_\infty(\mathcal{T})}\quad (x\in \ell_\infty(\mathcal{T})).
$$
We define $\sigma_\delta:G^\nu\to G$ by setting
\begin{equation}
\label{AP2}
\sigma_\delta=\pi_\delta\circ\bar{\mathfrak{m}}\circ J^\nu.
\end{equation}
The next two results are Lemma 4 and Corollary 2 of \cite{Hei04a}.
\begin{lemma}
\label{lem:2f}
Let $\mathcal{P}=(F,G,S,K,\Lambda)$ and $A$ any quantum algorithm for $\mathcal{P}$, let $\nu\in\N$, $\delta>0$, and let $\sigma_\delta(A^\nu)$ be the composition of $\nu$ copies of $A$ in the sense defined above. 
Then $n_\q(\sigma_\delta(A^\nu))=\nu n_\q(A)$ and for each $f\in F$,
$$
e(S,\sigma_\delta(A^\nu),f,G,e^{-\nu/8})\le (2+\delta)e(S,A,f,G).
$$
\end{lemma}
\begin{lemma}
\label{lem:1}
Let $\mathcal{P}'=(F,G,K,\Lambda)$, $k\in\N_0$
and $S_l:F\to G$ $(l=0,\dots,k)$, $S=\sum_{l=0}^k S_l$ be as in Lemma \ref{lem:2b}. 
 Assume $\nu_0,\dots,\nu_k\in \N$ satisfy
$$
\sum_{l=0}^k e^{-\nu_l/8}\le \frac{1}{4}.
$$ 
Let $n_0,\dots,n_k\in\N_0$ and put
$n=\sum_{l=0}^k \nu_l n_l$. Then
$$
e_n^\q(S,F,G)\le2\sum_{l=0}^k e_{n_l}^\q(S_l,F,G).
$$
If $G=\ell_\infty(\mathcal{T})$, then the relation holds with
constant $1$.
\end{lemma}

Let $\mathcal{P}'=(F,G,K,\Lambda)$ and $\wt{\mathcal{P}}'=(\wt{F},G,\wt{K},\wt{\Lambda})$
be tuples of the form \eqref{AP4}. We are now concerned with reducing problems based on $\mathcal{P}'$ to problems on  $\wt{\mathcal{P}}'$. First we consider a simple situation. Let $\Gamma:F\to\wt{F}$ be a mapping (the reduction mapping) such that there are $\eta: \wt{\Lambda}\to \Lambda$ and $\rho:K\to\wt{K}$ with 
\begin{equation}
\label{AU0}
(\Gamma (f))(\wt{\lambda})=\rho\circ f\circ\eta(\wt{\lambda})\quad (f\in F,\wt{\lambda}\in \wt{\Lambda}).
\end{equation}
The following was shown in \cite{Hei01}, Lemma 4 and Corollary 1.
\begin{lemma}
\label{lem:9}
Given $\mathcal{P}'=(F,G,K,\Lambda)$ and $\wt{\mathcal{P}}'=(\wt{F},G,\wt{K},\wt{\Lambda})$
of the form \eqref{AP4}, a mapping $\Gamma:F\to \wt{F}$ satisfying \eqref{AU0}, and a quantum algorithm $\wt{A}$ for  $\wt{\mathcal{P}}'$, 
$$
\wt{A}= ((\wt{A}_l)_{l=0}^{k-1},(\wt{h}_l)_{l=0}^{k-1},\wt{\phi}),\quad \wt{A}_l=(\wt{Q}_l,(\wt{X}_{lj})_{j=0}^{\wt{n}_l}),\quad \wt{Q}_l=(\wt{m}_l,\wt{m}_l',\wt{m}_l'',\wt{Z}_l,\wt{\tau}_l,\wt{\beta}_l),
$$
let  $A$ be the algorithm for $\mathcal{P}'$ which results from $\wt{A}$ by replacing the query $\wt{Q}_l$ by
\begin{equation*}
Q_l=(\wt{m}_l,\wt{m}_l',\wt{m}_l'',\wt{Z}_l,\tau_l,\beta_l),
\end{equation*}
where $\tau_l=\eta \circ\wt{\tau}_l$ and $\beta_l=\wt{\beta}_l\circ\rho$. Then $n_\q(A)=n_\q(\wt{A})$ and 
\begin{equation*}
A(f)=\wt{A}(\Gamma(f))\quad (f\in F).
\end{equation*}
Consequently, if $\wt{S}:\wt{F}\to G$ is any mapping and $S=\wt{S}\circ\Gamma$, 
then for each $n\in \N_0$
\begin{equation}
\label{AU1}
e_n^\q(S,F,G)\le e_n^\q(\wt{S},\wt{F},G).
\end{equation}
\end{lemma} 

Now we study a more complicated case of reduction. With $\mathcal{P}'$ and $\wt{\mathcal{P}}'$ as above let
$\Gamma:F\to \wt{F}$  be of the following type. There are $\kappa,m^*\in\N$ 
and mappings
\begin{eqnarray}
\eta_j&:& \wt{\Lambda}\to \Lambda \quad (j=0,\dots,\kappa-1)\label{AC5}\\
\beta&:& K\to\Z\big[0,2^{m^*}\big)\label{AC6}\\
\rho&:& \wt{\Lambda}\times\Z\big[0,2^{m^*}\big)^\kappa \to \wt{K}\label{AC7}
\end{eqnarray}
such that for $f\in F$ and $\wt{\lambda}\in \wt{\Lambda}$
\begin{equation}
\label{F1}
(\Gamma (f))(\wt{\lambda})=\rho(\wt{\lambda},\beta\circ f\circ\eta_0(\wt{\lambda}),\dots,\beta\circ f\circ\eta_{\kappa-1}(\wt{\lambda})).
\end{equation}
Here we first recall Lemma 1 of \cite{Hei01b}, just that some details of the proof given there are now included into the statement. 
\begin{lemma}
\label{lem:3}
For each quantum query $\wt{Q}=(\wt{m},\wt{m}',\wt{m}'',\wt{Z},\wt{\tau},\wt{\beta})$ on $\wt{\mathcal{P}''}$ and each mapping $\Gamma$ of the above 
form (\ref{AC5})--(\ref{F1})
there is a quantum algorithm without measurement $B$ on $\mathcal{P}''$
$$
B=(Q,
(Y_j)_{j=0}^{2\kappa}), \quad Q=(m,m',m'',Z,\tau,\beta),
$$
where 
\begin{eqnarray}
&&\kappa_0=\lceil\log \kappa\rceil,\quad     
m=\wt{m}+\kappa_0+\kappa m^*,\quad m'=\wt{m}'+\kappa_0, \quad
m''=m^*,
\label{AN1}\\
&&Z=\wt{Z}\times\Z[0,\kappa),\quad
\tau(i,j)=\eta_j (\wt{\tau}(i)) \quad\mbox{for}\quad (i,j)\in Z,
\label{AN2}
\end{eqnarray}
and for all $f\in F$ and all $\lg x\rs\in H_{\wt{m}}$,
\begin{equation}
\label{AN3}
\big(\wt{Q}_{\Gamma(f)}\lg x\rs\big)\lg 0\rs_{m-\wt{m}}=B_f\lg x\rs\lg 0\rs_{m-\wt{m}}.
\end{equation}
\end{lemma}

The following counterpart of Lemma \ref{lem:9} is essentially Corollary 1 of \cite{Hei01b}, a proof of which for the case $\kappa=1$ is contained in \cite{Hei01}. Again we need a more detailed version. For the sake of completeness, here the proof for arbitrary $\kappa$ is given.
\begin{lemma}
\label{lem:2}
Given $\mathcal{P}'=(F,G,K,\Lambda)$ and $\wt{\mathcal{P}}'=(\wt{F},G,\wt{K},\wt{\Lambda})$
of the form \eqref{AP4}, a mapping $\Gamma:F\to \wt{F}$ as in \eqref{AC5}--\eqref{F1}, and a quantum algorithm $\wt{A}$ for  $\wt{\mathcal{P}}'$, 
$$
\wt{A}= ((\wt{A}_l)_{l=0}^{k-1},(\wt{h}_l)_{l=0}^{k-1},\wt{\phi}),\quad \wt{A}_l=(\wt{Q}_l,(\wt{X}_{lj})_{j=0}^{\wt{n}_l}),\quad \wt{Q}_l=(\wt{m}_l,\wt{m}_l',\wt{m}_l'',\wt{Z}_l,\wt{\tau}_l,\wt{\beta}_l),
$$
there is a quantum algorithm $A$ for $\mathcal{P}'$
\begin{equation*}
A=((A_l)_{l=0}^{k-1},(h_l)_{l=0}^{k-1},\phi),\quad A_l=(Q_l,
(X_{l j})_{j=0}^{n_l}), \quad Q_l=(m_{l},m_{l}',m_l'',Z_l,\tau_l,\beta_l),
\end{equation*}
such that with  $\kappa_0=\lceil\log_2 \kappa\rceil$
\begin{eqnarray}
&&
m_l=\wt{m}_l+\kappa_0+\kappa m^*,\quad m_l'=\wt{m}_l'+\kappa_0, \quad
m_l''=m^*,
\label{AM7}\\
&&Z_l=\wt{Z}_l\times\Z[0,\kappa),\quad
\tau_l(i,j)=\eta_j (\wt{\tau}_l(i)) \quad\mbox{for}\quad (i,j)\in Z_l,\quad \beta_l=\beta,
\label{AM8}\\
&&n_l=2\kappa \wt{n}_l, \quad  n_\q(A)=2\kappa\,n_\q(\wt{A}),
\label{AM9} 
\end{eqnarray}
and for all $f\in F$
\begin{equation}
A(f)=\wt{A}(\Gamma(f)).\label{B7}
\end{equation}
Consequently, if $\wt{S}:\wt{F}\to G$ is any mapping and $S=\wt{S}\circ\Gamma$, 
then for each $n\in \N_0$
\begin{equation}
\label{AC4}
e_{2\kappa n}^\q(S,F,G)\le e_n^\q(\wt{S},\wt{F},G).
\end{equation}
\end{lemma} 
\begin{proof} 
Applying Lemma \ref{lem:3} to $\wt{Q}_l$, we obtain 
\begin{equation}
\label{AN4}
B_l=(Q_l,(Y_{lj})_{j=0}^{2\kappa}), \quad Q_l=(m_l,m_l',m_l'',Z_l,\tau_l,\beta),
\end{equation}
such that $Q_l$ satisfies \eqref{AM7} and \eqref{AM8} and
\begin{equation}
\label{AP0}
\big(\wt{Q}_{l,\Gamma(f)}\lg x\rs\big)\lg 0\rs_{m_l-\wt{m}_l}=B_{l,f}\lg x\rs\lg 0\rs_{m_l-\wt{m}_l}\quad \big(\lg x\rs\in H_{\wt{m}_l}\big).
\end{equation}
Furthermore, we set  $\lg h_0\rs =\big|\,\wt{h}_0\big> \lg 0\rs_{{m_l-\wt{m}_l}}$
and define for $1\le l\le k-1$ the mappings 
$$
h_l:\prod_{i=0}^{l-1}\big(\Z[0,2^{\wt{m}_i})\times\Z[0,2^{{m_i-\wt{m}_i}})\big) 
\to \Z[0,2^{\wt{m}_l})\times\Z[0,2^{{m_l-\wt{m}_l}}),
$$
by
$$
h_l((x_0,y_0),\dots,(x_{l-1},y_{l-1}))
=\big(\wt{h}_l(x_0,\dots,x_{l-1}),0\big).
$$
and
$$
\varphi:\prod_{l=0}^{k-1}\big(\Z[0,2^{\wt{m}_l})\times\Z[0,2^{{m_l-\wt{m}_l}})\big) \to G,
$$
by
\begin{equation}
\label{AN9}
\phi((x_0,y_0),\dots,(x_{k-1},y_{k-1}))=\wt{\phi}(x_0,\dots,x_{k-1}).
\end{equation}
To complete the definition of algorithm $A$ we have to specify the $(X_{l,j})_{j=0}^{2\kappa \wt{n}_l}$. For this purpose we note that for $f\in F$, $0\le l<k$,
\begin{equation}
\label{AN0}
\wt{A}_{l,\Gamma(f)} = \wt{X}_{l \wt{n}_l} \wt{Q}_{l,\Gamma(f)} \wt{X}_{ l,\wt{n}_l-1}\dots 
\wt{X}_{l,1} \wt{Q}_{l,\Gamma(f)} \wt{X}_{l,0}
\end{equation}
and by \eqref{AN4}
\begin{equation}
\label{AN5}
B_{l,f}=Y_{l,2\kappa}Q_{l,f}Y_{l,2\kappa-1}Q_{l,f}\dots Y_{l,1}Q_{l,f}Y_{l,0}.
\end{equation}
Setting  $\bar{X}_{lj}=\wt{X}_{lj}\otimes Id_{H_{{m_l-\wt{m}_l}}}$, where $Id_{H_{{m_l-\wt{m}_l}}}$
is the identity on $H_{{m_l-\wt{m}_l}}$, we conclude from \eqref{AN3}, \eqref{AP0}, \eqref{AN0}, and \eqref{AN5} for $\lg x\rs\in H_{\wt{m}_l}$
\begin{eqnarray}
\big(\wt{A}_{l,\Gamma(f)}\lg x\rs\big)\lg 0\rs_{{m_l-\wt{m}_l}}
&=&\left(\wt{X}_{l \wt{n}_l} \wt{Q}_{l,\Gamma(f)} \wt{X}_{ l,\wt{n}_l-1}\dots 
\wt{X}_{l,1} \wt{Q}_{l,\Gamma(f)} \wt{X}_{l,0}\lg x\rs\right)\lg 0\rs_{{m_l-\wt{m}_l}}
\notag\\
&=&\bar{X}_{l \wt{n}_l}\left(\wt{Q}_{l,\Gamma(f)} \wt{X}_{ l,\wt{n}_l-1}\dots 
\wt{X}_{l,1} \wt{Q}_{l,\Gamma(f)} \wt{X}_{l,0}\lg x\rs\right)\lg 0\rs_{{m_l-\wt{m}_l}}
\notag\\
&=&\bar{X}_{l \wt{n}_l} B_{l,f}\left( \wt{X}_{ l,\wt{n}_l-1}\dots 
\wt{X}_{l,1} \wt{Q}_{l,\Gamma(f)} \wt{X}_{l,0}\lg x\rs\right)\lg 0\rs_{{m_l-\wt{m}_l}}
\notag\\
&\dots&
\notag\\
&=&\bar{X}_{l \wt{n}_l} B_{l,f} \bar{X}_{ l,\wt{n}_l-1}\dots 
\bar{X}_{l,1} B_{l,f} \left(\wt{X}_{l,0}\lg x\rs\right)\lg 0\rs_{{m_l-\wt{m}_l}}   
\notag\\
&=&\bar{X}_{l \wt{n}_l} B_{l,f} \bar{X}_{ l,\wt{n}_l-1}\dots 
\bar{X}_{l,1} B_{l,f} \bar{X}_{l,0}\lg x\rs\lg 0\rs_{{m_l-\wt{m}_l}}   
\notag\\
&=&\bar{X}_{l \wt{n}_l}\underbrace{Y_{l,2\kappa}Q_{l,f}Y_{l,2\kappa-1}Q_{l,f}\dots Y_{l,1}Q_{l,f}Y_{l,0}}_{B_{l,f}}\bar{X}_{ l,\wt{n}_l-1}\dots
\notag\\
&&\dots \bar{X}_{l,1} \underbrace{Y_{l,2\kappa}Q_{l,f}Y_{l,2\kappa-1}Q_{l,f}\dots Y_{l,1}Q_{l,f}Y_{l,0}}_{B_{l,f}} \bar{X}_{l,0}\lg x\rs\lg 0\rs_{{m_l-\wt{m}_l}}.
\label{AN6}
\end{eqnarray}
Now we define for  $0\le j\le 2\kappa \wt{n}_l$ 
\begin{equation}
\label{AN7}
X_{l,j}=\left\{\begin{array}{lll}
  Y_{l,0}\bar{X}_{l,0} & \quad\mbox{if}\quad j=0   \\
  Y_{l,j_1} & \quad\mbox{if}\quad j=2\kappa j_0+j_1 \quad (0\le j_0 \le \wt{n}_l-1, \, 1\le j_1\le 2\kappa-1)\\
   Y_{l,0}\bar{X}_{l,j_0}Y_{l,2\kappa}& \quad\mbox{if}\quad j=2\kappa j_0 \quad (1\le j_0 \le \wt{n}_l-1)\\
  \bar{X}_{l,\wt{n}_l}Y_{l,2\kappa} & \quad\mbox{if}\quad j=2\kappa \wt{n}_l.
    \end{array}
\right. 
\end{equation}
Comparing \eqref{AN6} and \eqref{AN7}, we conclude
\begin{equation*}
\big(\wt{A}_{l,\Gamma(f)}\lg x\rs\big)\lg 0\rs_{{m_l-\wt{m}_l}}=A_{l,f}\lg x\rs\lg 0\rs_{{m_l-\wt{m}_l}}.
\end{equation*}
With  $x,y\in \Z[0,2^{\wt{m}_l})$ and  $z\in \Z[0,2^{{m_l-\wt{m}_l}})$ this implies (compare \eqref{Z8})
\begin{eqnarray*}
\wt{A}_{l,\Gamma(f)}(y,x)\ls z | 0\rs&=&\ls y \rg \wt{A}_{l,\Gamma(f)}\lg x\rs\ls z | 0\rs
=\ls z \rg\ls y \rg \big(\wt{A}_{l,\Gamma(f)}\lg x\rs\big)\lg 0\rs_{{m_l-\wt{m}_l}}
\\
&=&\ls z \rg\ls y \rg A_{l,f}\lg x\rs\lg 0\rs_{{m_l-\wt{m}_l}}=A_{l,f}((y,z),(x,0)),
\end{eqnarray*}
and therefore, by \eqref{B8},
\begin{eqnarray}
\lefteqn{ p_{A,f}((y_0,z_0)\dots,(y_{k-1},z_{k-1}))}
\nonumber\\ 
&=&
|A_{0,f}((y_0,z_0),h_0)|^2 |A_{1,f}((y_1,z_1),h_1((y_0,z_0)))|^2\dots
\nonumber\\
&&\dots |A_{k-1,f}((y_{k-1},z_{k-1}),h_{k-1}((y_0,z_0)\dots,(y_{k-2},z_{k-2})))|^2
\nonumber\\
&=&
|A_{0,f}((y_0,z_0),(\wt{h}_0,0))|^2 |A_{1,f}((y_1,z_1),(\wt{h}_1(y_0),0))|^2\dots
\nonumber\\
&&\dots |A_{k-1,f}((y_{k-1},z_{k-1}),\wt{h}_{k-1}(y_0,\dots,y_{k-2}),0\big))|^2
\nonumber\\
&=&
|\wt{A}_{0,\Gamma(f)}((y_0,\wt{h}_0))|^2 |\ls z_0 | 0\rs|^2|\wt{A}_{1,\Gamma(f)}(y_1,\wt{h}_1(y_0))|^2 |\ls z_1 | 0\rs|^2\dots
\nonumber\\
&&\dots |\wt{A}_{k-1,\Gamma(f)}((y_{k-1},z_{k-1}),\wt{h}_{k-1}(y_0,\dots,y_{k-2}))|^2|\ls z_{k-1} | 0\rs|^2
\nonumber\\
&=&\left\{\begin{array}{lll}
 p_{\wt{A},\Gamma(f)}(y_0,\dots,y_{k-1})  & \mbox{if} \quad z_0=\dots=z_{k-1}=0   \\
   0& \mbox{otherwise.}    \\
    \end{array}
\right. 
\label{AN8}
\end{eqnarray}
 Let $C\subseteq G$ be any subset. Then by \eqref{B9}, \eqref{AN9}, and \eqref{AN8}
\begin{eqnarray*}
A(f)(C)&=&\sum_{\phi((y_0,z_0)\dots,(y_{k-1},z_{k-1}))\in C}p_{A,f}((y_0,z_0)\dots,(y_{k-1},z_{k-1}))
\\
&=&\sum_{\phi((y_0,0)\dots,(y_{k-1},0))\in C}p_{A,f}((y_0,0)\dots,(y_{k-1},0))
\\
&=&\sum_{\wt{\phi}(x_0,\dots,x_{k-1})\in C}p_{\wt{A},\Gamma(f)}(y_0,\dots,y_{k-1})=\wt{A}(\Gamma(f))(C),
\end{eqnarray*}
which shows (\ref{B7}). To derive \eqref{AC4}, let $n\in\N$, $\delta>0$ and let $\wt{A}$ be a quantum algorithm for  $\wt{\mathcal{P}}'$ with $n^\q(A)\le n$ and 
\begin{equation}
\label{AU9}
e(\wt{S},\wt{A},\wt{F},G)\le e_n^\q(\wt{S},\wt{F},G)+\delta,
\end{equation}
and let $A$ be an algorithm for  $\mathcal{P}'$ satisfying \eqref{AM7}--\eqref{B7}. Let $\wt{S}:\wt{F}\to G$ be any mapping and $S=\wt{S}\circ\Gamma$. Then for $f\in F$
\begin{eqnarray*}
 \lefteqn{   A(f)\{g\in G:\,\|S(f)-g\|\le e(\wt{S},\wt{A},\wt{F},G) \} }
\\
&=&\wt{A}(\Gamma(f))\{g\in G:\,\|\wt{S}(\Gamma(f))-g\|\le e(\wt{S},\wt{A},\wt{F},G) \} \ge 3/4.
\end{eqnarray*}
Consequently, $e(S,A,f,G)\le e(\wt{S},\wt{A},\wt{F},G)$ and therefore, taking into account \eqref{AM9} and \eqref{AU9},
\begin{equation*}
e_{2\kappa n}^\q(S,F,G)\le \sup_{f\in F} e(S,A,f,G)\le e(\wt{S},\wt{A},\wt{F},G)\le e_n^\q(\wt{S},\wt{F},G)+\delta.
\end{equation*}
Relation \eqref{AC4} is an obvious consequence.

\end{proof}
Now we list some elementary properties, compare Lemma 6 of \cite{Hei01}.
\begin{lemma}
\label{lem:8}
Let $\mathcal{P}=(F,G,S,K,\Lambda)$ be as in \eqref{AP3}, $T:F\to G$ any mapping, and $n\in\N_0$.
Then the following hold:\\
(i)
$$
e_n^q(T,F,G)\le e_n^q(S,F,G)+\sup_{f\in F}\|T(f)-S(f)\|.
$$
(ii) For each $a\in\K$
$$
e_n^q(a S,F,G)=|a|e_n^q(S,F,G).
$$
(iii) If $K=\K$, $F$ is a subset of a linear space $F_1$ over $\K$, $S$ is the restriction to $F$ of a linear operator $S_1$ from $F_1$ to $G$, and each $\lambda\in \Lambda$ is the restriction to $F$ of a an element of $F_1^\#$, in particular, if $\mathcal{P}$ is a linear problem, then 
for all $a\in\K$
\begin{equation}
\label{AT11}
e_n^q(S,a F,G)=|a|e_n^q(S,F,G).
\end{equation}
\end{lemma}
\begin{proof}
Statement (i) is a simple consequence of the definitions and (ii) follows from Lemma \ref{lem:7}. Since (iii) slightly differs from the respective statement of Lemma 6 of \cite{Hei01}, we give the short proof. Trivially, for $a=0$ we have $e_n^q(S,{0},G)=0$. Now assume $a\ne 0$. We use Lemma \ref{lem:9}. So let $\wt{F}=a F$, $\Gamma:F\to \wt{F}$ be defined as
$\Gamma(f)=a f$, furthermore $\wt{\Lambda}=\Lambda$, $\wt{K}=\K$, $\eta=Id_{F_1^\#}$, $\rho(b)=ab\,(b\in\K)$. Using linearity, we get for $\lambda\in\Lambda$ 
\begin{equation*}
(\Gamma(f))(\lambda)=\lambda(af)=a\lambda(f)=af(\eta(\lambda))=\rho\circ f\circ \eta(\lambda),
\end{equation*}
so (\ref{AU0}) ifs satisfied. With $\wt{S}=a^{-1}S$ we have
$$
\wt{S}\Gamma=a^{-1} S\circ\Gamma=S,
$$
and hence, by Lemma \ref{lem:9} and statement (ii) above,
\begin{equation}
\label{AU2}
e_n^q(S,F,G)\le e_n^q(\wt{S},\wt{F},G)=e_n^q(a^{-1}S,aF,G)=|a|^{-1}e_n^q(S,a F,G).
\end{equation}
Replacing $F$  by $a_1 F$ and $a$ by $a_1^{-1}$, we get
$$
e_n^q(S,a_1 F,G)\le |a_1|e_n^q(S,F,G),
$$
which combined with \eqref{AU2} shows \eqref{AT11}.

\end{proof}

We need a simple relation to the classical deterministic setting, see e.g.\ \cite{Hei23a}, Section 1 for the definitions.
\begin{lemma}
\label{lem:5}
Let 
$$
A=((A_l)_{l=0}^{k-1},(h_l)_{l=0}^{k-1},\varphi), \quad A_l=(Q_l,(X_{l j})_{j=0}^{n_l}), \quad Q_l=(m_l,m'_l,m''_l,Z_l,\tau_l,\beta_l)
$$ 
be a quantum algorithm for $\mathcal{P}$ 
and let $0\le\theta<1/2$.
Then for each $\delta>0$ there is a classical deterministic algorithm $\wt{A}$ for $\mathcal{P}$ such that
\begin{eqnarray}
e(S,\wt{A},F,G)\le 2e(S,A,F,G,\theta)+\delta, \quad \ca(\wt{A})=\sum_{l=0}^{k-1}|Z_l|\le \sum_{l=0}^{k-1}2^{m_l'}. \nonumber
\end{eqnarray}
\end{lemma}
\begin{proof}
We define a classical deterministic non-adaptive algorithm 
$$
\wt{A}=\left((L_{l,i})_{0\le l<k, i\in Z_l},(t_{l,i})_{0\le l<k, i\in Z_l}, (\phi_{l,i})_{0\le l<k, i\in Z_l}\right)
$$ 
as follows. For $0\le l<k, i\in Z_l$, and   $f\in F$ set 
\begin{eqnarray*}
L_{l,i}(f)&=&f(\tau_l(i)),
\\[.2cm]
N(f)&:=&(f(\tau_l(i)))_{0\le l<k, i\in Z_l},
\\[.2cm]
t_{l,i}&=&\left\{\begin{array}{lll}
  1 & \quad\mbox{if}\quad l=k-1,\,i=\max Z_{k-1},   \\
  0 & \quad\mbox{otherwise,}    
    \end{array}
\right. 
\\[.2cm]
\phi_{l,i}&=&\begin{array}{lll}
  0 & \quad\mbox{if}\quad l<k-1\quad\mbox{or} \quad i<\max Z_{k-1}.   
    \end{array}
\end{eqnarray*}
Next we define the missing function $\phi_{k-1,\max Z_{k-1}}$. It suffices to do this on the set $N(F)$. So fix $f\in F$. Then according to \eqref{Z6}, $N(f)$ determines the matrices of the $Q_{l,f}$ $(0\le l<k)$. Since the $X_{lj}$ are known, by \eqref{Z7} and \eqref{Z8} the matrices of the operators $A_{l,f}$ are determined, as well. Furthermore, corresponding to \eqref{B8}, $p_{A,f}(x_0,\dots, x_{k-1})$ 
can be computed for all 
$(x_0,\dots, x_{k-1})\in \prod_{l=0}^{k-1}\Z[0,2^{m_l})$. This finally gives $A(f)(\{g\})$ for all 
\begin{equation*}
g\in G_0:=\left\{\phi(x_0,\dots,x_{k-1}):\, (x_0,\dots, x_{k-1})\in \prod_{l=0}^{k-1}\Z[0,2^{m_l})\right\},
\end{equation*}
by \eqref{B9}.
Note that $A(f)$ is a probability measure on $G_0$. Let $0<\delta\le 1-2\theta$. We define 
\begin{equation}
\label{AM2}
r=\min_{C\subseteq G_0,\, A(f)(C)\ge \theta+\delta} {\rm rad}(C),
\end{equation}
where ${\rm rad}$ means the radius. Since the minimum is taken over a finite set, it is attained on some $C_0$.  Let $y\in G$ be such that $\|y-z\|\le r+\delta$ for all $z\in C_0$. We define $y$ to be the output of $\wt{A}$ at input $f$, that is, 
\begin{equation*}
\wt{A}(f)=\phi_{k-1,\max Z_{k-1}}(N(f))=y.
\end{equation*}
Since $\theta+\delta\le 1-\theta$, it follows from definition \eqref{Z5} that 
\begin{equation*}
A(f)\{z\in G_0:\,\|S(f)-z\|\le e(S,A,f,G,\theta)\}\ge 1-\theta\ge \theta+\delta,
\end{equation*}
which together with \eqref{AM2} implies
\begin{equation*}
r\le e(S,A,f,G,\theta).
\end{equation*}
Moreover,  $ A(f)(C_0)\ge \theta+\delta$, therefore
\begin{equation}
\label{BA1}
C_0\cap \{z\in G_0:\,\|S(f)-z\|\le e(S,A,f,G,\theta)\}\ne \emptyset.
\end{equation}
Let $z_0$ be any element of the intersection \eqref{BA1}. Then 
\begin{equation*}
e(S,\wt{A},f,G)= \|S(f)-y\|\le \|S(f)-z_0\|+\|z_0-y\|\le 2e(S,A,f,G,\theta)+\delta.
\end{equation*}

\end{proof}

Finally we consider a random family of quantum algorithms and turn it into a single quantum algorithm by just incorporating the randomness. Let $m_0\in\N$ and let $\pi$ be a random variable with values in  $\Z[0,2^{m_0})$.
Let $(A_u)_{u\in \Z[0,2^{m_0})}$ be a family of quantum algorithms for $\mathcal{P}'=(F,G,K,\Lambda)$ of the form
\begin{equation}
A_{u}=((A_{lu})_{l=1}^k,(h_{lu })_{l=1}^k,\phi_u),\quad A_{lu}=(Q_{lu},
(X_{lju})_{j=0}^{n_l}), \quad Q_{lu}=(m_{l},m_{l}',m_l'',Z_{lu},\tau_{lu},\beta_l),\label{Y6}
\end{equation}
that is, the number of measurements of $A_u$, the number of queries of the $A_{lu}$, and the query parameters $m_{l}$, $m_{l}'$, $m_l''$, and $\beta_l$ do not depend on $u\in \Z[0,2^{m_0})$. Let $h_0\in\Z[0,2^{m_0})$, $X_0\in \mathcal{U}(H_{m_0})$ be such that
\begin{equation}
\label{AC8}
X_0\lg h_0\rs=\sum_{i=0}^{2^{m_0}-1} a_i\lg i\rs \quad\text{with}\quad\Prm(\{\pi=i\})=|a_i|^2.
\end{equation}
We define a new algorithm $B=[A_\pi]$ for $\mathcal{P}'$ as  
\begin{eqnarray}
B&=& \Big((B_{l})_{l=0}^k,(h_{l})_{l=0}^k,\phi\Big),\quad
B_0=(X_0),\quad B_{l}=(Q_{l},(X_{l j})_{j=0}^{n_l}),
\label{Y7}
\end{eqnarray}
where for $1\le l\le k$
\begin{eqnarray}
Q_{l}&=&(m_0+m_{l},m_0+m_{l}',m_l'',Z_{l},\tau_{l},\beta_l)
\label{Y5}\\
Z_l&=&\{(u,i):\, u\in\Z[0,2^{m_0}),\,i\in Z_{lu}\}
\label{B1}\\
 \tau_l(u,i)&=&\tau_{lu}(i)\quad((u,i)\in Z_l ),
\label{B2}
\end{eqnarray}
for $u\in\Z[0,2^{m_0})$
\begin{eqnarray}
X_{lj}\lg u\rs\lg z\rs&=&\lg u\rs(X_{l ju}\lg z\rs)\quad (z\in\Z[0,2^{m_l}))
\label{B3}\\
h_1(u)&=&(u,h_{1,u}),
\label{B10}
\end{eqnarray}
and for $2\le l\le k$,  and $u_{l_1}\in\Z[0,2^{m_0})$, $z_{l_1}\in \Z[0,2^{m_{l_1}})$ $(1\le l_1\le l)$,
\begin{eqnarray}
h_l\big(u,((u_{l_1},z_{l_1}))_{l_1=1}^{l-1}\big)&=&\big(u,h_{lu}\big((z_{l_1})_{l_1=1}^{l-1}\big)\big)
\label{B4}\\
\phi\big(u,((u_{l_1},z_{l_1}))_{l_1=1}^k\big)&=&\phi_u\big((z_{l_1})_{l_1=1}^k\big).\label{B5}
\end{eqnarray}
Observe that $B_0=(X_0)$ is a quantum algorithm with no measurement, which acts on $H_{m_0}$, with $n_\q(B_0)=0$, that is, no query is applied. Furthermore, for $l\ge 1$ the state space of $B_l$ is $H_{m_0+m_l}=H_{m_0}\otimes H_{m_l}$. Clearly,
\begin{equation*}
\label{AH8}
n_\q([A_\pi]):=\sum_{l=1}^k n_\q(B_l)=\sum_{l=1}^k n_l=\sum_{l=1}^k n_\q(A_{lu})=n_\q(A_u)\quad (u\in \Z[0,2^{m_0})).
\end{equation*}
\begin{lemma}
\label{lem:4}
Let $m_0,\pi,h_0,X_0$ and $(A_u)_{u\in \Z[0,2^{m_0})}$ be as above and let
$[A_\pi]$ be defined according to \eqref{AC8}--\eqref{B5} . Then for each subset $C\subseteq G$ 
\begin{equation}
\label{C0}
[A_\pi](f)(C)=\sum_{u\in \Z[0,2^{m_0})} \Prm(\{\pi=u\})A_u(f)(C).
\end{equation}
\end{lemma}
\begin{proof}
Let us look at the action of $B=[A_\pi]$. Let $f\in F$. At the start, we have by \eqref{AC8} 
\begin{equation*}
B_{0,f}\lg h_0\rs=X_0\lg h_0\rs=\sum_{u=0}^{2^{m_0}-1} a_u\lg u\rs,
\end{equation*}
so after measurement we obtain $\lg u\rs$ with probability $|a_u|^2=\Prm(\{\pi=u\})$, thus, $u$ can be considered as a realization of the random variable $\pi$. From then on $\lg u\rs$ remains in the first component $H_{m_0}$ throughout the rest of the algorithm and ensures that at stage $l$ the unitary transform $A_{lu,f}$ is applied to the second component $H_{m_l}$. It follows that the algorithm $A$ terminates with distribution $A_u(f)$ with probability  $\Prm(\{\pi=u\})$.

More formally, let $1\le l\le k$ and represent $\lg z\rs=\lg i\rs\lg x\rs\lg y\rs$ according to the splitting $H_{m_l}=H_{m'_l}\otimes H_{m''_l}\otimes H_{m_l-m'_l-m_l''}$. Then for $(u,i)\in Z_l$, using \eqref{B1} and \eqref{B2},
\begin{eqnarray*}
Q_{l,f}\lg u \rs\lg z\rs&=&\lg u\rs\lg i \rs\lg x\oplus \beta_l(f(\tau_l(u,i)))\rs\lg y\rs
\\
&=&\lg u\rs\lg i \rs\lg x\oplus \beta_l(f(\tau_{lu}(i)))\rs\lg y\rs
=\lg u \rs(Q_{lu,f}\lg z\rs),
\end{eqnarray*}
while for $(u,i)\not\in Z_l$ we have $i\not\in Z_{lu}$, thus
\begin{eqnarray*}
Q_{l,f}\lg u \rs\lg z\rs
&=&\lg u\rs\lg i \rs\lg x\rs\lg y\rs
=\lg u \rs(Q_{lu,f}\lg z\rs).
\end{eqnarray*}
This together with \eqref{B3} gives
\begin{equation*}
B_{l,f}\lg u \rs\lg z\rs=\lg u \rs(A_{lu,f}\lg z\rs).
\end{equation*}
By definition,
\begin{eqnarray}
\lefteqn{p_{B,f}(u,(u_1,z_1),\dots,(u_k,z_k))}
\nonumber\\
&=&|B_{0,f}(u,h_0)|^2|B_{1,f}((u_1,z_1),h_1(u))|^2\dots |B_{k,f}((u_k,z_k),h_k(u,(u_1,z_1),\dots,(u_{k-1},z_{k-1})))|^2.\nonumber
\end{eqnarray}
Furthermore, 
\begin{equation*}
B_{0,f}(u,h_0)=(\lg u \rs,B_{0,f}\lg h_0\rs)=a_u
\end{equation*}
and for $1\le l\le k$ 
\begin{eqnarray*}
&&B_{l,f}((u_l,z_l),h_l(u,(u_1,z_1),\dots,(u_{l-1},z_{l-1})))
\\
&=&  ( \lg u_l,z_l\rs, B_{l,f}\lg u, h_{lu}(z_1,\dots,z_{l-1})\rs)
\\
&=&
 ( \lg u_l,z_l\rs, \lg u\rs (A_{lu,f}\lg h_{lu}(z_1,\dots,z_{l-1})\rs))
\\
&=& \left\{\begin{array}{lll}
  ( \lg z_l\rs, A_{lu,f}\lg h_{lu}(z_1,\dots,z_{l-1})\rs)= 
A_{lu,f} (z_l, h_{lu}(z_1,\dots,z_{l-1}))& \quad\mbox{if}\quad  u_l=u  \\
  0 & \quad\mbox{otherwise,}    
    \end{array}
\right. 
\end{eqnarray*}
which implies
\begin{equation*}
p_{B,f}(u,(u_1,z_1),\dots,(u_k,z_k))=\left\{\begin{array}{lll}
  \Prm(\{\pi=u\})p_{A_u,f}(z_1,\dots, z_k) & \quad\mbox{if}\quad u=u_1=\dots u_k   \\
   0& \quad\mbox{otherwise.}    
    \end{array}
\right. 
\end{equation*}
It follows from \eqref{B5} and \eqref{B9} that for any subset $C\subseteq G$ 
\begin{eqnarray*}
B_f(C)&=&\sum_{\phi(u,(u_1,z_1),\dots,(u_k,z_k))\in C} p_{B,f}(u,(u_1,z_1),\dots,(u_k,z_k)) 
\\
&=& \sum_{u=0}^{2^{m_0}-1}\Prm(\{\pi=u\})\sum_{\phi_u(z_1,\dots, z_k)\in C} p_{A_u,f}(z_1,\dots, z_k), 
\end{eqnarray*}
thus \eqref{B9} implies \eqref{C0}. 

\end{proof}

Since by Lemma \ref{lem:4} the output of $[A_\pi]$ does not depend on $h_0$ and $X_0$, we drop them in the notation.

\begin{corollary}
\label{cor:1}
With the notation of Lemma \ref{lem:4} let $I_0\subseteq \Z[0,2^{m_0})$, $0\le \theta<1$, $f\in F$,  and  let $S:F\to G$ be any mapping. Then 
\begin{equation*}
\label{C5}
e(S,[A_\pi],f,G,1- (1-\theta)\Prm(\{\pi\in I_0\}))\le \max_{u\in I_0}e(S,A_u,f,G,\theta).
\end{equation*}
\end{corollary}

\begin{proof}
Indeed, let $\epsilon=\max_{u\in I_0}e(S,A_u,f,G,\theta)$ and put $C=\{g\in G:\, \|S(f)-g\|_G\le \epsilon\}$.
Using Lemma \ref{lem:4}, we conclude 
\begin{eqnarray*}
[A_\pi](f)(C)&\ge& \sum_{u\in I_0} \Prm(\{\pi=u\})A_u(f)(C)
\\
&\ge& \sum_{u\in I_0} \Prm(\{\pi=u\})A_u(f)(\{g\in G:\, \|S(f)-g\|_G\le e(S,A_u,f,G,\theta) \})
\\
&=&(1-\theta)\Prm(\{\pi\in I_0\}).
\end{eqnarray*}
\end{proof}

\section{A quantum discretization technique for parametric integration}
\label{sec:3}
Let $d \in \N$, $D=[0,1]^d$, and let
$C(D)$ denote the space of $\K$-valued continuous functions on $D$,
endowed with the supremum norm. For $1\leq p \le \infty$ let $L_p(D)$ be the
space of equivalence classes of $\K$-valued $p$-integrable functions 
equipped with the standard norm
\[
\| f \|_{L_p(D)} = \left\{\begin{array}{lll}
   &  \left( \int_D |f(t)|^p dt \right)^{1/p}\quad\mbox{if}\quad  p<\infty  \\[.2cm]
   &  \mbox{\rm ess-sup}_{t\in D} |f(t)|\quad\mbox{if}\quad p=\infty.
    \end{array}
\right.   
\]
The Sobolev space $W_p^r(D)$ is the set of all  $f\in L_p(D)$ such that
for all
 $\alpha =(\alpha_1, \dots , \alpha_d) \in \N_0^d$ with
$|\alpha|:= \sum_{j=1}^{d} \alpha_j \leq r$, the generalized partial derivative
$\partial^{\alpha} f$ belongs to $L_p(D)$. The norm on $W_p^r(D)$ is given by
\[
\| f \|_{W_p^r(D)} = \left\{\begin{array}{lll}
   & \left( \sum_{|\alpha|\leq r} \| \partial^{\alpha} f \|_{L_p(D)}^p \right)^{1/p}\quad\mbox{if}\quad p<\infty    \\
   & \max_{|\alpha|\leq r} \| \partial^{\alpha} f \|_{L_\infty(D) }\quad \mbox{if}\quad p=\infty.
    \end{array}
\right.
\]
The space $W_p^r(D)$ is continuously embedded into  $C(D)$ if and only if
\begin{equation}\label{PA2}
\left.\begin{array}{lll}
&p=1 &  \;\mbox{and} \quad r/d\ge 1   \\
\mbox{or}\\
  &1<p\le\infty & \;\mbox{and} \quad  r/d>1/p
    \end{array}\right\} 
\end{equation}
(embedding condition), see \cite{Ada75}, Ch.\ 5.

Now let $1\le p,q\le \infty$,  $d_1, d_2 \in \N_0$, with $d=d_1+d_2$, let $D_1=[0,1]^{d_1}, D_2=[0,1]^{d_2}$, thus
$D= D_1 \times D_2$. 
Consider the operator of parametric integration 
$S: W_p^r(D)\to  L_q(D_1)$ defined by
\begin{equation}
\label{AI3}
(Sf)(s) = \int_{D_2} f(s,t)\, dt \quad(s\in D_1).
\end{equation}
We assume $d\ge 1$, but we also include cases $d_1=0$ and $d_2=0$.
In the case of $d_\iota=0$ $(\iota=1,2)$ we let  $D_\iota=\{0\}$ be the one-point-set. The integral over $D_\iota$ as well as the space $L_q(D_\iota)$ are considered as defined  with respect to  the trivial probability measure on  $D_\iota=\{0\}$. Thus, for $d_1=0$ we have 
$L_q(D_1)=\K$ and $S: W_p^r(D)\to  \K$ is the integration operator $S=I$, given by
\begin{eqnarray}
\label{AE0}
 If = \int_D f(t)\, dt.
\end{eqnarray}
It is well-defined and continuous for all $r\in\N_0$ and $1\le p\le \infty$.

If $d_2=0$ we have  
$\int_{D_2} f(s)\, dt=f(s)$ and therefore $S: W_p^r(D)\to  L_q(D)$ is the approximation operator $S=J$, 
\begin{eqnarray}
\label{AI5}
 (Jf)(s) =f(s) \quad(s\in D).
\end{eqnarray}
If $d_1\ge 1$, the operator $S$ from \eqref{AI3} and \eqref{AI5} is well-defined and continuous iff the embedding of $W_p^r(D_1)$ to  $L_q(D_1)$ is continuous, that is, iff 
\begin{equation}\label{V1}
\left.\begin{array}{llll}
&1\le q<\infty & &  \;\mbox{and} \quad \frac{r}{d_1}\ge \left(\frac{1}{p}-\frac{1}{q}\right)_+  \\
\mbox{or}\\
&q=\infty,  &1<p<\infty, & \;\mbox{and} \quad  \frac{r}{d_1}>\frac{1}{p}\quad  
   \\\mbox{or}\\
&q=\infty, &p\in\{1,\infty\}, &  \;\mbox{and} \quad\frac{r}{d_1}\ge\frac{1}{p},
    \end{array}\right\} 
\end{equation}
with $a_+=\max(a,0)$ for $a\in\R$, see \cite{Ada75}, Th.\ 5.4. 
We also note that $S$ is compact, iff 
\begin{equation}\label{BB1}
 \frac{r}{d_1}>\left(\frac{1}{p}-\frac{1}{q}\right)_+ . 
\end{equation}
We consider standard information, that is, function values. This means we put $\Lambda=\{\delta_t:\,t\in D\}$, where  $\delta_t(f)=f(t)$  for $f\in \mathcal{F}(D,\K)$. However, if we consider the space $W_p^r(D)$, this needs some care since $W_p^r(D)$ consists of equivalence classes of functions. As in \cite{Hei23b}, we introduce the spaces $\mathcal{W}_p^r(D)$ as follows: 

If the embedding condtition \eqref{PA2} holds, then
each class $[f]\in W_p^r(D)$ contains a unique continuous representative and we set
$\mathcal{W}_p^r(D)=\{f\in C(D): [f]\in W_p^r(D)\}$, equipped with the norm $\|f\|_{\mathcal{W}_p^r(D)}:=\|[f]\|_{W_p^r(D)}$. In this case $W_p^r(D)$ and $\mathcal{W}_p^r(D)$ can be identified, and we use both notations.

If  \eqref{PA2}  does not hold, we let $\mathcal{W}_p^r(D)$ be the respective Sobolev space of functions (not equivalence classes), thus $f\in \mathcal{W}_p^r(D)$ iff $[f]\in W_p^r(D)$, where $[f]$ is the equivalence class of $f$ with respect to equality almost everywhere with respect to the Lebesgue measure. This is a linear space and  $\|f\|_{\mathcal{W}_p^r(D)}:=\|[f]\|_{W_p^r(D)}$ is a semi-norm on it. Similarly, for $r=0$ we use the notation $\mathcal{L}_p(D)$ along with $L_p(D)$. Clearly,  $S$ can also be viewed as an operator from $\mathcal{W}_p^r(D)$ to $L_q(D_1)$.
So we will be concerned with the problems
\begin{eqnarray}
&&(B_{\mathcal{W}_p^r(D)},L_q(D_1), S, \K, \Lambda) \quad \text{and}\quad(\mathcal{W}_p^r(D),L_q(D_1), S, \K, \Lambda)\label{AS1}
\end{eqnarray}
(compare Remark \ref{rem:1}).
Now we pass to the discrete versions of the spaces and operators above. Let $M,M_1,M_2$ be finite sets and define the space
$L_p^M$ to be the set of all functions $f: M \rightarrow \K$
with norm
\[
\| f \|_{L_p^M} = \left\{\begin{array}{lll}
 \displaystyle \left( \frac{1}{|M|} \sum_{i\in M} |f(i)|^p \right)^{1/p} & \quad\mbox{if}\quad  p<\infty  \\[.5cm]
  \displaystyle\max_{i\in M} |f(i)| & \quad\mbox{if}\quad p=\infty
    \end{array}
\right.
\]
and  the operator of discrete parametric integration (or vector-valued mean computation) as 
\begin{equation}
\label{AI1}
S^{M_1,M_2} : L_p^{M_1\times M_2} \rightarrow L_q^{M_1},\quad (S^{M_1,M_2}f)(i)  = \frac{1}{|M_2|} 
\sum_{j\in M_2} f(i,j) .
\end{equation}
Here we consider the problems
\begin{eqnarray}
&&\left( B_{L_p^{M_1\times M_2}},L_q^{M_1},S^{M_1,M_2},\K,\Lambda^{M_1,M_2}\right) \; \text{as well as}\; \left(L_p^{M_1\times M_2},L_q^{M_1},S^{M_1,M_2},\K,\Lambda^{M_1,M_2}\right),\label{AS2}
\end{eqnarray}
where
$\Lambda^{M_1,M_2}=\{\delta_{ij}:\, i\in M_1,\,j\in M_2\}$ with $\delta_{ij}(f)=f(i,j)$ for $f\in \mathcal{F}(M_1\times M_2,\K)$.

If $|M_1|=1$, $S^{M_1,M_2}$ turns into the operator of mean computation 
\begin{equation}
\label{AE8}
I^{M_2}:L_p^{M_2}\to \K,\quad I^{M_2}f  = \frac{1}{|M_2|} 
\sum_{j\in M_2} f(j),
\end{equation}
and similarly, if $|M_2|=1$, $S^{M_1,M_2}$ becomes the operator of finite dimensional approximation 
\begin{equation}
\label{AE9}
J^{M_1}: L_p^{M_1}\to L_q^{M_1}, \quad J^{M_1}f  = f.
\end{equation}
For $N_1,N_2\in\N$ we write $L_p^{N_1}$ for $L_p^{\Z[1,N_1]}$,  furthermore
$L_p^{N_1,N_2}$ for $L_p^{\Z[1,N_1]\times \Z[1,N_2]}$, $S^{N_1,N_2}$ for $S^{\Z[1,N_1],\Z[1,N_2]}$, and correspondingly  $I^N$ and $J^N$ ($N\in\N$).

In the sequel of this section we are concerned with discretization, that is, reducing the problem of parametric integration to a family of vector-valued summation problems. 
We present the needed ingredients of the discretization technique. This technique, developed in  \cite{Hei01b, Hei04b} and used in \cite{HNP04,GYW06,Wie06, YH08} is based on deterministic function evaluations, the availability of which requires that $W_p^r(D)$ is continuously embedded in $C(D)$. In \cite{Hei23b} a stochastic version of this technique was presented to cover the case of non-embedding. Here we develop a quantum variant based on the stochastic one. We therefore recall some needed ingredients from \cite{Hei23b} and refer to this paper for detailed explanations.

Let $r\in\N$ and $0<\delta_0<1$. If $r\ge 2$,  let $(t_k)_{k=1}^{\kappa}$ be the uniform grid of mesh size $(r-1)^{-1}(1-\delta_0)$ on $[0,1-\delta_0]^d$ and if $r=1$, set $\kappa=1$ and $t_1=0$, thus 
\begin{equation}
\kappa=r^d.\label{AT7}
\end{equation}
 Let $P:\mathcal{F}(D,\K)\to L_\infty(D)$ be for $d=1$ the respective Lagrange interpolation operator of degree $r-1$ and for $d>1$ its tensor product. 
Let $\mathcal{P}^{r-1}_{\rm max}(Q)$ denote the space of polynomials of maximum degree $r-1$ on a set $Q\subset \R^d$.
Then 
\begin{equation*}
(Pf)(t) = \sum_{k=1}^{\kappa} f(t_k) \varphi_k \quad (f\in \mathcal{F}(D,\K), t\in D)
\end{equation*}
with $(\phi_k)_{k=1}^{\kappa}$ being the $d$-dimensional tensor product Lagrange polynomials,  which form a basis in the space $\mathcal{P}^{r-1}_{\rm max}(\R^d)$. 

Let $\xi$ be a random variable over a probability space $(\Omega,\Sigma,\Prm)$ with values in $D(\delta_0):=[0,\delta_0)^d$ and set
\begin{equation}
(P_{0,\xi}f)(t) = \chi_{D}(t)\sum_{k=1}^{\kappa} f(t_k+\xi) \varphi_k(t-\xi) \quad (f\in \mathcal{F}(D,\K), t\in D).
\label{W6}
\end{equation}
\begin{remark}
\label{rem:2}
From now on a second source of randomness comes in through randomized approximation. All respective random variables will be considered as independent of the randomness related to quantum measurements.
\end{remark}
We represent 
$$
\varphi_k(t-\xi)=\sum_{j=1}^{\kappa}a_{jk}(\xi)\phi_j(t),
$$
with $a_{jk}\in \mathcal{P}^{r-1}_{\rm max}(\R^d)$, so  \eqref{W6} turns into
\begin{equation*}
P_{0,\xi}f = \chi_{D}\sum_{j=1}^{\kappa}\sum_{k=1}^{\kappa} a_{jk}(\xi)f(t_k+\xi) \varphi_j \quad (f\in \mathcal{F}(D,\K)).
\end{equation*}

For $l\in \N_0$ and $\gamma=1,2$ we partition $D_{\gamma}$ into $2^{d_{\gamma}l}$ congruent
cubes of disjoint interior with side length $2^{-l}$:
\[
D_\gamma  = \bigcup_{i_\gamma=1}^{2^{d_\gamma l}} D^{({\gamma})}_{li_\gamma}.
\]
 Identifying $\Z[1,2^{d_1l}]\times \Z[1,2^{d_2l}]$ with $\Z[1,2^{dl}]$ via the mapping $(i_1,i_2)\to 2^{d_2l}(i_1-1)+i_2$, we obtain a corresponding partition of $D$
\begin{equation}
\label{V9}
D = \bigcup_{i=1}^{2^{dl}}D_{li},\quad D_{li}=D_{li_1}^{(1)}\times D_{li_2}^{(2)}.
\end{equation}
Let $s_{li}\in D_{li}$ denote the point with minimal Euclidean norm.
We introduce the following operators 
$E_{li}$ and $R_{li}$ from $\mathcal{F}(D,\K)$ to $\mathcal{F}(D,\K)$ by setting for 
$s \in D$
\begin{eqnarray}
(E_{li} f)(s) &=& f(s_{li} +2^{-l}s)  
\label{AT2}\\
(R_{li} f)(s) &=& \left\{ \begin{array}{ll}
f(2^l(s-s_{li})) & \text{ if } s\in D_{li}\\
0 & \text{ otherwise.} 
\end{array}
\right.\label{WH4}
\end{eqnarray}
We also need the operators $R^{(\gamma)}_{li}$ for $1\le i\le 2^{d_\gamma l}$, 
which are defined analogously, with $D_\gamma$ instead of $D$ $(\gamma=1,2)$.

For $f\in \mathcal{F}(D,\K)$ and $l\in \N_0$ we set
\begin{equation}
\label{WF1}
P_{l\xi} f = \sum_{i=1}^{2^{dl}} R_{li} P_{0,\xi} E_{li} f
= \sum_{i=1}^{2^{dl}}\sum_{j=1}^{\kappa} \sum_{k=1}^{\kappa}a_{jk}(\xi)f(s_{li}+2^{-l}(t_k+\xi)) R_{li}\varphi_j.
\end{equation}
Next we define $P_{0,\xi}'  =  P_{1,\xi} - P_{0,\xi}$
which, as explained in \cite{Hei23b}, can be written as
\begin{equation*}
P_{0,\xi}'f = \sum_{j=1}^{\kappa'} \sum_{k=1}^{\kappa''} b_{jk}(\xi) f(t_{jk}(\xi)) 
 \psi_{j},
\end{equation*}
with 
\begin{equation}
\kappa'=2^{d}\kappa,\quad \kappa''=2\kappa \label{AT8}
\end{equation}
and for $1\le i_0\le 2^d$,  $1\le j_0\le \kappa$, $j=\kappa (i_0-1)+j_0$, $1\le k\le 2\kappa$
\begin{eqnarray}
\psi_j&=&R_{1,i_0} \varphi_{j_0} 
\notag\\
t_{jk}(\xi)&=&\left\{\begin{array}{lll}
   s_{1,i_0}+2^{-1}(t_k+\xi)& \quad\mbox{if}\quad  1\le k\le \kappa  \\
   t_{k-\kappa}+\xi& \quad\mbox{if}\quad \kappa+1\le k\le 2\kappa
    \end{array}
\right. 
\label{AP8}\\[.2cm]
b_{jk}&\in& \mathcal{P}^{r-1}_{\rm max}(\R^d).
\label{AG2}
\end{eqnarray}
Now we put
\begin{equation*}
\psi_{lij} = R_{li} \psi_j\quad (1\le i\le 2^{dl},\,1\le j\le \kappa')
\end{equation*}
and let
\begin{equation}
\label{eq:29}
P_{l\xi}'f  =  \sum_{i=1}^{2^{dl}} R_{li}P_{0,\xi}'E_{li}f =\sum_{i=1}^{2^{dl}}\sum_{j=1}^{\kappa'}
\sum_{{k}=1}^{\kappa''} b_{jk}(\xi)f(s_{li}+2^{-l}t_{jk}(\xi))\psi_{lij}.
\end{equation}
Denote 
\begin{equation}
\mathcal{I}_{1,l}=\Z[1,2^{d_1l}]\times\Z[1,\kappa'], \quad \mathcal{I}_{2,l}=\Z[1,2^{d_2l}]
\label{AS8}
\end{equation}
and set
\begin{equation}
\label{eq:26}
N_{1,l}:=|\mathcal{I}_{1,l}|=\kappa' 2^{d_1l},\quad N_{2,l}:=|\mathcal{I}_{2,l}|=2^{d_2l}. 
\end{equation}
Define the operator 
$
T_l:{\rm span}\left\{\psi_{lij}:i\in\Z[1,2^{dl}],
j\in\Z[1,\kappa']\right\}\to L_p^{\mathcal{I}_{1,l}\times \mathcal{I}_{2,l}}$ 
by  
\begin{equation}
\label{AK7}
T_l\psi_{lij} = \delta_{(i_1,j),i_2}\quad((i_1,j)\in \mathcal{I}_{1,l},\, i_2\in \mathcal{I}_{2,l},\, i=2^{d_2l}(i_1-1)+i_2).
\end{equation}
Then $T_l$ is correctly defined \cite{Hei23b} and 
\begin{equation}
\|T_lf\|_{L_p^{\mathcal{I}_{1,l}\times \mathcal{I}_{2,l}}}\le c\|f\|_{L_p(D)}\qquad \left(f\in {\rm span}\left\{\psi_{lij}\right\}\right). 
\label{S3}
\end{equation}
We define furthermore
\begin{equation}
\label{X1}
U_{l\xi}: \mathcal{F}(D,\K) \rightarrow L_p^{\mathcal{I}_{1,l}\times \mathcal{I}_{2,l}}, \quad U_{l\xi}=T_lP_{l\xi}'
\end{equation}
and have by \eqref{eq:29} and \eqref{AK7} 
\begin{equation}
\label{WF3}
(U_{l\xi}f)((i_1,j),i_2)=\sum_{{k}=1}^{\kappa''} b_{jk}(\xi) f(s_{li} +2^{-l}t_{jk}(\xi)).
\end{equation}
Let 
\begin{equation}
V_l:  L_q^{\mathcal{I}_{1,l}} \rightarrow L_q(D_1),\quad V_l g = \sum_{i_1=1}^{2^{d_1l}} \sum_{j=1}^{\kappa'}
g(i_1,j) R_{li_1}^{(1)}S\psi_j, \label{AT5}
\end{equation}
then
\begin{eqnarray*}
\|V_lg\|_{L_q(D_1)}&\le&
c \|g\|_{L_q^{\mathcal{I}_{1,l}}}.
\end{eqnarray*}
The following representation was shown in \cite{Hei23b}
\begin{equation}
\label{eq:41}
S =S-SP_{l_1\xi}+SP_{l_0\xi} +\sum_{l=l_0}^{l-1} V_l S^{\mathcal{I}_{1,l}, \mathcal{I}_{2,l}} U_{l\xi}.
\end{equation}

In the sequel, we will be interested in the following special cases of the random variable $\xi$ from \eqref{W6}: Let
\begin{eqnarray*}
\zeta&=&(\zeta^{(1)},\dots,\zeta^{(d)})
\end{eqnarray*}
be a uniformly distributed on  $D(\delta_0)=[0,\delta_0)^d$ random variable taking all values in $[0,\delta_0)^d$, let $m\in \N_0$ and set
\begin{eqnarray*}
\zeta_m&=&(\zeta_m^{(1)},\dots,\zeta_m^{(d)})=\delta_0 2^{-m}(\lfloor \delta_0^{-1}2^m\zeta^{(1)}\rfloor,\dots,\lfloor \delta_0^{-1} 2^m\zeta^{(d)}\rfloor), 
\end{eqnarray*}
hence $\zeta_m$ is uniformly distributed on 
\begin{equation}
\label{AD9}
D^{(m)}=\bigg\{\frac{i\delta_0}{2^m}:\; i\in \mathbb{Z}[0,2^m)\bigg\}^d\subset D(\delta_0).
\end{equation}
In particular, this includes the deterministic case $\zeta_0\equiv 0$ for $m=0$. Clearly,
\begin{equation}
\|\zeta(\omega)-\zeta_m(\omega)\|_{L_\infty^d}\le \delta_02^{-m}\quad(\omega\in \Omega).
\label{X8A}
\end{equation}
Furthermore, for  a
subset $\mathcal{E}\subset\mathcal{F}(D,\K)$ and $\tau>0$ we define
\begin{eqnarray*}
\epsilon(\mathcal{E},\tau)&=& \sup\{|f(s)-f(t)|\,|\,f\in\mathcal{E},\,s,t\in D, \|s-t\|_{L_\infty^d}\le \tau\}
\end{eqnarray*}
and write $\epsilon(f,\tau)$ for $\epsilon(\{f\},\tau)$. The set $\mathcal{E}$ is called uniformly equicontinuous, if
\begin{equation}
\label{AK8}
\lim_{\tau\to 0}\epsilon(\mathcal{E},\tau)=0.
\end{equation}
For convenience we introduce the following notation for a subset $M$ of  $\mathcal{F}(D,\K)$: 
\begin{equation}
\label{AK9}
{\rm UEq}(M)=\{\mathcal{E}\subset M:\, \mathcal{E} \text{  is uniformly equicontinuous}\}.
\end{equation}
For $f\in \mathcal{F}(D,\K)$ the following holds
\begin{eqnarray}
|f(t)-f(0)|\le 2^m\epsilon(f,2^{-m})\quad (t\in D).
\label{X5}
\end{eqnarray}
Consequently, for $f\in \mathcal{L}_1(D)$
\begin{eqnarray*}
\|f\|_{\mathcal{L}_1(D)}\ge |f(0)|-\|f-f(0)\|_{\mathcal{L}_1(D)}\ge  |f(0)|- 2^m\epsilon(f,2^{-m}),
\end{eqnarray*}
hence 
\begin{eqnarray*}
|f(0)|\le 2^m\epsilon(f,2^{-m})+\|f\|_{\mathcal{L}_1(D)}
\end{eqnarray*}
and, using again \eqref{X5}, we conclude 
\begin{eqnarray}
\|f\|_{\ell_\infty(D)}\le  2^{m+1}\epsilon(f,2^{-m})+\|f\|_{\mathcal{L}_1(D)}.\label{Y0}
\end{eqnarray}
From \eqref{W6}, \eqref{X8A}, and \eqref{Y0} we obtain for $f\in \mathcal{L}_1(D)$, $t\in D$, 
\begin{eqnarray*}
\lefteqn{|(P_{0,\zeta}f)(t)-(P_{0,\zeta_m}f)(t)|}
\nonumber\\
& \le &\sum_{k=1}^{\kappa} \Big(|f(t_k+\zeta)-f(t_k+\zeta_m)| |\varphi_k(t-\zeta)|
+ |f(t_k+\zeta_m)| |\varphi_k(t-\zeta)-\varphi_k(t-\zeta_m)|\Big)
\nonumber\\
&\le&\epsilon(f,\delta_0 2^{-m})\sum_{k=1}^{\kappa}\|\phi_k\|_{C([-\delta_0, 1]^d)} + d\delta_02^{-m}\|f\|_{\ell_\infty(D)}\sum_{k=1}^{\kappa}\|\phi_k\|_{C^1([-\delta_0, 1]^d)}
\nonumber\\
&\le& c(0)\big(\epsilon(f, \delta_02^{-m}) +2^{-m}\|f\|_{\ell_\infty(D)}\big)
\end{eqnarray*}
with 
\begin{equation*}
c(0)=\sum_{k=1}^{\kappa}(d\delta_0+1) \|\phi_k\|_{C^1([-\delta_0, 1]^d)}.
\end{equation*}
Therefore, using  \eqref{AT2}, \eqref{WH4}, \eqref{WF1}, and  \eqref{Y0},
\begin{eqnarray}
\lefteqn{\|P_{l,\zeta} f -P_{l,\zeta_m}f\|_{L_\infty(D)}}
\nonumber\\
&=& \bigg\|\sum_{i=1}^{2^{dl}} R_{li} (P_{0,\zeta}-P_{0,\zeta_m}) E_{li} f\bigg\|_{L_\infty(D)}
\le\max_{i\in\Z[1,2^{dl}]}\|(P_{0,\zeta}-P_{0,\zeta_m}) E_{li} f\|_{L_\infty(D)}
\nonumber\\
&\le& c(0)\max_{i\in\Z[1,2^{dl}]}(\epsilon(E_{li}f, \delta_0 2^{-m}) +2^{-m}\|E_{li} f\|_{\ell_\infty(D)})
\le c(0)(\epsilon(f, \delta_0 2^{-l-m}) + 2^{-m}\|f\|_{\ell_\infty(D)})
\nonumber\\
&\le&c(0)(3\epsilon(f,  2^{-m}) +  2^{-m}\|f\|_{L_1(D)}).
\label{W9}
\end{eqnarray}

The following result contains the crucial estimates for the discretization technique. We refer to Proposition 3.2 in \cite{Hei23b} for relations \eqref{R9} and \eqref{X7}.
\begin{proposition}
\label{pro:6}
Let $r,d\in\N$,  $1 \leq p,q \le \infty$, $r/d>1/p-1/q$, $1\le w\le q$, $w<\infty$.  Then there are constants $c_1,c_2,c_3>0$ such that for all $l\in \N_0$ 
\begin{eqnarray}
\label{R9}
\sup_{f\in \mathcal{B}_{\mathcal{W}_p^r(D)}}(\EE\|f-P_{l\zeta}f\|_{L_q(D)}^w)^{1/w}
&\le& c_1 2^{-rl+\left(\frac{1}{p}-\frac{1}{q}\right)_+dl}.
\end{eqnarray}
Moreover, for all subsets $\mathcal{E}\subset \mathcal{B}_{\mathcal{W}_p^r(D)} $ and $m\in\N_0$ 
\begin{eqnarray}
\label{X6}
\sup_{f\in \mathcal{E}}(\EE\|f-P_{l\zeta_{m}}f\|_{L_q(D)}^{w})^{1/w}
&\le& c_2 (2^{-rl+(1/p-1/q)_+dl}+\epsilon(\mathcal{E},  2^{-m}) + 2^{-m}).
\end{eqnarray}
Finally, if the embedding condition \eqref{PA2} holds, then
\begin{eqnarray}
\label{X7} 
\sup_{f\in \mathcal{B}_{\mathcal{W}_p^r(D)}}\|f-P_{l\zeta_0}f\|_{L_q(D)}
&\le& c_3 2^{-rl+\left(\frac{1}{p}-\frac{1}{q}\right)_+dl}.
\end{eqnarray}
\end{proposition}
\begin{proof} It remains to show \eqref{X6}, but since $\mathcal{B}_{\mathcal{W}_p^r(D)}\subset \mathcal{B}_{\mathcal{L}_1(D)}$, 
this follows from \eqref{W9}, \eqref{R9}, and the triangle inequality.
\end{proof}

\begin{lemma}
\label{lem:6}  
Let $r\in \N$, $1 \leq p,q \le \infty$, $1\le w\le p$, $w<\infty$,   and assume that \eqref{BB1} holds. Then the following holds for $\xi=\zeta$ and, if the embedding condition \eqref{PA2} is satisfied, also for $\xi=\zeta_0$.
There is a constant $c>0$ such that for all $l\in \N_0$ 
\begin{equation}
\label{AC2}
\sup_{f\in \mathcal{B}_{\mathcal{W}_p^r(D)}}\left(\EE\|Sf-SP_{l\xi}f\|_{L_q(D_1)}^w\right)^{1/w}\le c 2^{-rl+\left(\frac{1}{p}-\frac{1}{q}\right)_+d_1l}.
\end{equation}
Furthermore, for all subsets $\mathcal{E}\subset \mathcal{B}_{\mathcal{W}_p^r(D)} $ and $m\in\N_0$
\begin{equation}
\label{AC0}
\sup_{f\in \mathcal{E}}\left(\EE\|Sf-SP_{l\zeta_{m}}f\|_{L_q(D_1)}^w\right)^{1/w}\le c\Big( 2^{-rl+\left(\frac{1}{p}-\frac{1}{q}\right)_+d_1l}+\epsilon(\mathcal{E}, 2^{-m}) + 2^{-m}\Big).
\end{equation}
\end{lemma}
\begin{proof} Relation \eqref{AC2} is Lemma 3.4 of \cite{Hei23b}. Using \eqref{W9}, \eqref{AC0} is a direct consequence of \eqref{AC2}. 

\end{proof}

For $q=p$ it follows from \eqref{S3}, \eqref{X1}, and Proposition \ref{pro:6} that there is a constant $c(1)>0$ such that for all $l\in\N_0$
\begin{eqnarray}
\sup_{f\in \mathcal{B}_{\mathcal{W}_p^r(D)}}\EE\|U_{l\zeta}f\|_{L_p^{\mathcal{I}_{1,l}\times \mathcal{I}_{2,l}}}
&\le&
\|T_l\|\sup_{f\in \mathcal{B}_{\mathcal{W}_p^r(D)}}(\EE\|P_{l+1,\zeta}f-P_{l\zeta}f\|_{L_p(D)}
\le c(1)2^{-rl},
\label{R7}
\end{eqnarray}
moreover, for all $\mathcal{E}\subset \mathcal{B}_{\mathcal{W}_p^r(D)}$  and $m\in\N_0$
\begin{eqnarray}
\sup_{f\in \mathcal{E}}\EE\|U_{l\zeta_m}f\|_{L_p^{\mathcal{I}_{1,l}\times \mathcal{I}_{2,l}}}
&\le&
\|T_l\|\sup_{f\in \mathcal{E}}(\EE\|P_{l+1,\zeta_m}f-P_{l\zeta_m}f\|_{L_p(D)}
\nonumber\\
&\le& c(1)(2^{-rl}+\epsilon(\mathcal{E},  2^{-m}) + 2^{-m}),
\label{R8}
\end{eqnarray}
and if \eqref{PA2} holds,
\begin{eqnarray}
\sup_{f\in \mathcal{B}_{\mathcal{W}_p^r(D)}}\|U_{l\zeta_0}f\|_{L_p^{\mathcal{I}_{1,l}\times \mathcal{I}_{2,l}}}
&\le& c(1)2^{-rl}.
\label{AJ5}
\end{eqnarray}

We need discretizations of $\R$ and $\C$ and related mappings. For this purpose we define for $k\in\N$ the functions $\alpha_k:\R\to\Z[0,2^k)$ by
\begin{equation}\label{N1}
\alpha_k(z)=
\left\{\begin{array}{lll}
   0& \mbox{if} \quad z <-2^{k/2-1} \\
   \lfloor 2^{k/2}(z+2^{k/2-1})\rfloor       & \mbox{if} \quad  
   -2^{k/2-1}\le z <2^{k/2-1}\\
   2^{k}-1& \mbox{if} \quad z\ge 2^{k/2-1} 
   \end{array}
   \right.
\end{equation}
and $\mu_k:\R\to\R$ by
\begin{equation*}
\mu_k(y)=2^{-k/2}y-2^{k/2-1}.
\end{equation*}
It follows that for $|z|\le 2^{k/2-1}$ we have
\begin{equation}
\alpha_k(z)\le 2^{k/2}(z+2^{k/2-1})\le\alpha_k(z)+1,
\label{AT3}
\end{equation}
furthermore
\begin{eqnarray*}
\mu_k(2^{k/2}(z+2^{k/2-1}))&=& 2^{-k/2}2^{k/2}(z+2^{k/2-1})-2^{k/2-1}=z
\\
\mu_k(\alpha_k(z)+1)&=&\mu_k(\alpha_k(z))+2^{-k/2},
\end{eqnarray*}
and so, applying $\mu_k$ to \eqref{AT3}, we obtain
\begin{equation}
\mu_k(\alpha_k(z))\le z\le\mu_k(\alpha_k(z))+2^{-k/2}\quad (z\in\R, |z|\le 2^{k/2-1}).
\label{AT4}
\end{equation}
Now we define for $m^*\in \N$ the function $\beta:\K\to \Z[0,2^{m^*}) $  as
\begin{equation}\label{N3}
\beta(z)=
\left\{\begin{array}{lll}
   \alpha_{m^*}(z)& \mbox{if} \quad \K=\R \\
   \alpha_{\lfloor m^*/2\rfloor}(\Re(z))
   2^{\lfloor m^*/2\rfloor}+\alpha_{\lfloor m^*/2\rfloor}(\Im(z))&\mbox{if} \quad \K=\C. 
 \end{array}
\right.
\end{equation}
Furthermore,  let $\gamma:\Z[0,2^{m^*})\to\K$  for $y\in\Z[0,2^{m^*})$ be given as follows. If $\K=\R$, we set 
\begin{equation}
\gamma(y)=\mu_{m^*}(y),
\label{N2}
\end{equation}
while if $\K=\C$, we represent $y=y_1 2^{\lfloor m^*/2\rfloor}+y_2$ with $y_1\in \N_0$, $y_2\in \Z[0,2^{\lfloor m^*/2\rfloor})$ and put
\begin{eqnarray}
\gamma(y)&=&\mu_{\lfloor m^*/2\rfloor}(y_1)+\imath\mu_{\lfloor m^*/2\rfloor}(y_2).
\label{N4}
\end{eqnarray}
In case of $\K=\R$ we conclude from \eqref{N3}, \eqref{N2}, and \eqref{AT4} 
\begin{equation}
\label{E8}
\gamma(\beta(z))\le z\le \gamma(\beta(z))+2^{-m^*/2}\quad(z\in\R,|z|\le 2^{m^*/2-1}).
\end{equation}
If $\K=\C$, we get from \eqref{N3} and \eqref{N4} for $z\in \C$ 
\begin{eqnarray*}
\gamma(\beta(z))&=&\gamma(\alpha_{\lfloor m^*/2\rfloor}(\Re(z))
   2^{\lfloor m^*/2\rfloor}+\alpha_{\lfloor m^*/2\rfloor}(\Im(z)))
\\
&=&\mu_{\lfloor m^*/2\rfloor}(\alpha_{\lfloor m^*/2\rfloor}(\Re(z)))+\imath\mu_{\lfloor m^*/2\rfloor}(\alpha_{\lfloor m^*/2\rfloor}(\Im(z))),
\end{eqnarray*}
and therefore, again by \eqref{AT4},
\begin{eqnarray}
\Re(\gamma(\beta(z)))\le \Re(z)\le \Re(\gamma(\beta(z)))+2^{-\lfloor m^*/2\rfloor/2}\quad(z\in \C, |\Re(z)|\le 2^{\lfloor m^*/2\rfloor/2-1}),
\label{E6}\\[.2cm]
\Im(\gamma(\beta(z)))\le \Im(z)\le \Im(\gamma(\beta(z)))+2^{-\lfloor m^*/2\rfloor/2}\quad (z\in \C, |\Im(z)|\le 2^{\lfloor m^*/2\rfloor/2-1}).
\label{E7}
\end{eqnarray}
Using $\lfloor m^*/2\rfloor\ge (m^*-1)/2$, we obtain the following consequence of \eqref{E8}--\eqref{E7} valid for both $\K=\R$ and $\K=\C$
\begin{equation}
\label{E4}
|z-\gamma(\beta(z))|\le 2^{-(m^*-3)/4}\quad(z\in\K, \,|z|\le 2^{ (m^*-5)/4}).
\end{equation}

In the following we define a family of quantum algorithms $A_v^{(1)}$ for $v\in D(\delta_0)$ on $$
(\mathcal{F}(D,\K),L_q(D_1),\K,\Lambda).
$$ 
For this purpose we fix a number of parameters. Let 
\begin{eqnarray}
&&l_0,l_1\in\N_0,\quad l_0\le l_1, \quad m^*\in \N
\label{AR8}
\end{eqnarray}
and for $l=l_0,\dots,l_1-1$
\begin{eqnarray}
&&\Theta_l>0,\quad \nu_l\in \N,\quad n_l\in\N_0,
\label{AR9}\\
&&A_l  \text{ a quantum algorithm for } \left( L_p^{\mathcal{I}_{1,l}, \mathcal{I}_{2,l}},L_q^{\mathcal{I}_{1,l}},\K,\Lambda^{\mathcal{I}_{1,l}, \mathcal{I}_{2,l}}\right)\text{ with } n_\q(A_l)\le n_l. \label{AS0}
\end{eqnarray}
Furthermore, define $\eta_{lkv}:\mathcal{I}_{1,l}\times \mathcal{I}_{2,l}\to D$ for  $k=1,\dots,\kappa''$  by
\begin{equation*}
\eta_{lkv}((i_1,j),i_2)=s_{li}+2^{-l}t_{jk}(v)\quad (i=2^{d_2l}(i_1-1)+i_2,\;1\le j \le \kappa'),
\end{equation*}
$\rho_{lv}:\mathcal{I}_{1,l}\times \mathcal{I}_{2,l}\times\Z[0,2^{m^*})^{\kappa''}\to\K$ by
\begin{equation*}
\rho_{lv}((i_1,j),i_2,y_1,\dots,y_{\kappa''})=(2c(1))^{-1}\Theta_l2^{rl}\sum_{k=1}^{\kappa''}b_{jk}(v)\gamma(y_k),
\end{equation*}
with $c(1)$ being the constant from \eqref{R7}--\eqref{AJ5}, 
 and  $\Gamma_{lv}:\mathcal{F}(D,\K)\to L_p^{\mathcal{I}_{1,l}, \mathcal{I}_{2,l}}$   by
\begin{equation}
\label{AH4}
\Gamma_{lv}(f)((i_1,j),i_2)=\rho_{lv}(((i_1,j),i_2),(\beta\circ f\circ\eta_{lkv}((i_1,j),i_2))_{k=1}^{\kappa''})\quad (((i_1,j),i_2)\in \mathcal{I}_{1,l}\times \mathcal{I}_{2,l}),
\end{equation}
with $\beta$ and $\gamma$ from (\ref{N1})--(\ref{N4}).
Consequently, 
\begin{equation}
\label{AI0}
\Gamma_{lv}(f)((i_1,j),i_2)=(2c(1))^{-1}\Theta_l2^{rl}\sum_{k=1}^{\kappa''}b_{jk}(v)\gamma(\beta(f(s_{li}+2^{-l}t_{jk}(v))))
\end{equation}
(compare also \eqref{WF3}). Note that $\Gamma_{lv}$ is of the form \eqref{AC5}--\eqref{F1}.
Put 
\begin{equation}
\label{Y1}
\wt{A}_l=\sigma_1(A_l^{\nu_l}):L_p^{\mathcal{I}_{1,l},\mathcal{I}_{2,l}}\to L_q^{\mathcal{I}_{1,l}}\quad (l_0\le l<l_1),
\end{equation}
see \eqref{AP2} with $\delta=1$, then $n_\q(\wt{A}_l)\le \nu_ln_l$. Now
let $A_{lv}$ be the respective algorithm on $(\mathcal{F}(D,\K),L_q^{\mathcal{I}_{1,l}},\K,\Lambda)$ according to Lemma \ref{lem:2} such that 
\begin{eqnarray}
\label{D0}
  A_{lv}(f)=\wt{A}_l(\Gamma_{lv}(f)) \quad(f\in \mathcal{F}(D,\K)),\quad n_\q(A_{lv})\le 2\kappa'' \nu_ln_l.\label{C6}
\end{eqnarray}

Furthermore, let $A_{l_0v}^{(0)}$ be a quantum algorithm on $(\mathcal{F}(D,\K),L_q(D_1),\K,\Lambda)$ that simulates a classical computation producing $SP_{l_0v}f$ for $f\in \mathcal{F}(D,\K)$ up to sufficiently high precision. 
More precisely, let
\begin{equation}
\Upsilon=\{(i,k):\,1\le i\le 2^{dl_0}, 1\le k\le \kappa\}\label{AJ1}
\end{equation}
equipped with the lexicographical order and let $Id_{m^*}$ be the identity on $H_{m^*}$. We define a quantum algorithm
\begin{eqnarray}
&&A_{l_0v}^{(0)}= ((A_{l_0ikv})_{(i,k)\in \Upsilon}, (h_{l_0ik})_{(i,k)\in \Upsilon},\Phi_{l_0}),\quad 
A_{l_0ikv}=(Q_{l_0ikv},(X_j)_{j=0,1})
\label{AI7}\\
&&Q_{l_0ikv}=(m^*+1,1,m^*,\{0,1\},\tau_{l_0ikv},\beta), \label{AI8}
\\
&& \tau_{l_0ikv}(i_0)=s_{l_0i}+2^{-l_0}(t_k+v)\;(i_0=0,1),
\quad
X_j=Id_{m^*}\quad(j=0,1),
\notag\\
&& h_{l_0ik}: \prod_{(i_1,k_1)<(i,k)}\Z[0,2^{m^*})\to \Z[0,2^{m^*}),
\quad h_{l_0ik}((x_{i_1k_1})_{(i_1,k_1)<(i,k)})=0_{m^*},
\notag\\
&&\Phi_{l_0}((x_{ik})_{(i,k)\in \Upsilon})= \sum_{i=1}^{2^{dl_0}}\sum_{j=1}^{\kappa} \sum_{k=1}^{\kappa}a_{jk}(v)\gamma(x_{ik})SR_{l_0i}\phi_j,\label{AJ2}
\end{eqnarray}
(compare also \eqref{WF1}). 
We have
\begin{equation}
\label{Y9}
n_\q(A_{l_0v}^{(0)})=\kappa 2^{dl_0}
\end{equation}
and at input $f\in \mathcal{F}(D,\K)$ the deterministic output of algorithm $A_{l_0v}^{(0)}$ is 
\begin{equation}
\label{C8}
\sum_{i=1}^{2^{dl_0}}\sum_{j=1}^{\kappa} \sum_{k=1}^{\kappa}a_{jk}(v)\gamma \circ \beta(f(s_{l_0i}+2^{-l_0}(t_k+v))) SR_{l_0i}\phi_j\in L_q(D_1)
\end{equation}
(more formally, $A_{l_0v}^{(0)}(f)$ is the the Dirak probability concentrated at the point given by \eqref{C8}).

Finally we define the algorithm $A_v^{(1)}$ on $(\mathcal{F}(D,\K),L_q^{\mathcal{I}_{1,l}},\K,\Lambda)$ by composition according to Lemma \ref{lem:2a} with 
\begin{eqnarray*}
\psi:L_q(D_1)\times \prod_{l=l_0}^{l_1-1} L_q^{\mathcal{I}_{1,l}}\to L_q(D_1),\quad 
\psi(g^{(0)}, g_{l_0}, \dots,g_{l_1-1})=g^{(0)}+2c(1)\sum_{l=l_0}^{l_1-1} \Theta_l^{-1}2^{-rl}V_lg_l,
\end{eqnarray*}
with $V_l$ given by \eqref{AT5}, thus
\begin{equation}
A_v^{(1)}=A_{l_0v}^{(0)}+2c(1)\sum_{l=l_0}^{l_1-1} \Theta_l^{-1}2^{-rl}V_l A_{lv}.
\label{AH9}
\end{equation}
Then by  \eqref{D0} and \eqref{Y9}
\begin{equation}
n_\q(A_v^{(1)})\le\kappa 2^{dl_0} +2\kappa''\sum_{l=l_0}^{l_1-1}  \nu_ln_l \label{AH7}.
\end{equation}
For the definition of ${\rm UEq}$ we refer to \eqref{AK8} and \eqref{AK9}.

\begin{proposition} \label{pro:3} Let $r,d\in\N$,  $1 \leq p,q \le \infty$,  and assume \eqref{BB1}. Then there is a constant $c>0$ such that the following holds.
 
(i) For any $l_1\in\N_0$ and  $\mathcal{E}\in {\rm UEq}(\mathcal{B}_{W^r_p(D)})$ there are $M,M^*\in\N$ such that for $m\ge M$, $m^*\ge M^*$, $\Theta_{l_1}>0$, and $f\in \mathcal{E}$
\begin{eqnarray}
\lefteqn{\Prm\Bigg\{e\bigg(S,A_{\zeta_m}^{(1)},f,L_q(D_1),\sum_{l=l_0}^{l_1-1}e^{-\nu_l/8}\bigg)}
\notag\\
&\le& c\,\Theta_{l_1}^{-1}2^{-rl_1+\left(\frac{1}{p}-\frac{1}{q}\right)_+ d_1l_1}+
c\sum_{l=l_0}^{l_1-1} \Theta_l^{-1}2^{-rl}e(S^{\mathcal{I}_{1,l},\mathcal{I}_{2,l}},A_l,\mathcal{B}_{L_p^{\mathcal{I}_{1,l}, \mathcal{I}_{2,l}}},L_q^{\mathcal{I}_{1,l}})\Bigg\}\ge 1-\sum_{l=l_0}^{l_1} \Theta_l.\quad\;
\label{AQ4}
\end{eqnarray}

(ii) Moreover, for any $l_1\in\N_0$  and $\Theta>0$ there is an $M^*\in\N$ such that for $m^*\ge M^*$, $\Theta_{l_1}>0$, and $f\in \mathcal{B}_{\mathcal{W}^r_p(D)}$ 
\begin{eqnarray}
\lefteqn{\Prm\Bigg\{e\bigg(S,A_\zeta^{(1)},f,L_q(D_1),\sum_{l=l_0}^{l_1-1}e^{-\nu_l/8}\bigg)}
\notag\\
&\le& c\,\Theta_{l_1}^{-1}2^{-rl_1+\left(\frac{1}{p}-\frac{1}{q}\right)_+ d_1l_1}+
c\sum_{l=l_0}^{l_1-1} \Theta_l^{-1}2^{-rl}e(S^{\mathcal{I}_{1,l},\mathcal{I}_{2,l}},A_l,\mathcal{B}_{L_p^{\mathcal{I}_{1,l}, \mathcal{I}_{2,l}}},L_q^{\mathcal{I}_{1,l}})\Bigg\}
\notag\\
&\ge& 1-\Theta-\sum_{l=l_0}^{l_1} \Theta_l.
\label{AQ3}
\end{eqnarray}

(iii) Finally, if \eqref{PA2} holds, then for any $l_1\in\N_0$  there is an $M^*\in\N$ such that for $m^*\ge M^*$  and $\Theta_l= 1$ $(l_0\le l<l_1)$ we have
\begin{eqnarray}
\lefteqn{e\bigg(S,A_0^{(1)},\mathcal{B}_{\mathcal{W}_p^r(D)},L_q(D_1),\sum_{l=l_0}^{l_1-1}e^{-\nu_l/8}\bigg)}
\notag\\
&\le& c\,2^{-rl_1+\left(\frac{1}{p}-\frac{1}{q}\right)_+ d_1l_1}+
c\sum_{l=l_0}^{l_1-1} 2^{-rl}e(S^{\mathcal{I}_{1,l},\mathcal{I}_{2,l}},A_l,\mathcal{B}_{L_p^{\mathcal{I}_{1,l}, \mathcal{I}_{2,l}}},L_q^{\mathcal{I}_{1,l}}).
\label{AQ5}
\end{eqnarray}

\end{proposition}
\begin{proof} Let $m^*\in\N$,  $f\in \mathcal{B}_{\mathcal{W}^r_p(D)}$, and  
\begin{eqnarray}
&&v\in D(\delta_0,m^*,f):=\big\{v\in D(\delta_0):\,|f(s_{li}+2^{-l}t_{jk}(v))|\le 2^{ (m^*-5)/4}\;
\notag\\[.1cm]
&& \hspace{2cm}\text{for all}\; l\in\{l_0\}\cup\Z(l_0,l_1),i\in\Z[1,2^{dl}], j\in\Z[1,\kappa'],k\in\Z[1,\kappa'']\big\}.\label{AR4}
\end{eqnarray}
We recall that the $b_{jk}$ are polynomials on $\R^d$, see \eqref{AG2}, and set
\begin{equation}
\label{AH5}
c(2)=(2c(1))^{-1}\max_{1\le j\le\kappa'}\sum_{k=1}^{\kappa''}\|b_{jk}\|_{C(D)},
\end{equation}
with $c(1)$ being the constant from \eqref{R7}--\eqref{AJ5}. 
By Lemma \ref{lem:2b}, \eqref{eq:41}, and \eqref{AH9} 
\begin{eqnarray}
 \lefteqn{e(S, A_v^{(1)},f,L_q(D_1),\sum_{l=l_0}^{l_1-1} e^{-\nu_l/8} )}
\notag\\
&\le& \|(S-SP_{l_1v})f\|_{L_q(D_1)}+e( SP_{l_0v},A_{l_0v}^{(0)},f,L_q(D_1),0)
\nonumber\\
&&+\sum_{l=l_0}^{l_1-1} e(V_l S^{\mathcal{I}_{1,l},\mathcal{I}_{2,l}} U_{lv},2c(1)\Theta_l^{-1}2^{-rl}V_lA_{lv},f,L_q(D_1),e^{-\nu_l/8}).\label{D7}
\end{eqnarray}

Next we estimate the terms of \eqref{D7}. Using \eqref{WH4} and taking into account that the $\phi_j$ and $a_{jk}$ are polynomials, we readily derive 
\begin{equation*}
\|SR_{l_0i}\phi_j\|_{L_q(D_1)}=2^{-d_1l_0/q-d_2l_0}\|S\phi_j\|_{L_q(D_1)}\le c2^{-d_1l_0/q-d_2l_0}\quad  (1\le i\le 2^{dl_0}, 1\le j\le \kappa),
\end{equation*}
which together with \eqref{WF1}--\eqref{AP8}, \eqref{C8}, \eqref{E4}, and \eqref{AR4} gives
\begin{eqnarray}
\label{C9}
\lefteqn{e(SP_{l_0v},A_{l_0v}^{(0)},f,L_q(D_1),0)}
\nonumber\\
&=&\Bigg\|\sum_{i=1}^{2^{dl_0}}\sum_{j,k=1}^{\kappa} a_{jk}(v)(f(s_{l_0i}+2^{-l_0}(t_k+v)-\gamma \circ \beta(f(s_{l_0i}+2^{-l_0}(t_k+v))) SR_{l_0i}\phi_j\Bigg\|_{L_q(D_1)}
\nonumber\\
&\le&c2^{-d_1l_0/q-d_2l_0}\sum_{i=1}^{2^{dl_0}} \sum_{j,k=1}^{\kappa}|a_{jk}(v)(f(s_{l_0i}+2^{-l_0}t_{1,k+\kappa}(v))-\gamma \circ \beta(f(s_{l_0i}+2^{-l_0}t_{1,k+\kappa}(v))| 
\nonumber\\
&\le&c2^{d_1l_0(1-1/q)-m^*/4}.
\end{eqnarray}
By (\ref{WF3}), \eqref{AI0}, (\ref{E4}), \eqref{AR4}, and \eqref{AH5} we have for $l_0\le l<l_1$
\begin{eqnarray}
\label{H1}
\lefteqn{|(2c(1))^{-1}\Theta_l2^{rl}(U_{lv} f)((i_1,j),i_2)-\Gamma_{lv}(f)((i_1,j),i_2)|}\nonumber\\
&\le& (2c(1))^{-1}\Theta_l2^{rl}\sum_{k=1}^{\kappa''}|b_{jk}(v)|
|f(s_{li}+2^{-l}t_{jk}(v))-\gamma(\beta(f(s_{li}+2^{-l}t_{jk}(v))))|\nonumber\\
&\le& (2c(1))^{-1}\Theta_l 2^{rl-(m^*-3)/4}\sum_{k=1}^{\kappa''}|b_{jk}(v)|\le c(2)\Theta_l 2^{rl-(m^*-3)/4},
\end{eqnarray}
and consequently, 
\begin{eqnarray}
\|(2c(1))^{-1}\Theta_l2^{rl}U_{lv} f-\Gamma_{lv}(f)\|_{L_p^{\mathcal{I}_{1,l}, \mathcal{I}_{2,l}}}
&\le&  c(2)\Theta_l2^{rl-(m^*-3)/4}. \label{E5}
\end{eqnarray}
Moreover, by Lemma \ref{lem:7},
\begin{eqnarray}
 \lefteqn{e\big(V_l S^{\mathcal{I}_{1,l},\mathcal{I}_{2,l}} U_{lv},2c(1)\Theta_l^{-1}2^{-rl}V_lA_{lv},f,L_q(D_1),e^{-\nu_l/8}\big)}
\nonumber\\[.2cm]
&\le&
 2c(1)\Theta_l^{-1}2^{-rl}\|V_l\|e\big((2c(1))^{-1}\Theta_l2^{rl}S^{\mathcal{I}_{1,l},\mathcal{I}_{2,l}} U_{lv},A_{lv},f,L_q^{\mathcal{I}_{1,l}},e^{-\nu_l/8}\big),\label{D8}
\end{eqnarray}
According to Lemma \ref{lem:2f} and \eqref{Y1} we have   
\begin{equation}
\label{AS3}
e(S^{\mathcal{I}_{1,l}, \mathcal{I}_{2,l}},\wt{A}_l,\mathcal{B}_{L_p^{\mathcal{I}_{1,l}, \mathcal{I}_{2,l}}},L_q^{\mathcal{I}_{1,l}},e^{-\nu_l/8})\le   3e(S^{\mathcal{I}_{1,l}, \mathcal{I}_{2,l}},A_l,\mathcal{B}_{L_p^{\mathcal{I}_{1,l}, \mathcal{I}_{2,l}}},L_q^{\mathcal{I}_{1,l}}),
\end{equation}
and using furthermore Lemma \ref{lem:8}(i), Lemma \ref{lem:2},  \eqref{D0}, \eqref{E5}, and \eqref{AS3}, we obtain
\begin{eqnarray}
\lefteqn{e( (2c(1))^{-1}\Theta_l2^{rl}S^{\mathcal{I}_{1,l},\mathcal{I}_{2,l}} U_{lv},A_{lv},f,L_q^{\mathcal{I}_{1,l}},e^{-\nu_l/8})}
\nonumber\\
&\le& e( S^{\mathcal{I}_{1,l},\mathcal{I}_{2,l}} \Gamma_{lv},A_{lv},f,L_q^{\mathcal{I}_{1,l}},e^{-\nu_l/8})+\|(2c(1))^{-1}\Theta_l2^{rl}S^{\mathcal{I}_{1,l},\mathcal{I}_{2,l}} U_{lv}f-S^{\mathcal{I}_{1,l},\mathcal{I}_{2,l}} \Gamma_{lv}(f)\|
\nonumber\\
&\le&  e( S^{\mathcal{I}_{1,l},\mathcal{I}_{2,l}} ,\wt{A}_l,\Gamma_{lv}(f),L_q^{\mathcal{I}_{1,l}},e^{-\nu_l/8})+c(2)\Theta_l2^{rl-(m^*-3)/4}
\|S^ {\mathcal{I}_{1,l},\mathcal{I}_{2,l}}\|
\nonumber\\
&\le& 3e(S^{\mathcal{I}_{1,l}, \mathcal{I}_{2,l}},A_l,\Gamma_{lv}(f),L_q^{\mathcal{I}_{1,l}})+c(2)\Theta_l2^{rl-(m^*-3)/4} N_{1,l}^{(1/p-1/q)_+}.\label{D9}
\end{eqnarray}
We conclude from \eqref{D7}, \eqref{C9}, \eqref{D8}, and \eqref{D9} that for $m^*\in\N$,  $f\in \mathcal{B}_{\mathcal{W}^r_p(D)}$, and  $v\in D(\delta_0,m^*,f)$
\begin{eqnarray}
 \lefteqn{e(S, A_v^{(1)},f,L_q(D_1),\sum_{l=l_0}^{l_1-1} e^{-\nu_l/8} )}
\notag\\
&\le& \|(S-SP_{l_1v})f\|_{L_q(D_1)}+c2^{d_1l_0(1-1/q)-m^*/4}
\nonumber\\
&&+c\sum_{l=l_0}^{l_1-1} \Big(\Theta_l^{-1}2^{-rl}e(S^{\mathcal{I}_{1,l}, \mathcal{I}_{2,l}},A_l,\Gamma_{lv}(f),L_q^{\mathcal{I}_{1,l}})+2^{-m^*/4} N_{1,l}^{(1/p-1/q)_+} \Big).\label{AR5}
\end{eqnarray}

Now we complete the proof separately for the three cases 
and start with (iii). By assumption $\Theta_l\equiv 1$ and  $\mathcal{B}_{\mathcal{W}_p^r(D)}$ is a bounded subset of $C(D)$, so we choose $m^*$ such that 
\begin{equation}
\label{AQ0}
\sup_{f\in \mathcal{B}_{\mathcal{W}_p^r(D)}}\|f\|_{C(D)}\le 2^{ (m^*-5)/4},
\end{equation}
hence $D(\delta_0,m^*,f)=D(\delta_0)$ for all  $f\in\mathcal{B}_{\mathcal{W}_p^r(D)}$.
By \eqref{AJ5} and \eqref{H1}
\begin{eqnarray}
\|\Gamma_{l,0}f\|_{L_p^{N_l}} &\le& (2c(1))^{-1}2^{rl}  \|U_{l{\zeta_0}}f\|_{L_p^{N_l}}+c(2)2^{rl-(m^*-3)/4}
\notag\\
&\le& 1/2 +c(2) 2^{rl-(m^*-3)/4}  \le 1,
\label{AQ1}
\end{eqnarray}
provided $m^*$ also satisfies
\begin{equation}
\label{AJ7}
c(2) 2^{rl_1-(m^*-3)/4}\le \frac{1}{2}.
\end{equation}
By \eqref{AC2} of Lemma \ref{lem:6} 
\begin{eqnarray}
\|(S-SP_{l_1,0})f\|_{L_q(D_1)}
&\le&  c 2^{-rl_1+(1/p-1/q)_+d_1l_1}.
\label{D6a}
\end{eqnarray}
Combining \eqref{AR5}, \eqref{AQ1}, and \eqref{D6a}, we obtain for $A_0^{(1)}$
\begin{eqnarray*}
\lefteqn{  e\bigg(S, A_0^{(1)},f,L_q(D_1),\sum_{l=l_0}^{l_1-1} e^{-\nu_l/8} \bigg)}
\nonumber\\
&\le& c 2^{-rl_1+(1/p-1/q)_+d_1l_1}+c2^{d_1l_0(1-1/q)-m^*/4}
\nonumber\\
&&+c\sum_{l=l_0}^{l_1-1} 
\Big(2^{-rl}e\big(S^{\mathcal{I}_{1,l}, \mathcal{I}_{2,l}},A_l,\mathcal{B}_{L_p^{\mathcal{I}_{1,l}, \mathcal{I}_{2,l}}},L_q^{\mathcal{I}_{1,l}}\big)+2^{-m^*/4} N_{1,l}^{(1/p-1/q)_+}\Big)
\nonumber\\
&\le& c 2^{-rl_1+(1/p-1/q)_+d_1l_1}
+c\sum_{l=l_0}^{l_1-1} 
2^{-rl}e\big(S^{\mathcal{I}_{1,l}, \mathcal{I}_{2,l}},A_l,\mathcal{B}_{L_p^{\mathcal{I}_{1,l}, \mathcal{I}_{2,l}}},L_q^{\mathcal{I}_{1,l}}\big),
\end{eqnarray*}
if $m^*$ also satisfies 
\begin{equation}
\label{AP1}
2^{d_1l_0(1-1/q)-m^*/4} +2^{-m^*/4} \sum_{l=l_0}^{l_1-1} N_{1,l}^{(1/p-1/q)_+}\le 2^{-rl_1+(1/p-1/q)_+d_1l_1}.
\end{equation}
It remains to choose $M^*\in \N$ in such a way that \eqref{AQ0}, \eqref{AJ7}, and \eqref{AP1} hold for all $m^*\ge M^*$, completing the proof of (iii).

In the case (i) we are given $\mathcal{E}\in {\rm UEq}(\mathcal{B}_{\mathcal{W}_p^r(D)})$ and can choose $m^*$ in such a way that 
\begin{equation}
\label{E3}
\sup_{f\in \mathcal{E}}\|f\|_{C(D)}\le 2^{ (m^*-5)/4},
\end{equation}
thus $D(\delta_0,m^*,f)=D(\delta_0)$ for all  $f\in\mathcal{E}$. Relations \eqref{E5} and \eqref{R8} imply for $f\in \mathcal{E}$, $l_0\le l<l_1$
\begin{eqnarray}
\lefteqn{\EE\|\Gamma_{l{\zeta_m}}f\|_{L_p^{N_l}} \le (2c(1))^{-1}\Theta_l2^{rl} \EE \|U_{l{\zeta_m}}f\|_{L_p^{N_l}}+c(2)\Theta_l2^{rl-(m^*-3)/4}}
\nonumber\\
&\le&\Theta_l2^{rl-1}  (2^{-rl}+\epsilon(\mathcal{E},  2^{-m}) + 2^{-m})+c(2)\Theta_l2^{rl-(m^*-3)/4} 
\nonumber\\[.1cm]
&\le&\Theta_l/2+\Theta_l2^{rl-1}  (\epsilon(\mathcal{E},  2^{-m}) + 2^{-m})+c(2)\Theta_l2^{rl-(m^*-3)/4} \le \Theta_l
\label{C4}
\end{eqnarray}
provided $m$ and $m^*$ satisfy
\begin{equation}
\label{AI2}
2^{rl_1-1}  (\epsilon(\mathcal{E},  2^{-m}) + 2^{-m})+c(2) 2^{rl_1-(m^*-3)/4}\le \frac{1}{2}.
\end{equation}
By \eqref{AC0} of Lemma \ref{lem:6}, denoting the involved constant by $c(3)$, we have
\begin{eqnarray}
\EE\|Sf-SP_{l_1\zeta_m} f\|_{L_q(D_1)}\le c(3)(2^{-rl_1+(1/p-1/q)_+d_1l_1}+\epsilon(\mathcal{E}, 2^{-m}) + 2^{-m}).
\label{AQ9}
\end{eqnarray}
Let 
\begin{eqnarray}
\Omega_{lf}&=&\{\omega\in\Omega:\, \|\Gamma_{l\zeta_m(\omega)}(f)\|_{L_p^{N_l}}\le 1\}\quad (l_0\le l<l_1)
\label{D4}\\
\Omega_{l_1f}&=&\big\{\omega\in\Omega:\, \|Sf-SP_{l_1\zeta_m(\omega)} f\|_{L_q(D_1)}
\notag\\
&&\hspace{1.9cm}\le c(3)\Theta_{l_1}^{-1}(2^{-rl_1+(1/p-1/q)_+d_1l_1}+\epsilon(\mathcal{E}, 2^{-m}) + 2^{-m})\big\}.
\label{D5}
\end{eqnarray}
It follows from \eqref{C4}  and \eqref{AQ9} that
\begin{eqnarray*}
\Prm(\Omega_{lf})\ge 1-\Theta_l\quad(l_0\le l<l_1), \quad\Prm(\Omega_{l_1f})\ge 1-\Theta_{l_1},
\end{eqnarray*}
hence 
\begin{equation}
\Prm\bigg(\bigcap_{l=l_0}^{l_1} \Omega_{l,f}\bigg)\ge 1-\sum_{l=l_0}^{l_1} \Theta_l.\label{D2}
\end{equation}
Joining \eqref{AR5}, \eqref{D4}, and \eqref{D5}, we obtain for $\omega\in \bigcap_{l=l_0}^{l_1} \Omega_{l,f}$
\begin{eqnarray}
\lefteqn{  e\bigg(S, A_{\zeta_m(\omega)}^{(1)},f,L_q(D_1),\sum_{l=l_0}^{l_1-1} e^{-\nu_l/8} \bigg)}
\nonumber\\
&\le& c \Theta_{l_1}^{-1}(2^{-rl_1+(1/p-1/q)_+d_1l_1}+\epsilon(\mathcal{E}, 2^{-m}) + 2^{-m})+c2^{d_1l_0(1-1/q)-m^*/4}
\nonumber\\
&&+c\sum_{l=l_0}^{l_1-1} 
\Big(\Theta_l^{-1}2^{-rl}e\big(S^{\mathcal{I}_{1,l}, \mathcal{I}_{2,l}},A_l,\mathcal{B}_{L_p^{\mathcal{I}_{1,l}, \mathcal{I}_{2,l}}},L_q^{\mathcal{I}_{1,l}}\big)+2^{-m^*/4} N_{1,l}^{(1/p-1/q)_+}\Big)
\nonumber\\
&\le& c\Theta_{l_1}^{-1} 2^{-rl_1+(1/p-1/q)_+d_1l_1}
+c\sum_{l=l_0}^{l_1-1} 
\Theta_l^{-1}2^{-rl}e\big(S^{\mathcal{I}_{1,l}, \mathcal{I}_{2,l}},A_l,\mathcal{B}_{L_p^{\mathcal{I}_{1,l}, \mathcal{I}_{2,l}}},L_q^{\mathcal{I}_{1,l}}\big),
\label{E0}
\end{eqnarray}
provided $m$ and $m^*$ also satisfy
\begin{equation}
\label{AI4}
\epsilon(\mathcal{E}, 2^{-m}) + 2^{-m}+2^{d_1l_0(1-1/q)-m^*/4} +2^{-m^*/4}\sum_{l=l_0}^{l_1-1} N_{1,l}^{(1/p-1/q)_+}\le 2^{-rl_1+(1/p-1/q)_+d_1l_1}.
\end{equation}
Now we choose  $M,M^*\in\N$ in such a way that for all $m\ge M$, $m^*\ge M^*$ relations 
\eqref{E3}, \eqref{AI2}, and \eqref{AI4} hold. Then the desired estimate \eqref{AQ4} follows from \eqref{D2} and \eqref{E0}.

In case of (ii) we are given the additional parameter $\Theta>0$.
Let $f\in \mathcal{B}_{\mathcal{W}^r_p(D)}$, then $\|f\|_{L_1(D)}\le 1$, and we define for $m^*\in\N$
\begin{equation*}
D(m^*,f)=\{t\in D: |f(t)|> 2^{ (m^*-5)/4}\}. 
\end{equation*}
It follows that
\begin{equation*}
|D(m^*,f)|\le 2^{-(m^*-5)/4},
\end{equation*}
with $|D(m^*,f)|$ the Lebesgue measure of $D(m^*,f)$. According to \eqref{AP8}  we have for $1\le k\le \kappa$
\begin{eqnarray*}
\lefteqn{\Prm\{s_{li}+2^{-l}t_{jk}(\zeta)\in D(m^*,f)\}}
\notag\\
&=&\Prm\{s_{li}+2^{-l}(s_{1,i_0}+2^{-1}(t_k+\zeta))\in D(m^*,f)\}
\notag\\
&=&
\Prm\{\zeta\in 2(2^l(D(m^*,f)-s_{li})-s_{1,i_0})-t_k\}
\le \delta_0^{-d}2^{d(l+1)}\big|D(m^*,f)\big|\le \delta_0^{-d}2^{d(l+1)-(m^*-5)/4},
\end{eqnarray*}
and similarly for $\kappa+1\le k\le 2\kappa$ 
\begin{eqnarray*}
\lefteqn{\Prm\{s_{li}+2^{-l}t_{jk}(\zeta)\in D(m^*,f)\}}
\notag\\
&=&\Prm\{s_{li}+2^{-l}(t_{k-\kappa}+\zeta)\in D(m^*,f)\}
\\
&=&
\Prm\{\zeta\in 2^l(D(m^*,f)-s_{li})-t_{k-\kappa}\}
\le \delta_0^{-d}2^{dl}\big|D(m^*,f)\big|\le \delta_0^{-d}2^{dl-(m^*-5)/4}.
\end{eqnarray*}
Consequently, setting 
\begin{equation*}
\Omega_f:= \{\omega\in\Omega: \zeta(\omega)\in D(\delta_0,m^*,f)\},
\end{equation*}
we obtain, compare \eqref{AR4},
\begin{eqnarray}
\Prm(\Omega_f)
&=&\Prm\big\{|f(s_{li}+2^{-l}t_{jk}(\zeta))|\le 2^{ (m^*-5)/4} 
\notag\\
&&\quad\text{for all}\; l\in\{l_0\}\cup\Z(l_0,l_1),i\in\Z[1,2^{dl}],
j\in\Z[1,\kappa'],k\in\Z[1,\kappa'']\big\}   
\notag\\
&\ge& 1- \kappa'\kappa'' \delta_0^{-d}2^{-(m^*-5)/4+d(2l_1+1)}\ge 1-\Theta\label{AP9}
\end{eqnarray}
provided
\begin{equation}
\label{AR7}
 \kappa'\kappa'' \delta_0^{-d}2^{-(m^*-5)/4+d(2l_1+1)}\le \Theta.
\end{equation}
The rest of the proof of (ii) goes much along the lines of that of (i). 
From \eqref{R7} and \eqref{E5} we conclude
\begin{eqnarray}
\int_{\Omega_f}\|\Gamma_{l{\zeta}}f\|_{L_p^{N_l}}d\Prm(\omega) &\le& (2c(1))^{-1}\Theta_l2^{rl} \EE \|U_{l{\zeta}}f\|_{L_p^{N_l}}+c(2)\Theta_l2^{rl-(m^*-3)/4}
\nonumber\\
&\le&\Theta_l/2+c(2)\Theta_l2^{rl-(m^*-3)/4} \le \Theta_l,
\label{AQ6}
\end{eqnarray}
provided  $m^*$ also satisfies
\begin{equation}
\label{AQ7}
c(2) 2^{rl_1-(m^*-3)/4}\le \frac{1}{2}.
\end{equation}
Relation \eqref{AC2} of Lemma \ref{lem:6}, with the constant renamed to $c(4)$, gives
\begin{eqnarray}
\EE\|Sf-SP_{l_1\zeta} f\|_{L_q(D_1)}\le c(4)2^{-rl_1+(1/p-1/q)_+d_1l_1}.
\label{AQ8}
\end{eqnarray}
With 
\begin{eqnarray}
\Omega_{lf}&=&\{\omega\in\Omega:\, \|\Gamma_{l\zeta(\omega)}(f)\|_{L_p^{N_l}}\le 1\}\quad(l_0\le l<l_1)
\\
\Omega_{l_1f}&=&\big\{\omega\in\Omega:\, \|Sf-SP_{l_1\zeta(\omega)} f\|_{L_q(D_1)}\le c(4)\Theta_{l_1}^{-1}2^{-rl_1+(1/p-1/q)_+d_1l_1}\}
\label{AR0}
\end{eqnarray}
we have from \eqref{AQ6}, \eqref{AQ8}, and \eqref{AP9}
\begin{eqnarray*}
\Prm(\Omega_{lf}^c\cap \Omega_f)\le \Theta_l\quad(l_0\le l< l_1),\quad \Prm(\Omega_{l_1f}^c)\le\Theta_{l_1},\quad\Prm(\Omega_f^c)\le \Theta.
\end{eqnarray*}
Therefore
\begin{eqnarray}
 &&\Prm\bigg(\Omega_f\cap\bigcap_{l=l_0}^{l_1} \Omega_{lf}\bigg)
=\Prm\bigg(\Omega_f\cap\bigcap_{l=l_0}^{l_1-1} \big(\Omega_{lf}\cup\Omega_f^c\big)\cap \Omega_{l_1f}\bigg)
\ge 1-\Theta-\sum_{l=l_0}^{l_1} \Theta_l,
\label{AR1}
\end{eqnarray}
and \eqref{AR5}, \eqref{AR0}, and \eqref{AR1} give this time for $\omega\in \Omega_f\cap\bigcap_{l=l_0}^{l_1} \Omega_{l,f}$
\begin{eqnarray}
\lefteqn{  e\bigg(S, A_{\zeta(\omega)}^{(1)},f,L_q(D_1),\sum_{l=l_0}^{l_1-1} e^{-\nu_l/8} \bigg)}
\nonumber\\
&\le& c \Theta_{l_1}^{-1}2^{-rl_1+(1/p-1/q)_+d_1l_1}+c2^{d_1l_0(1-1/q)-m^*/4}
\nonumber\\
&&+c\sum_{l=l_0}^{l_1-1} 
\Big(\Theta_l^{-1}2^{-rl}e\big(S^{\mathcal{I}_{1,l}, \mathcal{I}_{2,l}},A_l,\mathcal{B}_{L_p^{\mathcal{I}_{1,l}, \mathcal{I}_{2,l}}},L_q^{\mathcal{I}_{1,l}}\big)+2^{-m^*/4} N_{1,l}^{(1/p-1/q)_+}\Big)
\nonumber\\
&\le& c\Theta_{l_1}^{-1} 2^{-rl_1+(1/p-1/q)_+d_1l_1}
+c\sum_{l=l_0}^{l_1-1} 
\Theta_l^{-1}2^{-rl}e\big(S^{\mathcal{I}_{1,l}, \mathcal{I}_{2,l}},A_l,\mathcal{B}_{L_p^{\mathcal{I}_{1,l}, \mathcal{I}_{2,l}}},L_q^{\mathcal{I}_{1,l}}\big),
\label{AR3}
\end{eqnarray}
provided  $m^*$ also fulfills
\begin{equation}
\label{AR2}
2^{d_1l_0(1-1/q)-m^*/4} +2^{-m^*/4}\sum_{l=l_0}^{l_1-1} N_{1,l}^{(1/p-1/q)_+}\le 2^{-rl_1+(1/p-1/q)_+d_1l_1}.
\end{equation}
If we choose  $M^*\in\N$ so that for all  $m^*\ge M^*$ relations 
\eqref{AR7}, \eqref{AQ7}, and \eqref{AR2} hold, then  \eqref{AQ3} follows from \eqref{AP9}, \eqref{AR1} and \eqref{AR3}.

\end{proof}

From now on we identify $S^{\mathcal{I}_{1,l},\mathcal{I}_{2,l}}$, $L_p^{\mathcal{I}_{1,l},\mathcal{I}_{2,l}}$, and $L_q^{\mathcal{I}_{1,l}}$ with $S^{N_{1,l},N_{2,l}}$,  $L_p^{N_{1,l},N_{2,l}}$, and $L_q^{N_{1,l}}$, respectively, see \eqref{AS8} and \eqref{eq:26}. For the next result we formulate the following conditions on $\nu_l\in \N$ and  $\Theta_l>0$:
\begin{eqnarray}
\sum_{l=l_0}^{l_1-1}e^{-\nu_l/8}&\le& \frac{1}{12},
\label{X8}\\
\sum_{l=l_0}^{l_1-1} \Theta_l&\le& \frac{1}{12}. 
\label{AS6}
\end{eqnarray}
\begin{corollary} \label{cor:2} Let $r,d\in\N$,  $1 \leq p,q \le \infty$,  and assume that \eqref{BB1} holds. 
There is a constant $c>0$ such that for $l_0,l_1\in \N_0$, $l_0\le l_1$, $n_l\in \N$, $\nu_l\in \N$ $(l_0\le l<l_1)$ satisfying \eqref{X8}, and  
\begin{equation}
\wt{n}=\kappa 2^{dl_0} +2\kappa''\sum_{l=l_0}^{l_1-1}  \nu_ln_l,
\label{AS9}
\end{equation}
the following hold: If $\Theta_l>0$ fulfill \eqref{AS6}, then
\begin{eqnarray}
\lefteqn{\sup_{\mathcal{E}\in {\rm UEq}(\mathcal{B}_{\mathcal{W}_p^r(D)})}e_{\wt{n}}^\q(S,\mathcal{E},L_q(D_1))}
\notag\\
&\le& c\,2^{-rl_1+\left(\frac{1}{p}-\frac{1}{q}\right)_+ d_1l_1}+
c\sum_{l=l_0}^{l_1-1} \Theta_l^{-1}2^{-rl}e_{n_l}^\q(S^{N_{1,l},N_{2,l}},\mathcal{B}_{L_p^{N_{1,l},N_{2,l}}},L_q^{N_{1,l}}).
\label{AS4}
\end{eqnarray}
Moreover, if the embedding condition \eqref{PA2} holds, then 
\begin{eqnarray}
e_{\wt{n}}^\q(S,\mathcal{B}_{W^r_p(D)},L_q(D_1))
&\le& c\,2^{-rl_1+\left(\frac{1}{p}-\frac{1}{q}\right)_+ d_1l_1}+
c\sum_{l=l_0}^{l_1-1}2^{-rl}e_{n_l}^\q(S^{N_{1,l},N_{2,l}},\mathcal{B}_{L_p^{N_{1,l},N_{2,l}}},L_q^{N_{1,l}}).\quad
\label{AS5}
\end{eqnarray}
\end{corollary}
\begin{proof} Let $\delta>0$ and let 
$A_l$  be a quantum algorithm for $\big(L_p^{N_{1,l},N_{2,l}},L_q^{N_{1,l}},S^{N_{1,l},N_{2,l}},\K,\Lambda^{N_{1,l},N_{2,l}}\big)$  with  $n_\q(A_l)\le n_l$, see \eqref{AS0}, and 
\begin{equation}
\label{AT0}
e(S^{N_{1,l},N_{2,l}},A_l,\mathcal{B}_{L_p^{N_{1,l}, N_{2,l}}},L_q^{N_{1,l}})\le e_{n_l}^\q(S^{N_{1,l},N_{2,l}},\mathcal{B}_{L_p^{N_{1,l},N_{2,l}}},L_q^{N_{1,l}})+\delta.
\end{equation}

Relation \eqref{AS5} directly follows from (iii) of Proposition \ref{pro:3} together with \eqref{AH7}, \eqref{X8} and \eqref{AT0}.
To show \eqref{AS4}, we use (i) of  Proposition \ref{pro:3} and turn the (finite) random family of quantum algorithms into one by the construction \eqref{Y6}--\eqref{B5} in Section 2. 
So let $c(5)$ denote the constant $c$ of Proposition \ref{pro:3}, let $\mathcal{E}\in {\rm UEq}(\mathcal{B}_{\mathcal{W}_p^r(D)})$, let $M$ and $M^*$ be according to (i) of  Proposition \ref{pro:3}, and fix $m\ge M$, $m^*\ge M^*$, $\Theta_{l_1}=1/12$,  and $f\in \mathcal{E}$. Furthermore, let $\Xi:D^{(m)}\to \Z[0,2^{dm})$ (see \eqref{AD9}) be a bijection and set $\pi=\Xi\zeta_m$, thus $\pi$ is uniformly distributed on $\Z[0,2^{dm})$. Define
\begin{eqnarray}
\lefteqn{I_0=\Xi\Bigg\{v\in D^{(m)}:\,e\bigg(S,A_v^{(1)},f,L_q(D_1),\sum_{l=l_0}^{l_1-1} e^{-\nu_l/8}\bigg)}
\notag\\
&\le& c(5)\Theta_{l_1}^{-1}2^{-rl_1+\left(\frac{1}{p}-\frac{1}{q}\right)_+ d_1l_1}+
c(5)\sum_{l=l_0}^{l_1-1} \Theta_l^{-1}2^{-rl}e(S^{N_{1,l},N_{2,l}},A_l,\mathcal{B}_{L_p^{N_{1,l}, N_{2,l}}},L_q^{N_{1,l}})\Bigg\}.
\label{AS7}
\end{eqnarray}
Then \eqref{AQ4} shows that
\begin{eqnarray}
\Prm(\{\pi\in I_0\})&\ge& 1-\sum_{l=l_0}^{l_1-1} \Theta_l-\frac{1}{12}.
\label{}
\end{eqnarray}
We put $m_0=dm$, $h_0=\lg 0\rs\in\Z[0,2^{m_0})$ and let $X_0\in \mathcal{U}(H_{m_0})$ be the Walsh-Hadamard transform on $m_0$ qubits, meaning that $X_0h_0=2^{-m_0/2}\sum_{i=0}^{2^{m_0}-1}\lg i\rs$. It follows from definitions \eqref{AJ1}--\eqref{AI8} that the family $(A_{l_0v}^{(0)})_{v\in D^{(m)}}$ is of the form \eqref{Y6}. Similarly, by \eqref{AH4}, \eqref{AI0}, Lemma \ref{lem:2}, and \eqref{D0}, for each $l$ with $l_0\le l<l_1$ the family $(A_{lv})_{v\in D^{(m)}}$ is of this form.  Therefore \eqref{AH9} implies that also $(A_v^{(1)})_{v\in D^{(m)}}$ is of the form \eqref{Y6}.
So we can define
\begin{equation*}
A^{(1)}=[A_\pi^{(1)}]=[A_{\Xi\zeta_m}^{(1)}]
\end{equation*}
according to \eqref{Y6}--\eqref{B5}. It follows from \eqref{AH7} that 
\begin{equation}
\label{C7}
n_\q(A^{(1)})=\kappa 2^{dl_0}+2\kappa''\sum_{l=l_0}^{l_1-1}\nu_l n_l.
\end{equation}
Using  \eqref{X8}, and \eqref{AS6}, we obtain
\begin{eqnarray}
\lefteqn{ 1-\bigg(1-\sum_{l=l_0}^{l_1-1} e^{-\nu_l/8}\bigg)\,\Prm\big(\{\pi\in I_0\}\big)}
\nonumber\\[.2cm]
&\le& 1-\bigg(1-\sum_{l=l_0}^{l_1-1} e^{-\nu_l/8}\bigg)\bigg(1-\sum_{l=l_0}^{l_1-1}\Theta_l-\frac{1}{12}\bigg)
\nonumber\\[.2cm]
&=& \frac{1}{12}+\sum_{l=l_0}^{l_1-1}\Theta_l+\bigg(\sum_{l=l_0}^{l_1-1} e^{-\nu_l/8}\bigg)\bigg(1-\sum_{l=l_0}^{l_1-1}\Theta_l-\frac{1}{12}\bigg)
\le \frac{1}{12}+\sum_{l=l_0}^{l_1-1}\Theta_l+\sum_{l=l_0}^{l_1-1} e^{-\nu_l/8}\le \frac{1}{4}.\quad
\label{AI6}
\end{eqnarray}
From \eqref{AI6}, Corollary \ref{cor:1} with $\theta=\sum_{l=l_0}^{l_1-1} e^{-\nu_l/8}$, and  \eqref{AS7} 
we conclude
\begin{eqnarray}
 \lefteqn{e(S, A^{(1)},f,L_q(D_1))}
\notag\\
&\le& e\left(S, A^{(1)},f,L_q(D_1),1-\bigg(1-\sum_{l=l_0}^{l_1-1} e^{-\nu_l/8}\bigg)\,\Prm\big(\{\pi\in I_0\}\big)\right)
\nonumber\\
&\le&\max_{u\in I_0}e\bigg(S, A_{\Xi^{-1}u}^{(1)},f,L_q(D_1),\sum_{l=l_0}^{l_1-1} e^{-\nu_l/8} \bigg)
=\max_{v\in \Xi^{-1} I_0}e\bigg(S, A_v^{(1)},f,L_q(D_1),\sum_{l=l_0}^{l_1-1} e^{-\nu_l/8} \bigg)
\notag\\
&\le&c(5)\Theta_{l_1}^{-1}2^{-rl_1+\left(\frac{1}{p}-\frac{1}{q}\right)_+ d_1l_1}+
c(5)\sum_{l=l_0}^{l_1-1} \Theta_l^{-1}2^{-rl}e(S^{N_{1,l},N_{2,l}},A_l,\mathcal{B}_{L_p^{N_{1,l}, N_{2,l}}},L_q^{N_{1,l}})
\notag\\
&\le&12c(5)2^{-rl_1+\left(\frac{1}{p}-\frac{1}{q}\right)_+ d_1l_1}+
c(5)\sum_{l=l_0}^{l_1-1} \Theta_l^{-1}2^{-rl}\big(e_{n_l}^\q(S^{N_{1,l},N_{2,l}},\mathcal{B}_{L_p^{N_{1,l},N_{2,l}}},L_q^{N_{1,l}})+\delta\big).\label{X9}
\end{eqnarray}
Since $\delta>0$ was arbitrary, relations  \eqref{C7}  and \eqref{X9} imply the desired result.

\end{proof}
Next we show a lower bound. For $l\in\N_0$ define 
\begin{equation}
\label{AF7}
M_{1,l}=2^{d_1l},\quad M_{2,l}=2^{d_2l}.
\end{equation}
\begin{proposition} 
\label{pro:4}  Let $r\in\N_0$, $d\in\N$,  $1 \leq p,q \le \infty$, and assume that $d_1=0$ or \eqref{V1} holds.
Then there is a constant $c>0$ such that for $l\in\N_0$ and $n\in\N$
\begin{equation*}
  \sup_{\mathcal{E}\in {\rm UEq}(\mathcal{B}_{\mathcal{W}_p^r(D)})} e_n^\q(S,\mathcal{E},L_q(D_1))\ge c2^{-rl} e_{2n}^\q(S^{M_{1,l},M_{2,l}},\mathcal{B}_{L_p^{M_{1,l},M_{2,l}}},L_q^{M_{1,l}}).
\end{equation*}
\end{proposition}
\begin{proof}
Let $l\in \N_0$, $n\in\N$. For $\iota=1,2$ let $\theta^{({\iota})}$ be a $C^{\infty}$ function on $\R^{d_{\iota}}$ 
with $\text{supp}\; \theta^{({\iota})} \subset (0,1)^{d_{\iota}}$ and
$$
\tau_{\iota}:= \int_{D_{\iota}} \theta^{({\iota})}(t)\,dt\ne 0.
$$ 
We put $\theta(s,t)=\theta^{(1)}(s)\theta^{(2)}(t)$ $(s\in D_1,t\in D_2)$,
$\sigma=\|\theta \|_{\mathcal{W}_p^r(D)}$,
and
for $1\le i\le 2^{dl}$, $1\le i_1\le 2^{d_1l}$, and  $1\le i_2\le2^{d_2l}$
\begin{equation}
\label{AB9}
\theta_{li} = R_{li}\theta,\quad\theta_{li_1}^{(1)} = R_{l i_1}^{(1)}\theta^{(1)},\quad \theta_{li_2}^{(2)} = R_{l i_2}^{(2)}\theta^{(2)},
\end{equation}
with  $R_{li}$, $R_{li_1}^{(1)}$, and $R_{li_2}^{(2)}$  given by (\ref{WH4}).
Then from \eqref{V9}
\begin{eqnarray}
\theta_{li}(s,t)&=& \theta_{li_1}^{(1)}(s)\theta_{li_2}^{(2)}(t)\quad (i=2^{d_2l}(i_1-1)+i_2,\,s\in [0,1]^{d_1},\,t\in [0,1]^{d_2})
\nonumber\\
(S\theta_{li})(s)&=&\int_{D_{l i_2}^{(2)}} \theta_{li}(s,t)\,dt 
=  \tau_2M_{2,l}^{-1}\theta_{li_1}^{(1)}(s) \quad(s\in D_1),
\label{eq:84}\\
\| \theta_{li} \|_{\mathcal{W}_p^r(D)} &\le& 2^{rl-(d_1+d_2)l/p}\|\theta \|_{\mathcal{W}_p^r(D)}=\sigma 2^{rl}(M_{1,l}M_{2,l})^{-1/p}.\notag
\end{eqnarray}
Consequently, taking into account the disjointness of the supports
of the $\theta_{li}$, we have for all $a_i\in\K\quad (i=1,\dots,2^{dl})$,
\begin{equation}
\label{eq:85}
\Big\|\sum_{i=1}^{2^{dl}} a_i\theta_{li}\Big\|_{\mathcal{W}_p^r(D)}
=\left(\sum_{i=1}^{2^{dl}} |a_i|^p\|\theta_{li}\|_{\mathcal{W}_p^r(D)}^p\right)^{1/p}
\le \sigma 2^{rl}\big\|(a_i)_{i=1}^{2^{dl}}\big\|_{L_p^{M_{1,l},M_{2,l}}} 
\end{equation}
(which, with the corresponding modifications, also holds for $p=\infty$).
Fix any $m^*\in\N$ with 
\begin{equation*}
(m^*-5)/4\ge dl/p.
\end{equation*}
Let $\beta:\K\to\Z[0,2^{m^*})$ and $\gamma:\Z[0,2^{m^*})\to\K$ be defined as in
(\ref{N1})--(\ref{N4}). For $f\in \mathcal{B}_{L_p^{M_{1,l},M_{2,l}}}$ we have 
$$
|f(i)|\le (M_{1,l}M_{2,l})^{1/p}=2^{dl/p}\le 2^{ (m^*-5)/4}.
$$
Hence by (\ref{E4}), for $1\le i \le M_{1,l}M_{2,l}$,
\begin{equation}
\label{N9}
|f(i)-\gamma(\beta(f(i)))|\le 2^{-(m^*-3)/4}\qquad (f\in \mathcal{B}_{L_p^{M_{1,l},M_{2,l}}}).
\end{equation}
Define 
\begin{eqnarray}
&&\Gamma_l: L_p^{M_{1,l},M_{2,l}} \to \mathcal{W}_p^r(D), \quad \Gamma_l f =\sum_{i=1}^{M_{1,l}M_{2,l}}  \gamma(\beta(f(i)))\theta_{li}, 
\label{T9}\\
&&\Psi_l: L_q(D_1) \rightarrow L_q^{M_{1,l}},\quad (\Psi_l g)(i_1) = M_{1,l} \int_{D_{li_1}^{(1)}} g(s)ds \quad (1\le i_1\le M_{1,l}).
\label{U0}
\end{eqnarray}
By (\ref{eq:85}) and \eqref{N9}, for  $f\in \mathcal{B}_{L_p^{M_{1,l},M_{2,l}}}$
\begin{eqnarray}
\|\Gamma_l f\|_{\mathcal{W}_p^r(D)}&\le&  \sigma  2^{rl}\| \gamma\circ\beta\circ f\|_{L_p^{M_{1,l},M_{2,l}}}\le  \sigma 2^{rl}\left(1+2^{ -(m^*-3)/4}\right),
\label{AB8}
\end{eqnarray}
and by H\"older's inequality for $q<\infty$,
\begin{eqnarray}
\| \Psi_l g \|_{L_q^{M_{1,l}}}^q & = &
M_{1,l}^{q-1} \sum_{i_1=1}^{M_{1,l}}\bigg| \int_{D_{li_1}^{(1)}} g(s)ds \bigg|^{q}
 \leq 
M_{1,l}^{q-1} \sum_{i_1=1}^{M_{1,l}}\int_{D_{li_1}^{(1)}} |g(s)|^q ds \big|D_{li_1}^{(1)}\big|^{q-1} 
         = \| g\|_{L_q(D_1)}^q, \qquad\label{AB5}
\end{eqnarray}
which with the obvious changes also holds for $q=\infty$.
Furthermore,  writing $i=(i_1,i_2)$, relations (\ref{eq:84}),  \eqref{T9}, and \eqref{U0} imply
\begin{eqnarray}
\label{eq:89}
\Psi_l  S  \Gamma_l f 
& = &
\Psi_l  S \sum_{i_1=1}^{M_{1,l}}\sum_{i_2=1}^{M_{2,l}}  \gamma(\beta(f(i_1,i_2)))\theta_{li}=
\Psi_l  \sum_{i_1=1}^{M_{1,l}}\left(\tau_2  M_{2,l}^{-1}\sum_{i_2=1}^{M_{2,l}} 
\gamma(\beta(f(i_1,i_2))) \right) \theta_{li_1}^{(1)} 
\notag \\
& = & \tau_1 \tau_2\sum_{i_1=1}^{M_{1,l}}\left( M_{2,l}^{-1}\sum_{i_2=1}^{M_{2,l}}
 \gamma(\beta(f(i_1,i_2))) \right) e_{li_1} = \tau_1\tau_2 S^{M_{1,l},M_{2,l}}(\gamma\circ\beta\circ f),
\end{eqnarray}
with $e_{li_1}$ the unit vectors in $\K^{M_{1,l}}$.

Define 
\begin{eqnarray*}
&&\eta_l:D\to \Z[1,2^{dl}],\hspace{1.5cm}\eta_l(s)=\min\{i\,|\,s\in D_{li}\}
\\
&&\rho_l:D\times\Z[0,2^{m^*})\to\K, \quad\rho_l(s,z)=\gamma(z)\theta_{\eta_l(s)}(s).
\end{eqnarray*}
Then
\begin{eqnarray*}
\Gamma_l(f)(s)&=&\sum_{i=1}^{M_{1,l}M_{2,l}} \gamma\circ\beta\circ f(i)\,\theta_{li}(s)
=\gamma\circ\beta\circ f(\eta_l(s))\,\theta_{l\eta_l(s)}(s)
=\rho_l(s,\beta\circ f\circ\eta_l(s)).
\end{eqnarray*}
So $\Gamma_l$ is of the needed form \eqref{AC5}--\eqref{F1}, 
with $\kappa=1$ and by \eqref{AB8},  
$$
\Gamma_l\big(\mathcal{B}_{L_p^{M_{1,l},M_{2,l}}}\big)\subset\sigma 2^{rl}\left(1+2^{ -(m^*-3)/4}\right)\mathcal{B}_{\mathcal{W}_p^r(D)}.
$$
We set 
\begin{equation*}
\mathcal{E}'=\sigma^{-1} 2^{-rl}\left(1+2^{ -(m^*-3)/4}\right)^{-1}\Gamma_l\big(\mathcal{B}_{L_p^{M_{1,l},M_{2,l}}}\big)\subset \mathcal{B}_{\mathcal{W}_p^r(D)}.
\end{equation*}
Since $\theta$ is a $C^\infty$ function, the definitions \eqref{AB9} of $\theta_{li}$ and \eqref{T9} of $\Gamma_l$ imply that $\mathcal{E}'$ is a bounded subset of a finite dimensional subspace of $C(D)$, hence uniformly equicontinuous. 
By \eqref{AB5} and Lemmas \ref{lem:7}, \ref{lem:2}, 
and \ref{lem:8}(iii),
\begin{eqnarray}
\lefteqn{e_{2n}^\q\left(\Psi_l\circ S\circ\Gamma_l,\mathcal{B}_{L_p^{M_{1,l},M_{2,l}}},L_q^{M_{1,l}}\right)}
\notag\\
&\le& e_{2n}^\q\left(S\circ\Gamma_l,\mathcal{B}_{L_p^{M_{1,l},M_{2,l}}},L_q(D_1)\right)
\le e_n^\q\left(S,\Gamma_l\big(\mathcal{B}_{L_p^{M_{1,l},M_{2,l}}}\big),L_q(D_1)\right)
\nonumber\\
&=& e_n^\q\left(S,\sigma 2^{rl}\left(1+2^{ -(m^*-3)/4}\right)
\mathcal{E}',L_q(D_1)\right)
=\sigma 2^{rl}\left(1+2^{ -(m^*-3)/4}\right)e_n^\q\left(S,\mathcal{E}',L_q(D_1)\right).\quad
\label{AH1}
\end{eqnarray}
From (\ref{N9}) we get
$$
\sup_{f\in \mathcal{B}_{L_p^{M_{1,l},M_{2,l}}}} \|S^{M_{1,l},M_{2,l}}f-S^{M_{1,l},M_{2,l}}(\gamma\circ\beta\circ f)\|_{L_q^{{M_{1,l}}}}
\le 2^{ -(m^*-3)/4},
$$
and hence, by Lemma \ref{lem:8} (i) and (ii),
\eqref{eq:89}, and (\ref{AH1}),
\begin{eqnarray*} 
&&e_{2n}^\q(S^{M_{1,l},M_{2,l}},\mathcal{B}_{L_p^{M_{1,l},M_{2,l}}},L_q^{M_{1,l}})
\le e_{2n}^\q(S^{M_{1,l},M_{2,l}}\circ\overline{\gamma\circ\beta},\mathcal{B}_{L_p^{M_{1,l},M_{2,l}}},L_q^{M_{1,l}})+2^{ -(m^*-3)/4}
\notag\\
&=& (\tau_1\tau_2)^{-1}e_{2n}^\q(\Psi_l\circ S\circ\Gamma_l,\mathcal{B}_{L_p^{M_{1,l},M_{2,l}}},L_q^{M_{1,l}})+2^{ -(m^*-3)/4}\\
&\le& (\tau_1\tau_2)^{-1}\sigma 2^{rl}\left(1+2^{-(m^*-3)/4}\right)
e_n^\q(S,\mathcal{E}',L_q(D_1))+2^{-(m^*-3)/4}
\\
&\le& (\tau_1\tau_2)^{-1}\sigma 2^{rl}\left(1+2^{-(m^*-3)/4}\right)
\sup_{\mathcal{E}\in {\rm UEq}(\mathcal{B}_{\mathcal{W}_p^r(D)})}e_n^\q(S,\mathcal{E},L_q(D_1))+2^{-(m^*-3)/4},
\end{eqnarray*}
where $\overline{\gamma\circ\beta}$ stands for 
$$
\overline{\gamma\circ\beta}:L_p^{M_{1,l},M_{2,l}}\to L_p^{M_{1,l},M_{2,l}}, \quad (\overline{\gamma\circ\beta}(f))(i,j)=\gamma(\beta(f(i,j))).
$$
Since $m^*$ can be made arbitrarily large, we get
\begin{equation*}
e_{2n}^\q(S^{M_{1,l},M_{2,l}},\mathcal{B}_{L_p^{M_{1,l},M_{2,l}}},L_q^{M_{1,l}})\le (\tau_1\tau_2)^{-1}\sigma 2^{rl} \sup_{\mathcal{E}\in {\rm UEq}(\mathcal{B}_{\mathcal{W}_p^r(D)})}e_n^\q(S,\mathcal{E},L_q(D_1)).
\end{equation*}

\end{proof}
\section{Approximation}
\label{sec:4}
In this section we consider  approximation of functions from $\mathcal{W}_p^r(D)$ in the norm of $L_q(D)$. This is the special case  $d_2=0$, thus $d_1=d$  of parametric integration with $S=J$, see \eqref{AI5}. We will use the respective results of Section \ref{sec:3}.
Neglecting logarithmic factors, the following is the main result of this section. We include the already known case $r/d>1/p$, see \cite{Hei04b}. 
\begin{theorem}
\label{theo:2}
Let $r\in\N_0$, $d\in\N$, $1\le p,q \le\infty$, and assume that \eqref{V1} holds.
Then for $r/d\ge 2/p-2/q$
\begin{equation}
\label{AB1} 
\sup_{\mathcal{E}\in {\rm UEq}(\mathcal{B}_{\mathcal{W}_p^r(D)})} e_n^\q(J,\mathcal{E},L_q(D))\asymp_{\log} n^{-r/d},
\end{equation}
while for $r/d< 2/p-2/q$
\begin{equation}
\label{AB2}
\sup_{\mathcal{E}\in {\rm UEq}(\mathcal{B}_{\mathcal{W}_p^r(D)})} e_n^\q(J,\mathcal{E},L_q(D))\asymp n^{-2r/d+2/p-2/q}.
\end{equation}
If the embedding condition \eqref{PA2} is satisfied, then in \eqref{AB1} and \eqref{AB2} the left-hand sides can be replaced by $e_n^\q(J,\mathcal{B}_{W_p^r(D)},L_q(D))$. If the embedding condition \eqref{PA2} does not hold, then 
\begin{equation*}
 e_n^\q(J,\mathcal{B}_{\mathcal{W}_p^r(D)}\cap C(Q),L_q(D))\asymp 1.
\end{equation*}
\end{theorem}

 Theorem \ref{theo:2} is a direct 
consequence of 
Propositions \ref{pro:8} and \ref{pro:11} below, which also contain the logarithmic factors. 
We start with the upper bounds, for which we need the following result on discrete approximation from  \cite{Hei04a}, Proposition 2.
\begin{proposition}
\label{pro:2} 
Let $1\le p, q \le \infty$. In the case $p<q$ there is a constant $c>0$ such that 
for all $n,N\in\N$ 
$$
e_n^\q(J^{N},\mathcal{B}_{L_p^N},L_q^N)\le 
c\,\min\left(\left(\frac{N}{n}\log \left(n/\sqrt{N}+2\right)\right)^{2/p-2/q},N^{1/p-1/q}
\right).
$$
In the case $p\ge q$, we have
$$
e_n^\q(J^{N},\mathcal{B}_{L_p^N},L_q^N)\le 1.
$$
\end{proposition}

\begin{proposition}
\label{pro:8}
Let $r\in \N_0$, $d\in\N$, $1\le p,q \le\infty$, and assume that \eqref{V1} holds.  
Then there exists a constant $c>0$ such that for all $n\in\N$ with $n>2$
the following hold: For $p\ge q$ we have
\begin{equation}
\label{XA1}
\sup_{\mathcal{E}\in {\rm UEq}(\mathcal{B}_{\mathcal{W}_p^r(D)})}e_n^\q(J,\mathcal{E},L_q(D))\le c n^{-r/d}.
\end{equation}
Now assume $p < q$. If $r/d>2/p-2/q$, then
\begin{equation}
\label{XA2}
\sup_{\mathcal{E}\in {\rm UEq}(\mathcal{B}_{\mathcal{W}_p^r(D)})}e_n^\q(J,\mathcal{E},L_q(D))
\le c n^{-r/d}(\log n)^{2/p-2/q},
\end{equation}
if $r/d=2/p-2/q$, then
\begin{equation}
\label{XA3}
\sup_{\mathcal{E}\in {\rm UEq}(\mathcal{B}_{\mathcal{W}_p^r(D)})}e_n^\q(J,\mathcal{E},L_q(D))
\le c n^{-r/d}(\log n)^{4/p-4/q+2}(\log\log n)^{2/p-2/q},
\end{equation}
and if $r/d<2/p-2/q$, then
\begin{equation}
\label{XA4}
\sup_{\mathcal{E}\in {\rm UEq}(\mathcal{B}_{\mathcal{W}_p^r(D)})}e_n^\q(J,\mathcal{E},L_q(D))
\le c n^{-2r/d+2/p-2/q}.
\end{equation}
If the embedding condition \eqref{PA2} is satisfied, then the left-hand sides of \eqref{XA1}, \eqref{XA2}, and \eqref{XA4}, 
can be replaced by $e_n^\q(J,\mathcal{B}_{W_p^r(D)},L_q(D))$, while  \eqref{XA3} improves to 
\begin{equation}
\label{XA3a}
e_n^\q(J,\mathcal{B}_{W_p^r(D)},L_q(D))
\le c n^{-r/d}(\log n)^{4/p-4/q+1}(\log\log n)^{2/p-2/q}.
\end{equation}
\end{proposition}
\begin{proof} We start with the case 
\begin{equation}\label{AM6}
 \frac{r}{d}=\left(\frac{1}{p}-\frac{1}{q}\right)_+ , 
\end{equation}
meaning that \eqref{V1} holds but \eqref{BB1} does not. If $p\ge q$, this implies $r=0$
and we have 
\begin{equation*}
\sup_{\mathcal{E}\in {\rm UEq}(\mathcal{B}_{\mathcal{L}_p(D)})}e_n^\q(J,\mathcal{E},L_q(D))
\le \|J:L_p(D)\to L_q(D)\|=1,
\end{equation*}
thus \eqref{XA1}.  If $p<q$, then \eqref{AM6} turns into $
\frac{r}{d}=\frac{1}{p}-\frac{1}{q}>0$.
Consequently, neither the condition of \eqref{XA2} nor that of \eqref{XA3} can hold. Furthermore, the exponent on the right-hand side of \eqref{XA4} is zero, hence that relation follows from \eqref{V1}. 

For the rest of the proof we suppose 
\begin{equation*}
 \frac{r}{d}>\left(\frac{1}{p}-\frac{1}{q}\right)_+\,, 
\end{equation*}
thus, \eqref{BB1} is satisfied.
Clearly, we can assume
\begin{equation}
\label{L1}
n\ge \max(\kappa2^d,5),
\end{equation}
with $\kappa$ given by \eqref{AT7}.
Define 
\begin{equation}
\label{D1}
l_0=\lfloor d^{-1}\log(n/\kappa)\rfloor,
\end{equation}
thus $l_0\ge 1$. Furthermore, set 
\begin{equation}
\label{H9}
l_1=\left\{\begin{array}{lll}
  l_0 & \mbox{if} \quad p\ge q   \\
  2l_0 & \mbox{if} \quad p<q.   \\
    \end{array}
\right.
\end{equation}
Relation \eqref{eq:26} turns into $N_{1,l}=\kappa'2^{dl}$, $N_{2,l}=1$, where $\kappa'$ was defined in \eqref{AT8}, and in the sequel we write $N_l$ instead of $N_{1,l}$. Let $\nu_l,n_l\in \N$ $(l=l_0,\dots,l_1-1)$ with $(\nu_l)$ satisfying \eqref{X8}, and let $\wt{n}$ given by \eqref{AS9}. If $p\ge q$, we have $l_1=l_0$ and \eqref{XA1} is a direct consequence of \eqref{AS4} and \eqref{AS5} of Corollary \ref{cor:2}. 

From now on we assume $p <q$. 
Then Corollary \ref{cor:2} and 
Proposition \ref{pro:2} imply the following. If the embedding condition does not hold and  $\Theta_l>0$ satisfy  \eqref{AS6}, then
\begin{eqnarray}
\lefteqn{\sup_{\mathcal{E}\in {\rm UEq}(\mathcal{B}_{\mathcal{W}_p^r(D)})}e_{\wt{n}}^\q(J,\mathcal{E},L_q(D))}
\nonumber\\
&\le&
c\,2^{-rl_1+\left(\frac{1}{p}-\frac{1}{q}\right)_+ dl_1} +c\sum_{l=l_0}^{l_1-1} \Theta_l^{-1}2^{-rl}
e_{n_l}^\q(J^{N_l},\mathcal{B}_{L_p^{N_l}},L_q^{N_l})
\nonumber\\
&\le&
c\,2^{-rl_1+\left(\frac{1}{p}-\frac{1}{q}\right)_+ dl_1} +c\sum_{l=l_0}^{l_1-1} \Theta_l^{-1}2^{-rl}
n_l^{-(\frac{2}{p}-\frac{2}{q})}N_l^{\frac{2}{p}-\frac{2}{q}}
(\log(n_l/\sqrt{N_l}+2))^{\frac{2}{p}-\frac{2}{q}},\label{H8a} 
\end{eqnarray}
and if the embedding condition \eqref{PA2} holds, then  we have  similarly
\begin{eqnarray}
\lefteqn{e_{\wt{n}}^\q(J,\mathcal{B}_{W_p^r(D)},L_q(D))}
\nonumber\\
&\le&
c\,2^{-rl_1+\left(\frac{1}{p}-\frac{1}{q}\right)_+ dl_1} +c\sum_{l=l_0}^{l_1-1} 2^{-rl}
n_l^{-(\frac{2}{p}-\frac{2}{q})}N_l^{\frac{2}{p}-\frac{2}{q}}
(\log(n_l/\sqrt{N_l}+2))^{\frac{2}{p}-\frac{2}{q}}.\label{H8b} 
\end{eqnarray}
The rest of the proof is carried out for the case that \eqref{PA2} does not hold, the case  that \eqref{PA2} does hold follows in the same way with the obvious changes, using \eqref{H8b}, see the comments given at the end of the proof. 

First suppose that 
\begin{equation*}
\frac{r}{d}>\frac{2}{p}-\frac{2}{q}.
\end{equation*}
Fix any $\delta>0$ (not depending on $n$) with
\begin{equation}
\label{N5}
r>\left(\frac{2}{p}-\frac{2}{q}\right)(d+\delta)
\end{equation}
and put for $l=l_0,\dots,l_1-1$
\begin{eqnarray}
n_l=\left\lceil 2^{dl_0-\delta(l-l_0)}\right\rceil,\quad
\nu_l=\left\lceil 8(2\ln(l-l_0+1)+ \ln 24)\right\rceil,\quad
\Theta_l&=&\frac{1}{24}(l-l_0+1)^{-2}\label{Y3}.
\end{eqnarray}
Then
\begin{eqnarray}
\sum_{l=l_0}^{l_1-1}e^{-\nu_l/8}\le \sum_{l=l_0}^{l_1-1}\Theta_l=\frac{1}{24}\sum_{l=l_0}^{l_1-1}(l-l_0+1)^{-2}<\frac{1}{24}\bigg(1+\int_1^\infty x^{-2}dx\bigg)= \frac{1}{12},
\label{I2D}
\end{eqnarray}
hence the assumptions \eqref{X8} and (\ref{AS6}) are satisfied, and
 for $\wt{n}$ given by (\ref{AS9}) we have
\begin{eqnarray}
\label{I3}
\wt{n}&\le& n+2\kappa''\sum_{l=l_0}^{l_1-1} 
\left\lceil 8(2\ln(l-l_0+1)+ \ln 24)\right\rceil 
\left\lceil2^{dl_0-\delta(l-l_0)}\right\rceil\nonumber\\
&\le& c\,2^{dl_0} \le cn.
\end{eqnarray}
Furthermore, 
\begin{eqnarray*}
n_l/\sqrt{N_l}+2\le 2^{d(l_0-l/2)} +2\le 2\cdot2^{d(l_0-l/2+1)}
\end{eqnarray*}
and therefore 
\begin{eqnarray}
\log(n_l/\sqrt{N_l}+2)\le d(l_0-l/2+2)\le 2 d(l_0-l/2+1). \label{AT6}
\end{eqnarray}
while 
(\ref{L1})--(\ref{H8a}), \eqref{N5}, \eqref{Y3}, and \eqref{AT6} give
\begin{eqnarray}
\lefteqn{\sup_{\mathcal{E}\in {\rm UEq}(\mathcal{B}_{\mathcal{W}_p^r(D)})}e_{\wt{n}}^\q(J,\mathcal{E},L_q(D))}\nonumber\\
&\le& c\,2^{(-\frac{2r}{d}+\frac{2}{p}-\frac{2}{q})dl_0}+c\sum_{l=l_0}^{l_1-1} 
2^{-rl-(\frac{2}{p}-\frac{2}{q})dl_0+(\frac{2}{p}-\frac{2}{q})\delta(l-l_0)
+(\frac{2}{p}-\frac{2}{q})dl}
(l_0-l/2+1)^{\frac{2}{p}-\frac{2}{q}}(l-l_0+1)^2\nonumber\\
&\le& c\,2^{-rl_0}+c\,2^{-rl_0}
\sum_{l=l_0}^{l_1-1}2^{(-r+(\frac{2}{p}-\frac{2}{q})(d+\delta))(l-l_0)}(l_0-l/2+1)^{\frac{2}{p}-\frac{2}{q}}(l-l_0+1)^2\nonumber\\
&\le& c\,2^{-rl_0}(l_0+1)^{\frac{2}{p}-\frac{2}{q}}
\le cn^{-\frac{r}{d}}(\log n)^{\frac{2}{p}-\frac{2}{q}}.\label{V2}
\end{eqnarray}
Now (\ref{XA2}) is a consequence of  
(\ref{I3}) and  (\ref{V2}).

Next assume 
$$
\frac{r}{d}<\frac{2}{p}-\frac{2}{q},
$$
choose $\delta>$ so that
\begin{equation}
\label{N5a}
r<\left(\frac{2}{p}-\frac{2}{q}\right)(d-\delta),
\end{equation}
and put for $l=l_0,\dots,l_1-1$
\begin{eqnarray}
n_l=\left\lceil 2^{dl_0-\delta(l_1-l)}\right\rceil,\quad
\nu_l=\left\lceil 8(2\ln(l_1-l)+ \ln 24)\right\rceil,\quad
\Theta_l=\frac{1}{24}(l_1-l)^{-2}\label{Y4}.
\end{eqnarray}
Similarly to the above,   \eqref{X8} and (\ref{AS6}) are fulfilled, and
\begin{eqnarray}
\label{I3B}
\wt{n}&\le& n+2\kappa''\sum_{l=l_0}^{l_1-1} 
\left\lceil 8(2\ln(l_1-l)+ \ln 24)\right\rceil 
\left\lceil2^{dl_0-\delta(l_1-l)}\right\rceil
\le c\,2^{dl_0} \le cn.
\end{eqnarray}
With the help of  
(\ref{L1})--(\ref{H8a}),  (\ref{N5a}), and (\ref{Y4}), we get
\begin{eqnarray}
\lefteqn{\sup_{\mathcal{E}\in {\rm UEq}(\mathcal{B}_{\mathcal{W}_p^r(D)})}e_{\wt{n}}^\q(J,\mathcal{E},L_q(D))}
\notag\\
&\le& c\,2^{(-\frac{2r}{d}+\frac{2}{p}-\frac{2}{q})dl_0}+c\sum_{l=l_0}^{l_1-1} 
2^{-rl-(\frac{2}{p}-\frac{2}{q})dl_0+(\frac{2}{p}-\frac{2}{q})\delta(l_1-l)
+(\frac{2}{p}-\frac{2}{q})dl}(l_0-l/2+1)^{\frac{2}{p}-\frac{2}{q}}(l_1-l)^2
\notag\\
&=& c\,2^{(-\frac{2r}{d}+\frac{2}{p}-\frac{2}{q})dl_0}
+c\,2^{-rl_1+(\frac{2}{p}-\frac{2}{q})dl_0}
\sum_{l=l_0}^{l_1-1}
2^{(r-(\frac{2}{p}-\frac{2}{q})(d-\delta))(l_1-l)}(l_1-l)^{\frac{2}{p}-\frac{2}{q}+2}
\notag\\
&\le& c\,2^{(-\frac{2r}{d}+\frac{2}{p}-\frac{2}{q})dl_0}
\le cn^{-\frac{2r}{d}+\frac{2}{p}-\frac{2}{q}},\label{AL4}
\end{eqnarray}
which combined with (\ref{I3B}) yields (\ref{XA4}).

Finally we deal with the case
\begin{equation}
\label{AE7}
\frac{r}{d}=\frac{2}{p}-\frac{2}{q}.
\end{equation}
We define for $l=l_0,\dots,l_1-1$
\begin{eqnarray}
n_l=\left\lceil 2^{dl_0}(l_0+1)^{-1}(\ln(l_0+2))^{-1}\right\rceil,\quad
\nu_l=\left\lceil 8(\ln(l_0+2)+\ln 12)\right\rceil,\quad
\Theta_l=\frac{1}{12}(l_0+1)^{-1}\label{Y8}.
\end{eqnarray}
Obviously, relations \eqref{X8}  and (\ref{AS6}) are satisfied again and
\begin{eqnarray*}
\wt{n}&\le& n+2\kappa''\sum_{l=l_0}^{l_1-1} 
\left\lceil 8(\ln(l_0+2)+\ln 12) \right\rceil 
\left\lceil 2^{dl_0}(l_0+1)^{-1}(\ln(l_0+2))^{-1}\right\rceil\nonumber\\
&\le& c\,2^{dl_0} \le cn.
\end{eqnarray*}
Using 
(\ref{L1})--(\ref{H8a}), \eqref{AE7}, and (\ref{Y8}) we conclude that
\begin{eqnarray}
\lefteqn{\sup_{\mathcal{E}\in {\rm UEq}(\mathcal{B}_{\mathcal{W}_p^r(D)})}e_{\wt{n}}^\q(J,\mathcal{E},L_q(D))}
\nonumber\\
&\le& c\,2^{-rl_1/2}\nonumber\\
&&+c\sum_{l=l_0}^{l_1-1} 
2^{-rl-(\frac{2}{p}-\frac{2}{q})dl_0+(\frac{2}{p}-\frac{2}{q})dl}(l_0-l/2+1)^{\frac{2}{p}-\frac{2}{q}}
(l_0+1)^{\frac{2}{p}-\frac{2}{q}+1}(\ln(l_0+2))^{\frac{2}{p}-\frac{2}{q}}
\label{AK1}\\
&=& c\,2^{-rl_0}+c\,(l_0+1)^{\frac{2}{p}-\frac{2}{q}+1}
(\ln(l_0+2))^{\frac{2}{p}-\frac{2}{q}}2^{-(\frac{2}{p}-\frac{2}{q})dl_0}\sum_{l=l_0}^{l_1-1}(l_0-l/2+1)^{\frac{2}{p}-\frac{2}{q}}
\label{AK2}\\
&\le& c\,2^{-rl_0}(l_0+1)^{\frac{4}{p}-\frac{4}{q}+2}(\ln(l_0+2))^{\frac{2}{p}-\frac{2}{q}}
\label{AK3}\\
&\le& cn^{-\frac{r}{d}}(\log n)^{\frac{4}{p}-\frac{4}{q}+2}
(\log\log n)^{\frac{2}{p}-\frac{2}{q}},\label{AJ9}
\end{eqnarray}
proving (\ref{XA3}). If the embedding condition \eqref{PA2} holds, we argue in the same way using $\Theta_l\equiv 1$ and  \eqref{H8b}. Due to geometric series summation this affects neither \eqref{V2} nor \eqref{AL4}. However, it reduces the exponents of $(l_0+1)$ in \eqref{AK1}--\eqref{AK3} and that of $\log n$ in \eqref{AJ9} by 1. This shows \eqref{XA3a} and completes the proof.

\end{proof}

Now we turn to the lower bounds. We use the following result for discrete approximation, which is a combination of \cite{Hei04a}, Proposition 4  and \cite{Hei06}, Proposition 2.
\begin{proposition}
\label{pro:1} Let $1\le p,q \le\infty$. There are constants $c_0,c_1>0$
such that for all $n,N\in \N$, with $n\le c_0N$ and $N>4$ 
\begin{eqnarray}
\lefteqn{e_n^\q(J^N,\mathcal{B}_{L_p^N},L_q^N)   }
\nonumber\\
&\ge& c_1\max\left(\min\left(\left(\frac{N}{n}\right)^{2/p-2/q}
\left(\log \left(n/\sqrt{N}+2\right)\right)^{-2/q},N^{1/p-1/q}\right),1\right).
\label{AL1}
\end{eqnarray}
\end{proposition}
\begin{proof}
Indeed, Proposition 2 of \cite{Hei06}  shows that 
\begin{equation}
\label{AL2}
e_n^\q(J^N,\mathcal{B}_{L_p^N},L_q^N)\ge c,
\end{equation}
and Proposition 4 of \cite{Hei04a} gives for $p\le q$
\begin{equation*}
e_n^\q(J^N,\mathcal{B}_{L_p^N},L_q^N)\ge c\min\left(\left(\frac{N}{n}\right)^{2/p-2/q}
\left(\log \left(n/\sqrt{N}+2\right)\right)^{-2/q},N^{1/p-1/q}\right).
\end{equation*}
This together with \eqref{AL2} implies \eqref{AL1} for $p\le q$.
On the other hand, if $p>q$, we have 
\begin{equation*}
\min\left(\left(\frac{N}{n}\right)^{2/p-2/q}
\left(\log \left(n/\sqrt{N}+2\right)\right)^{-2/q},N^{1/p-1/q}\right)\le c,
\end{equation*}
which together with \eqref{AL2} proves \eqref{AL1} also for this case.

\end{proof}

\begin{proposition}
\label{pro:11}
Let $r\in\N_0$, $d\in\N$, $1\le p,q \le\infty$, and assume that \eqref{V1} holds.  
Then there exists a  constant $c>0$ such that for all $n\in\N$ with $n>4$
the following holds:
If $r/d\ge 2/p-2/q$, then 
\begin{equation}
\label{AG3}
\sup_{\mathcal{E}\in {\rm UEq}(\mathcal{B}_{\mathcal{W}_p^r(D)})} e_n^\q(J,\mathcal{E},L_q(D))\ge c n^{-r/d}
\end{equation}
and if $r/d< 2/p-2/q$, then
\begin{equation}
\label{XA4a}
\sup_{\mathcal{E}\in {\rm UEq}(\mathcal{B}_{\mathcal{W}_p^r(D)})} e_n^\q(J,\mathcal{E},L_q(D))
\ge c n^{-2r/d+2/p-2/q}.
\end{equation}
If the embedding condition \eqref{PA2} does not hold, then for all $n\in\N$
\begin{equation}
\label{AB4}
 e_n^\q(J,\mathcal{B}_{\mathcal{W}_p^r(D)}\cap C(Q),L_q(D))\ge c.
\end{equation}
\end{proposition}
\begin{proof} We start with relation \eqref{AB4}. It was shown in \cite{Hei08a}, Theorem 4.3, that if \eqref{PA2} does not hold, then there is a constant $c>0$ such that for all $n\in \N$
\begin{equation*}
e_n^\de(J,\mathcal{B}_{\mathcal{W}_p^r(D)}\cap C(Q),L_q(D))\ge c.
\end{equation*}
This together with  Lemma  \ref{lem:5} implies \eqref{AB4}.

To prove the remaining statements, we use Proposition \ref{pro:4} with  $d_1=d$, $d_2=0$, hence by \eqref{AF7}, $M_{1,l}=2^{dl}$, $M_{2,l}=1$, and Proposition \ref{pro:1} with $N=M_{1,l}$. 
For the case $r/d\ge 2/p-2/q$  we choose
$l=\lceil d^{-1}(\log(n/c(6))+1)\rceil$, where $c(6)$ is the constant $c_0$ from 
Proposition \ref{pro:1}, which can be assumed to satisfy $0<c(6)\le 1$.
 It follows that 
\begin{equation}
\label{AU7}
\log(n/c(6))+1\le dl<\log(n/c(6))+d+1
\end{equation}
and therefore
\begin{equation}
\label{N6}
c(6) 2^{-d}M_{1,l} = c(6) 2^{-d}2^{dl}< 2n\le c(6) 2^{dl} = c(6) M_{1,l}.
\end{equation}
Now we use
Proposition \ref{pro:4}, Proposition \ref{pro:1}, and (\ref{N6}) to obtain 
\begin{eqnarray*}
\lefteqn{\sup_{\mathcal{E}\in {\rm UEq}(\mathcal{B}_{\mathcal{W}_p^r(D)})} e_n^\q(J,\mathcal{E},L_q(D))\ge c2^{-rl} e_{2n}^\q\Big(J^{M_{1,l}},\mathcal{B}_{L_p^{M_{1,l}}},L_q^{M_{1,l}}\Big)}
\\
&\ge& c2^{-rl}\max\left(\min\left(\left(\frac{M_{1,l}}{2n}\right)^{2/p-2/q}\left(\log \left(2n/\sqrt{M_{1,l}}+2\right)\right)^{-2/q},M_{1,l}^{1/p-1/q}\right),1\right)\ge c n^{-r/d},
\end{eqnarray*}
which is the desired lower bound (\ref{AG3}).
In the case $r/d< 2/p-2/q$, which implies $p<q$, we set 
$l=\lceil d^{-1}(\log(n^2/c(6))+1)\rceil$ and conclude similarly
\begin{equation}
\label{N6a}
c(6) 2^{-d}M_{1,l}=c(6) 2^{-d}2^{dl}< 2n^2\le  c(6) 2^{dl} =c(6) M_{1,l}.
\end{equation}
Consequently, Propositions \ref{pro:4}, \ref{pro:1}, and (\ref{N6a}) lead to
\begin{eqnarray*}
\lefteqn{\sup_{\mathcal{E}\in {\rm UEq}(\mathcal{B}_{\mathcal{W}_p^r(D)})} e_n^\q(J,\mathcal{E},L_q(D))\ge c2^{-rl} e_{2n}^\q\Big(J^{M_{1,l}},\mathcal{B}_{L_p^{M_{1,l}}},L_q^{M_{1,l}}\Big)}
\\
&\ge& c2^{-rl}\max\left(\min\left(\left(\frac{M_{1,l}}{2n}\right)^{2/p-2/q}\left(\log \left(2n/\sqrt{M_{1,l}}+2\right)\right)^{-2/q},M_{1,l}^{1/p-1/q}\right),1\right)
\\
&\ge&  c 2^{-rl}M_{1,l}^{1/p-1/q}\ge c n^{-2r/d+2/p-2/q},
\end{eqnarray*}
showing \eqref{XA4a}.
\end{proof}

\section{Integration}\label{sec:5}
This section is devoted to  integration of functions from $\mathcal{W}_p^r(D)$, so  we consider  the special case  $d_1=0$ of parametric integration with $S=I$, see \eqref{AE0}. Since here the target space of $S=I$ is always the scalar field $\K$, we drop it in the notation of minimal errors. With logarithmic factors suppressed, the following is the main result of this section. As in Section \ref{sec:4} we include the known results for the case $r/d>1/p$, see \cite{Nov01,Hei01b}. 

\begin{theorem}
\label{theo:3}
Let $r\in\N_0$, $d\in\N$, $1\le p \le\infty$.
Then for $r/d\ge 2/p-1$
\begin{equation}
\label{AE1} 
\sup_{\mathcal{E}\in {\rm UEq}(\mathcal{B}_{\mathcal{W}_p^r(D)})} e_n^\q(I,\mathcal{E})\asymp_{\log} n^{-r/d-1},
\end{equation}
and for $r/d< 2/p-1$
\begin{equation}
\label{AE2}
\sup_{\mathcal{E}\in {\rm UEq}(\mathcal{B}_{\mathcal{W}_p^r(D)})} e_n^\q(I,\mathcal{E})\asymp n^{-2r/d+2/p-2}.
\end{equation}
If the embedding condition \eqref{PA2} is satisfied, then \eqref{AE1} holds with $e_n^\q(I,\mathcal{B}_{W_p^r(D)})$ in place of $\sup_{\mathcal{E}\in {\rm UEq}(\mathcal{B}_{\mathcal{W}_p^r(D)})} e_n^\q(I,\mathcal{E})$. If the embedding condition  does not hold, then 
\begin{equation}
\label{AE6}
 e_n^\q(I,\mathcal{B}_{\mathcal{W}_p^r(D)}\cap C(Q))\asymp 1.
\end{equation}
\end{theorem}
Note that the restriction  $r/d< 2/p-1$ of \eqref{AE2} contradicts the embedding condition. As in Section \ref{sec:4} we prove two propositions, \ref{pro:5} and \ref{pro:9}, which contain also the logarithmic factors and imply the theorem.
We need some  results about the mean computation operator $I^N$, see \eqref{AE9}. Relations \eqref{K1} and \eqref{K2} are from \cite{Hei01}, Theorems 1 and 2, while \eqref{K3} is Corollary 2 of  \cite{HN01}.
\begin{proposition}
\label{pro:1a} Let $1\le p\le \infty$.
There are constants $c_0,c_1,c_2>0$ 
such that for all $n,N\in\N$ with $2<n\le c_0 N$,
\begin{eqnarray}
c_1 n^{-1}&\le& e_n^\q(I^N,\mathcal{B}_{L_p^N})\,\le\, c_2 n^{-1}\quad\mbox{if}\quad 2<p\le\infty,
\label{K1}\\
c_1 n^{-1}&\le& e_n^\q(I^N,\mathcal{B}_{L_2^N})\,\le\, c_2 n^{-1}(\log n)^{3/2}\log\log n,\label{K2}
\end{eqnarray}
and 
\begin{eqnarray}
\lefteqn{
c_1 \min(n^{-2(1-1/p)},n^{-2/p}N^{2/p-1})\le e_n^\q(I^N,\mathcal{B}_{L_p^N})}\nonumber\\
&&
\le c_2\min(n^{-2(1-1/p)},n^{-2/p}N^{2/p-1})\max(\log(n/\sqrt{N}),1)^{2/p-1}\quad\mbox{if}\quad 1\le p<2. \label{K3}
\end{eqnarray}
\end{proposition}
The next is Lemma 6  of \cite{Hei01b}, which improves the upper bound of \eqref{K2} for $n$ close to $N$.
\begin{lemma}\label{lem:6a}
There is a constant $c>0$ such that for all $n,N\in\N$
with $n \le N$,
$$
e_n^\q(I^N,\mathcal{B}_{L_2^N})\le cn^{-1}\lambda(n,N)^{3/2}\log\lambda(n,N),
$$
where 
\begin{equation}
\label{CF1}
\lambda(n,N)=\log(N/n)+\log\log (n+1)+2.  
\end{equation}
\end{lemma}
\begin{proposition}
\label{pro:5}
Let $r\in\N_0$, $d\in\N$, and $1\le p \le\infty$. 
Then there is a constant $c>0$
such that for all $n\in\N$ with $n>4$
\begin{eqnarray}
\sup_{\mathcal{E}\in {\rm UEq}(\mathcal{B}_{\mathcal{W}_p^r(D)})} e_n^\q(I,\mathcal{E})&\le& c n^{-r/d-1}
\quad\mbox{if}\quad 2<p\le\infty,
\label{AF1}\\[.2cm]
\sup_{\mathcal{E}\in {\rm UEq}(\mathcal{B}_{W^r_2(D)})}e_n^\q(I,\mathcal{E})&\le& c n^{-r/d-1}\left\{\begin{array}{lll}
 (\log n)^{3/2}\log\log n  & \quad\mbox{if}\quad  r=0  \\
  (\log\log n)^{3/2}\log\log\log n & \quad\mbox{if}\quad  r>0.
    \end{array}
\right. 
\label{AF2}
\end{eqnarray}
Now assume $p<2$. Then
\begin{eqnarray}
\sup_{\mathcal{E}\in {\rm UEq}(\mathcal{B}_{\mathcal{W}_p^r(D)})} e_n^\q(I,\mathcal{E})&\le& c n^{-r/d-1}(\log n)^{2/p-1}
\quad \mbox{if}\quad r/d>2/p-1,
\label{AF3}\\[.2cm]
\sup_{\mathcal{E}\in {\rm UEq}(\mathcal{B}_{\mathcal{W}_p^r(D)})} e_n^\q(I,\mathcal{E})&\le& c n^{-r/d-1}(\log n)^{4/p+1}(\log\log n)^{2/p}
\quad\mbox{if}\quad r/d=2/p-1,
\label{AF4}\\[.3cm]
\sup_{\mathcal{E}\in {\rm UEq}(\mathcal{B}_{\mathcal{W}_p^r(D)})} e_n^\q(I,\mathcal{E})&\le& c n^{-2r/d+2/p-2}
\quad\mbox{if}\quad r/d<2/p-1.
\label{AF5}
\end{eqnarray}
If the embedding condition \eqref{PA2} is satisfied, then the left-hand sides in \eqref{AF1}, \eqref{AF2}, and \eqref{AF3} 
can be replaced by $e_n^\q(I,\mathcal{B}_{W_p^r(D)})$, while  \eqref{AF4} is replaced by
\begin{equation}
\label{AF4a}
e_n^\q(I,\mathcal{B}_{W_1^r(D)})
\le c n^{-r/d-1}(\log n)^4(\log\log n)^2.
\end{equation}
\end{proposition}
Note that $p=1$  is the only case that the embedding condition can hold under the restriction  $r/d= 2/p-1$ of \eqref{AF4}, and it does not hold at all if $r/d< 2/p-1$.

\medskip

\noindent {\it Proof of Proposition \ref{pro:5}.} The case $r=0$ is contained in Theorem 3 of \cite{Hei01}, where we note that obviously in this case the condition $(1\le p<2)\wedge(r/d\ge 2/p-1)$ of \eqref{AF3} and \eqref{AF4} cannot hold. Thus in the sequel we suppose $r\in \N$. Furthermore, we can assume 
\begin{equation}
\label{AL0}
n\ge \max(\kappa,16),
\end{equation}
with $\kappa$ from \eqref{AT7}, since the case $n< \max(\kappa,16)$ trivially follows from $\|I:\mathcal{W}_p^r(D)\to \K\|=1$, so the zero algorithm can be applied.
Let  
\begin{equation}
\label{D1a}
l_0=\lfloor d^{-1}\log(n/\kappa)\rfloor,
\end{equation}
so by \eqref{AL0} we have $l_0\ge 0$, and moreover, 
\begin{equation}
\label{AU8}
\kappa 2^{dl_0}\le n< \kappa 2^{d(l_0+1)}.
\end{equation}
Let furthermore $l_1\in \N_0$ with $l_1\ge l_0$, to be fixed later-on.

For the integration problem, relation \eqref{eq:26} becomes $N_{1,l}=\kappa'$, $N_{2,l}=2^{dl}$, with $\kappa'$ from \eqref{AT8}, and we write $N_l$ instead of $N_{2,l}$. Furthermore, since $d_1=0$, $D_1=\{0\}$ is the one-point set, so $L_q(D_1)$ is just $\K$, and all $L_q$-norms are the same on $\K$. It is convenient to choose $q=p$. With $\nu_l,\Theta_l>0$  $(l=l_0,\dots,l_1-1)$  satisfying \eqref{X8} and \eqref{AS6}, $\wt{n}_l\in \N$, and $\wt{n}$ given by 
\begin{equation}
\label{C7a}
\wt{n}=\kappa 2^{dl_0}+2\kappa''\sum_{l=l_0}^{l_1-1}\nu_l \wt{n}_l,
\end{equation}
Corollary \ref{cor:2} gives
\begin{eqnarray}
\sup_{\mathcal{E}\in {\rm UEq}(\mathcal{B}_{\mathcal{W}_p^r(D)})}e_{\wt{n}}^\q(I,\mathcal{E})
&\le&
c\,2^{-rl_1} +c\sum_{l=l_0}^{l_1-1} \Theta_l^{-1}2^{-rl}
e_{\wt{n}_l}^\q(S^{\kappa',N_l},\mathcal{B}_{L_p^{\kappa',N_l}},L_p^{\kappa'})\label{H8},
\end{eqnarray}
and if the embedding condition \eqref{PA2} holds, then 
\begin{eqnarray}
e_{\wt{n}}^\q(I,\mathcal{B}_{W_1^r(D)})
&\le&
c\,2^{-rl_1} +c\sum_{l=l_0}^{l_1-1} 2^{-rl}
e_{\wt{n}_l}^\q(S^{\kappa',N_l},\mathcal{B}_{L_p^{\kappa',N_l}},L_p^{\kappa'}).\label{AL9} 
\end{eqnarray}
Let $e_k$ be the $k$-th unit vector in $L_\infty^{\kappa'}$ and define
\begin{eqnarray*}
I_k^{N_l}:L_p^{\kappa',N_l}\to L_\infty^{\kappa'},\quad I_k^{N_l}f  &=& \frac{1}{N_l} 
\sum_{j=1}^{N_l} f(k,j)e_k \quad(1\le k\le \kappa'),
\end{eqnarray*}
thus 
\begin{equation*}
S^{\kappa',N_l}=\sum_{k=1}^{\kappa'}I_k^{N_l}.
\end{equation*}
Furthermore, let
\begin{eqnarray*}
&&\Gamma_k: L_p^{\kappa',N_l}\to L_p^{N_l},\quad \Gamma_k(f) =(\kappa')^{-1/p}f(k,\,\cdot\,),\qquad \Psi_k:\K\to L_p^{\kappa'},\quad \Psi_k a=(\kappa')^{1/p}ae_k,
\end{eqnarray*}
then 
$$
I_k^{N_l}=\Psi_k I^{N_l}\Gamma_k. 
$$ 
We can represent $\Gamma_k(f)=\rho\circ f\circ \eta_k$ with 
$$
\eta_k:\Z[1,N_l]\to \Z[1,\kappa']\times\Z[1,N_l], \quad\eta_k(j)=(k,j),\quad \rho:\K\to\K, \quad \rho(a)=(\kappa')^{-1/p}a.
$$
 Since $\|\Gamma_k\|=\|\Psi_k\|=1$, we obtain from Lemma \ref{lem:7} and \eqref{AC4} of Lemma \ref{lem:2}, 
\begin{equation*}
e_m^\q(I_k^{N_l},\mathcal{B}_{L_p^{\kappa',N_l}},L_p^{\kappa'})\le e_m^\q(I^{N_l},\mathcal{B}_{L_p^{N_l}})\quad (m\in\N_0).
\end{equation*}
Let $n_l\in \N$  $(l=l_0,\dots,l_1-1)$. With $\nu(1)=\lceil 8\ln(4\kappa')\rceil$ we have $\kappa'e^{-\nu(1)/8}\le 1/4$, and setting
$\wt{n}_l=\kappa'\nu(1)n_l$, Lemma \ref{lem:1} gives
$$
e_{\wt{n}_l}^\q(S^{\kappa',N_l},\mathcal{B}_{L_p^{\kappa',N_l}},L_p^{\kappa'})\le 2 \sum_{k=1}^{\kappa'} e_{n_l}^\q(I_k^{N_l},\mathcal{B}_{L_p^{\kappa',N_l}},L_p^{\kappa'})\le 2\kappa'e_{n_l}^\q(I^{N_l},\mathcal{B}_{ L_p^{N_l}}).
$$
Combined with \eqref{H8} we arrive at
\begin{eqnarray}
\sup_{\mathcal{E}\in {\rm UEq}(\mathcal{B}_{\mathcal{W}_p^r(D)})}e_{\wt{n}}^\q(I,\mathcal{E})
&\le&
c\,2^{-rl_1} +c\sum_{l=l_0}^{l_1-1} \Theta_l^{-1}2^{-rl}
e_{\wt{n}_l}^\q(S^{\kappa',N_l},\mathcal{B}_{L_p^{\kappa',N_l}},L_p^{\kappa'})
\notag\\
&\le&
c\,2^{-rl_1} +c\sum_{l=l_0}^{l_1-1} \Theta_l^{-1}2^{-rl}
e_{n_l}^\q(I^{N_l},\mathcal{B}_{ L_p^{N_l}}),\label{AF9} 
\end{eqnarray}
and correspondingly, under the embedding condition  \eqref{PA2}, combined with \eqref{AL9},
\begin{eqnarray}
e_{\wt{n}}^\q(I,\mathcal{B}_{W_p^r(D)})
&\le&
c\,2^{-rl_1} +c\sum_{l=l_0}^{l_1-1} 2^{-rl}
e_{\wt{n}_l}^\q(S^{\kappa',N_l},\mathcal{B}_{L_p^{\kappa',N_l}},L_p^{\kappa'})
\notag\\
&\le&
c\,2^{-rl_1} +c\sum_{l=l_0}^{l_1-1} 2^{-rl}
e_{n_l}^\q(I^{N_l},\mathcal{B}_{ L_p^{N_l}}),\label{AM0} 
\end{eqnarray}
where according to \eqref{AU8} and \eqref{C7a}
\begin{equation}
\label{AG5}
\wt{n}=\kappa 2^{dl_0}+2\kappa''\sum_{l=l_0}^{l_1-1}\nu_l \wt{n}_l
\le n+2\kappa'\kappa''\nu(1)\sum_{l=l_0}^{l_1-1}\nu_l n_l.
\end{equation}

Similarly to the proof of Proposition \ref{pro:8} we assume from now on that the embedding condition  \eqref{PA2} does not hold. If the embedding condition \eqref{PA2} holds, then  we use $\Theta_l\equiv 1$, see \eqref{AS6} and \eqref{AM0}, which requires no essential changes. Again, some comments are given at the end of the proof. 
We set 
\begin{equation}
\label{H9a}
l_1=\lceil (1+d/r)l_0 \rceil,
\end{equation}
hence $l_1\ge  l_0$. Let $\delta>0$ to be fixed later on (independent of $n$), and put for $l=l_0,\dots,l_1-1$
\begin{eqnarray}
n_l=\left\lceil 2^{dl_0-\delta(l-l_0)}\right\rceil,\quad
\nu_l=\left\lceil 8(2\ln(l-l_0+1)+ \ln 24)\right\rceil,\quad
\Theta_l&=&\frac{1}{24}(l-l_0+1)^{-2}\label{I1a},
\end{eqnarray}
so obviously \eqref{X8} and \eqref{AS6} hold (compare \eqref{Y3} and \eqref{I2D}). By \eqref{D1a} and \eqref{AG5}--(\ref{I1a}), 
\begin{eqnarray}
\label{I3a}
\wt{n}&\le& n+2\kappa'\kappa''\nu(1)\sum_{l=l_0}^{l_1-1} 
\left\lceil 8(2\ln(l-l_0+1)+ \ln 24)\right\rceil 
\left\lceil2^{dl_0-\delta(l-l_0)}\right\rceil\le c\,2^{dl_0} \le cn.
\end{eqnarray}

Now we prove the upper bound in the case $2\le p\le\infty$. Here we choose $\delta<1$.
If $p>2$, relations (\ref{D1a}),  (\ref{AF9}), \eqref{H9a}, (\ref{I1a}), and Proposition \ref{pro:1a}  give
\begin{eqnarray}
\sup_{\mathcal{E}\in {\rm UEq}(\mathcal{B}_{\mathcal{W}_p^r(D)})}e_{\wt{n}}^\q(I,\mathcal{E})
&\le& c\,2^{-rl_1}+c\sum_{l=l_0}^{l_1-1} 2^{-rl}n_l^{-1}(l-l_0+1)^2
\notag\\
&\le& c\,2^{-rl_1}+c\,2^{-(r+d)l_0}\sum_{l=l_0}^{l_1-1} 2^{-(r-\delta)(l-l_0)}(l-l_0+1)^2
\notag\\
&\le& c\,2^{-(r+d)l_0}\le cn^{-r/d-1},
\label{AL5}
\end{eqnarray}
thus, also taking into account \eqref{I3a}, relation \eqref{AF1} follows. 
If $p=2$, we get from (\ref{AF9}) and Lemma \ref{lem:6a}
\begin{eqnarray}
\lefteqn{\sup_{\mathcal{E}\in {\rm UEq}(\mathcal{B}_{\mathcal{W}_p^r(D)})}e_{\wt{n}}^\q(I,\mathcal{E})}
\notag\\
&\le& c\,2^{-rl_1}+c\sum_{l=l_0}^{l_1-1} 2^{-rl}
n_l^{-1}\lambda(n_l,N_l)^{3/2}\log\lambda(n_l,N_l)(l-l_0+1)^2
\notag\\
&\le& c\,2^{-rl_1}+c\,2^{-(r+d)l_0}
\sum_{l=l_0}^{l_1-1} 
2^{-(r-\delta)(l-l_0)}\lambda(n_l,N_l)^{3/2}\log\lambda(n_l,N_l)(l-l_0+1)^2.
\label{CF3}
\end{eqnarray}
By \eqref{CF1} and \eqref{AL0}, 
\begin{eqnarray*}
\lambda(n_l,N_l)&=&\log(N_l/n_l)+\log\log (n_l+1)+2
\\
&\le&c(l-l_0+\log\log n+1)\le c(l-l_0+1)\log\log n
\\[.2cm]
\log\lambda(n_l,N_l)&\le&c(\log(l-l_0+1)+\log\log\log n+1)
\\
&\le& c\log(l-l_0+2)\log\log\log n.
\end{eqnarray*}
Consequently,
\begin{eqnarray}
\lefteqn{\sum_{l=l_0}^{l_1-1} 
2^{-(r-\delta)(l-l_0)}\lambda(n_l,N_l)^{3/2}\log\lambda(n_l,N_l)(l-l_0+1)^2}
\notag\\
&\le& c(\log\log n)^{3/2}\log\log\log n\sum_{l=l_0}^{l_1-1} 
2^{-(r-\delta)(l-l_0)}(l-l_0+1)^{7/2}\log(l-l_0+2)
\notag\\
&\le& c(\log\log n)^{3/2}\log\log\log n. \label{AF0}
\end{eqnarray}
 Combining \eqref{CF3} and \eqref{AF0} and using \eqref{H9a}, we obtain
\begin{eqnarray}
\sup_{\mathcal{E}\in {\rm UEq}(\mathcal{B}_{\mathcal{W}_p^r(D)})}e_{\wt{n}}^\q(I,\mathcal{E})
&\le& c\,2^{-rl_1}+c\,2^{-(r+d)l_0}(\log\log n)^{3/2}\log\log\log n
\notag\\
&\le& cn^{-r/d-1}(\log\log n)^{3/2}\log\log\log n,
\label{AL6}
\end{eqnarray}
which together with \eqref{I3a} shows \eqref{AF2}.

Next we treat the case $1\le p <2$. It follows from (\ref{AF9}) and  
Proposition \ref{pro:1a} that
\begin{eqnarray}
\lefteqn{\sup_{\mathcal{E}\in {\rm UEq}(\mathcal{B}_{\mathcal{W}_p^r(D)})}e_{\wt{n}}^\q(I,\mathcal{E})}
\notag\\
&& \le c\,2^{-rl_1}+c\sum_{l=l_0}^{l_1-1} \Theta_l^{-1}2^{-rl}
\min(n_l^{-2(1-1/p)},n_l^{-2/p}N_l^{2/p-1})\max(\log(n_l/\sqrt{N_l}),1)^{2/p-1}.\quad
\label{P1}
\end{eqnarray}
We consider three sub-cases. First we assume 
$$
\frac{r}{d}>\frac{2}{p}-1
$$
and fix $\delta>0$ with 
\begin{equation}
\label{I5}
r-\left(\frac{2}{p}-1\right)d-\frac{2}{p}\delta>0.
\end{equation}
It follows from (\ref{D1a}), \eqref{H9a},  \eqref{I1a}, (\ref{P1}), 
and (\ref{I5}) that
\begin{eqnarray}
\lefteqn{\sup_{\mathcal{E}\in {\rm UEq}(\mathcal{B}_{\mathcal{W}_p^r(D)})}e_{\wt{n}}^\q(I,\mathcal{E})}
\notag\\
&\le& c\,2^{-rl_1}+c\sum_{l=l_0}^{l_1-1} 2^{-rl}
n_l^{-2/p}N_l^{2/p-1}\max(\log(n_l/\sqrt{N_l}),1)^{2/p-1}(l-l_0+1)^2
\notag\\
&\le& c\,2^{-rl_1}+c\sum_{l=l_0}^{l_1-1} 
2^{-rl-\frac{2}{p}dl_0+\frac{2}{p}\delta(l-l_0)+(\frac{2}{p}-1)dl}
(l_0+1)^{\frac{2}{p}-1}(l-l_0+1)^2
\notag\\
&\le& c\,2^{-rl_1}+c\,2^{-(r+d)l_0}(l_0+1)^{\frac{2}{p}-1}
\sum_{l=l_0}^{l_1-1}2^{\left(-r+\left(\frac{2}{p}-1\right)d+\frac{2}{p}\delta\right)(l-l_0)}(l-l_0+1)^2
\notag\\
&\le& c\,2^{-(r+d)l_0}(l_0+1)^{\frac{2}{p}-1}\le cn^{-r/d-1}(\log n)^{\frac{2}{p}-1},
\label{AL7}
\end{eqnarray}
which together with \eqref{I3a} gives \eqref{AF3}.

Now suppose that $r/d=2/p-1$. Instead of \eqref{I1a} 
we  define for $l=l_0,\dots,l_1-1$
\begin{eqnarray}
&&n_l=\left\lceil 2^{dl_0}(l_1-l_0+1)^{-1}(\ln(l_1-l_0+2))^{-1}\right\rceil,\quad
\nu_l=\left\lceil 8(\ln(l_1-l_0+2)+\ln 12)\right\rceil,
\label{Y8a}\\
&&\Theta_l=\frac{1}{12}(l_1-l_0+1)^{-1}\label{Y8b}.
\end{eqnarray}
Obviously, relations \eqref{X8}  and (\ref{AS6}) are satisfied again and by \eqref{AG5}
\begin{eqnarray}
\label{I3D}
\wt{n}&\le& n+2\kappa'\kappa''\nu(1)\sum_{l=l_0}^{l_1-1} 
\left\lceil 8(\ln(l_1-l_0+2)+\ln 12)\right\rceil 
\left\lceil 2^{dl_0}(l_1-l_0+1)^{-1}(\ln(l_1-l_0+2))^{-1}\right\rceil\nonumber\\
&\le& c\,2^{dl_0} \le cn.
\end{eqnarray}
As a consequence of (\ref{D1a}), \eqref{H9a}, (\ref{P1}), (\ref{Y8a}),
and (\ref{Y8b}) we obtain
\begin{eqnarray}
\lefteqn{\sup_{\mathcal{E}\in {\rm UEq}(\mathcal{B}_{\mathcal{W}_p^r(D)})}e_{\wt{n}}^\q(I,\mathcal{E})}
\notag\\
&\le& c\,2^{-rl_1}+c\sum_{l=l_0}^{l_1-1} 2^{-rl}
n_l^{-2/p}N_l^{2/p-1}\max(\log(n_l/\sqrt{N_l}),1)^{2/p-1}(l_1-l_0+1)
\label{AK4}\\
&\le& c\,2^{-rl_1}+c\sum_{l=l_0}^{l_1-1} 
2^{-rl-\frac{2}{p}dl_0+(\frac{2}{p}-1)dl}
(l_1-l_0+1)^{2/p+1}(\ln(l_1-l_0+2))^{2/p}(l_0+1)^{2/p-1}\quad
\label{AK5}\\
&\le& c\,2^{-rl_1}+c\,2^{-(r+d)l_0}(l_0+1)^{4/p+1}(\log(l_0+2))^{2/p}
\label{AK6}\\
&\le& c\,2^{-(r+d)l_0}(l_0+1)^{4/p+1}(\log(l_0+2))^{2/p}
\le cn^{-r/d-1}(\log n)^{4/p+1}(\log\log n)^{2/p}, \label{AK0}
\end{eqnarray}
so \eqref{AF4} follows from \eqref{I3D} and \eqref{AK0}.

Finally assume $0<r/d<2/p-1$. Here we replace \eqref{H9a} by  
\begin{equation*}
l_1=\left\lceil 2l_0+\frac{2-2/p}{r}dl_0\right\rceil,
\end{equation*}
choose $\delta>0$ so that
\begin{equation}
\label{AF6}
2\left(1-\frac{1}{p}\right)\delta<r<\left(\frac{2}{p}-1\right)d-\frac{2}{p}\delta,
\end{equation}
define for  $l_0\le l< 2l_0$
\begin{eqnarray}
n_l=\left\lceil 2^{dl_0-\delta(2l_0-l)}\right\rceil,\quad
\nu_l=\left\lceil 8(2\ln(2l_0-l+1)+ \ln 48)\right\rceil,\quad
\Theta_l&=&\frac{1}{48}(2l_0-l+1)^{-2},\quad\label{I1b}
\end{eqnarray}
and for $2l_0\le l<l_1$
\begin{eqnarray}
n_l=\left\lceil 2^{dl_0-\delta(l-2l_0)}\right\rceil,\quad
\nu_l=\left\lceil 8(2\ln(l-2l_0+1)+ \ln 48)\right\rceil,\quad
\Theta_l&=&\frac{1}{48}(l-2l_0+1)^{-2}.\quad\label{I1c}
\end{eqnarray}
Then, arguing analogously to \eqref{I2D}, 
$$
\sum_{l=l_0}^{2l_0-1}\Theta_l+\sum_{l=2l_0}^{l_1-1}\Theta_l\le \frac{1}{12},
\qquad
\sum_{l=l_0}^{2l_0-1}e^{-\nu_l/8}+\sum_{l=2l_0}^{l_1-1}e^{-\nu_l/8}\le \frac{1}{12},
$$
so \eqref{X8} and \eqref{AS6} hold again. Moreover, for $\wt{n}$ given by \eqref{AG5},
\begin{eqnarray}
\label{I3b}
\wt{n}&\le& n+2\kappa'\sum_{l=l_0}^{2l_0-1}
\left\lceil 8(2\ln(2l_0-l+1)+ \ln 48)\right\rceil 
\left\lceil2^{dl_0-\delta(2l_0-l)}\right\rceil
\notag\\
&&+2\kappa'\kappa''\nu(1)\sum_{l=2l_0}^{l_1-1}
\left\lceil 8(2\ln(l-2l_0+1)+ \ln 48)\right\rceil 
\left\lceil2^{dl_0-\delta(l-2l_0)}\right\rceil
\le c\,2^{dl_0} \le cn.
\end{eqnarray}
Then by (\ref{P1}) and  \eqref{AF6}--\eqref{I1c}
\begin{eqnarray}
\lefteqn{\sup_{\mathcal{E}\in {\rm UEq}(\mathcal{B}_{\mathcal{W}_p^r(D)})}e_{\wt{n}}^\q(I,\mathcal{E})}
\notag\\
&=& c\,2^{-rl_1}+c\sum_{l=l_0}^{2l_0-1} 2^{-rl}
	n_l^{-2/p}N_l^{2/p-1}\max(\log(n_l/\sqrt{N_l}),1)^{2/p-1}(2l_0-l+1)^2
\notag\\
&& +c\sum_{l=2l_0}^{l_1-1} 2^{-rl}n_l^{-2(1-1/p)}\max(\log(n_l/\sqrt{N_l}),1)^{2/p-1}
(l-2l_0+1)^2
\notag\\
&\le& c\,2^{-rl_1}+c\sum_{l=l_0}^{2l_0-1} 
2^{-rl-\frac{2}{p}dl_0+\frac{2}{p}\delta(2l_0-l)+\left(\frac{2}{p}-1\right)dl}
(l_0-l/2+1)^{\frac{2}{p}-1}(2l_0-l+1)^2
\notag\\
&& +c\sum_{l=2l_0}^{l_1-1}  2^{-rl
-2(1-1/p)(dl_0-\delta(l-2l_0))}(l-2l_0+1)^2
\notag\\
&\le& c\,2^{-rl_1}+c\,2^{-2rl_0-\frac{2}{p}dl_0+2\left(\frac{2}{p}-1\right)dl_0}
\sum_{l=l_0}^{2l_0-1}2^{(r-(\frac{2}{p}-1)d+\frac{2}{p}\delta)(2l_0-l)}(2l_0-l+1)^{\frac{2}{p}+1}
\notag\\
&& +c2^{-2rl_0-2\left(1-\frac{1}{p}\right)dl_0}\sum_{l=2l_0}^{l_1-1}
2^{\left(-r+2\left(1-\frac{1}{p}\right)\delta\right)(l-2l_0)}(l-2l_0+1)^2
\notag\\
&\le& c\,2^{-rl_1}+2^{-2rl_0-2\left(1-\frac{1}{p}\right)dl_0}\le c n^{-2r/d+2/p-2}.
\label{AL8}
\end{eqnarray}
This together with \eqref{I3b} proves \eqref{AF5}.

If the embedding condition \eqref{PA2} holds, we use $\Theta_l\equiv 1$, see \eqref{AS6}, and \eqref{AM0} and argue in the same way, with obvious changes. This does not alter the estimates \eqref{AL5}, \eqref{AL6}, and \eqref{AL7}. On the other hand, in the case, $r/d=2/p-1$, the removal of a factor of $(l_1-l_0+1)$ from \eqref{AK4} and \eqref{AK5} reduces the exponent of $(l_0+1)$ in \eqref{AK6}, \eqref{AK0}, and therefore also that of $\log n$ in \eqref{AK0} by 1. This shows (\ref{AF4a}) and completes the proof.

\qed
\begin{proposition}
\label{pro:9}
Let $r\in\N_0$, $d\in\N$, and $1\le p \le\infty$. 
Then there is a constant $c>0$
such that for all $n\in\N$ with $n>4$,  
\begin{eqnarray}
\sup_{\mathcal{E}\in {\rm UEq}(\mathcal{B}_{\mathcal{W}_p^r(D)})} e_n^\q(I,\mathcal{E})&\ge& c n^{-r/d-1}
\quad\mbox{if}\quad r/d\ge 2/p-1,
\label{AG9}\\[.2cm]
\sup_{\mathcal{E}\in {\rm UEq}(\mathcal{B}_{\mathcal{W}_p^r(D)})} e_n^\q(I,\mathcal{E})&\ge& c n^{-2r/d+2/p-2}
\quad\mbox{if}\quad r/d< 2/p-1.\label{AH0}
\end{eqnarray}
If the embedding condition \eqref{PA2} does not hold, then for all $n\in\N$
\begin{equation}
\label{AG6}
 e_n^\q(I,\mathcal{B}_{\mathcal{W}_p^r(D)}\cap C(Q))\ge c.
\end{equation}
\end{proposition}

\begin{proof} We argue similarly to the proof of Proposition \ref{pro:11}. If the embedding condition \eqref{PA2} does not hold, the proof of Theorem 4.3 in \cite{Hei08a} is easily modified to show that there is a constant $c>0$ such that for all $n\in \N$
\begin{equation*}
e_n^\de(I,\mathcal{B}_{\mathcal{W}_p^r(D)}\cap C(Q))\ge c.
\end{equation*}
Using  Lemma  \ref{lem:5} again gives \eqref{AG6}.

Next we apply Proposition \ref{pro:4} with  $d_1=0$, $d_2=d$, thus by \eqref{AF7} $M_{1,l}=1$, $M_{2,l}=2^{dl}$, and Proposition \ref{pro:1a} with $N=M_{2,l}$. Let $n\in \N$, $n>4$.
In the case  $r/d\ge 2/p-1$ we set
$l=\lceil d^{-1}(\log(n/c(7))+1)\rceil$, where $c(7)$ is the constant $c_0$ from 
Proposition \ref{pro:1a}. Clearly, it can be assumed that $0<c(7)\le 1$.
Then arguing as in \eqref{AU7} and \eqref{N6}, we have
\begin{equation}
\label{AG7}
c(7) 2^{-d}M_{2,l} = c(7) 2^{-d}2^{dl}\le 2n< c(7) 2^{dl} = c(7) M_{2,l}.
\end{equation}
Since $\mathcal{B}_{\mathcal{W}_p^r(D)}\subset \mathcal{B}_{\mathcal{W}^r_q(D)}$ for $p>q$, we can assume $2<p\le\infty$. From Proposition \ref{pro:4}, \eqref{K1} of Proposition \ref{pro:1a}, and (\ref{AG7}) we get
\begin{eqnarray*}
\sup_{\mathcal{E}\in {\rm UEq}(\mathcal{B}_{\mathcal{W}_p^r(D)})} e_n^\q(I,\mathcal{E})\ge c2^{-rl} e_{2n}^\q\Big(I^{M_{2,l}},\mathcal{B}_{L_p^{M_{2,l}}}\Big)
\ge c2^{-rl}n^{-1}\ge c n^{-r/d-1},
\end{eqnarray*}
thus (\ref{AG9}).
Now let $r/d< 2/p-1$, hence $1\le p<2$, and put 
$l=\lceil d^{-1}(\log(n^2/c(7))+1)\rceil$.
It follows as in \eqref{N6a} that
\begin{equation}
\label{AG8}
c(7) 2^{-d}M_{2,l} = c(7) 2^{-d}2^{dl}\le 2n^2\le c(7) 2^{dl} = c(7) M_{2,l}.
\end{equation}
Applying Proposition \ref{pro:4}, \eqref{K3} of Proposition \ref{pro:1a}, and (\ref{AG8}) leads to 
\begin{eqnarray*}
\sup_{\mathcal{E}\in {\rm UEq}(\mathcal{B}_{\mathcal{W}_p^r(D)})} e_n^\q(I,\mathcal{E})&\ge& c2^{-rl} e_{2n}^\q\Big(I^{M_{2,l}},\mathcal{B}_{L_p^{M_{2,l}}}\Big)
\ge c2^{-rl}\min(n^{-2(1-1/p)},n^{-2/p}M_{2,l}^{2/p-1})
\\
&\ge& c2^{-rl}n^{-2(1-1/p)}\ge c n^{-2r/d+2/p-2},
\end{eqnarray*}
showing (\ref{AH0}).

\end{proof}

\section{Comments}\label{sec:6}

We summarize the results 
in two tables and compare them with
the respective known quantities in the classical deterministic and randomized settings. For the results in the classical settings see \cite{Hei10} and references therein. We are interested in determining the extend of superiority of the quantum over the classical settings. The entries of the table give the order of the minimal errors, in the quantum setting logarithmic factors are suppressed, and (EC) means the embedding condition  \eqref{PA2} holds. To exclude trivial quantum rates, we assume 
\begin{equation}
\label{AM5}
\frac{r}{d}>\left(\frac{1}{p}-\frac{1}{q}\right)_+\quad \text{for approximation},\quad r>0 \vee p>1 \quad \text{for integration}.
\end{equation}
\[\begin{array}{l|l|l|l}
\mbox{Approximation}\;\mathcal{W}_p^r(D)\to L_q(D)
& \ \mbox{deterministic}\  & \, \mbox{random}\, & \, \mbox{quantum}\,\\ \hline 
&&&\\
(i)\quad \frac{r}{d}\ge\frac{2}{p}-\frac{2}{q}\;\wedge\mbox{(EC)}& n^{-r/d+(1/p-1/q)_+} & \, n^{-r/d +(1/p-1/q)_+}  
& \, n^{-r/d}\\
 &&&\\
(ii)\quad \frac{r}{d}\ge\frac{2}{p}-\frac{2}{q}\;\wedge\neg\mbox{(EC)}&n^0 & \, n^{-r/d +(1/p-1/q)_+}  
& \, n^{-r/d}\\
&&&\\
(iii)\quad \frac{1}{p}-\frac{1}{q}<\frac{r}{d}<\frac{2}{p}-\frac{2}{q}\;\wedge\mbox{(EC)}& n^{-r/d+1/p-1/q} & \, n^{-r/d +1/p-1/q}  
& \, n^{-2r/d+2/p-2/q}\\
 &&&\\
(\romannumeral 4)\quad \frac{1}{p}-\frac{1}{q}<\frac{r}{d}<\frac{2}{p}-\frac{2}{q}\;\wedge\neg\mbox{(EC)}& n^0 & \, n^{-r/d +1/p-1/q}  
& \, n^{-2r/d+2/p-2/q}
\end{array}
\]

\medskip\medskip

\[
\hspace*{-.5cm}\begin{array}{l|l|l|l}
 \mbox{Integration} \;\mathcal{W}_p^r(D)& \ \mbox{det}\  & \, \mbox{random}\, & \, \mbox{quantum}\,\\ \hline 
&&&\\
(\romannumeral 5)\quad \frac{r}{d}\ge\frac{2}{p}-1\;\wedge\mbox{(EC)} &\; n^{-r/d } \;& \, n^{-r/d -1 + 1/{\rm min}(p,2) }
&   n^{-r/d -1}
\\
&&&\\
 (\romannumeral 6)\quad \frac{r}{d}\ge\frac{2}{p}-1\;\wedge\neg\mbox{(EC)} &\; n^0 & \, n^{-r/d -1 + 1/{\rm min}(p,2) }
&   n^{-r/d -1}
\\
&&&\\
(\romannumeral 7)\quad \frac{r}{d}<\frac{2}{p}-1\quad \big(\mbox{implies }\neg\mbox{(EC)}\big)   &\; n^0 & \, n^{-r/d -1 + 1/p }
&   n^{-2r/d-2+2/p}
\end{array}
\]

To see the speedups of quantum computation (in other words, the gaps between the rates in the different settings), we define similar to \cite{Hei23b}, Section 6, for approximation
\begin{eqnarray*}
\gamma^\de(n;J,r,p,q,d)&=& \frac{e_n^\de(J, \mathcal{B}_{\mathcal{W}_p^r(D)}\cap C(D)),L_q(D))}{\sup_{\mathcal{E}\in {\rm UEq}(\mathcal{B}_{\mathcal{W}_p^r(D)})} e_n^\q(J,\mathcal{E},L_q(D))}
\\
\gamma^\ran(n;J,r,p,q,d)&=& \frac{e_n^\ran(J, \mathcal{B}_{\mathcal{W}_p^r(D)},L_q(D))}{\sup_{\mathcal{E}\in {\rm UEq}(\mathcal{B}_{\mathcal{W}_p^r(D)})} e_n^\q(J,\mathcal{E},L_q(D))}
\end{eqnarray*}
and correspondingly, for integration,
\begin{eqnarray*}
\gamma^\de(n;I,r,p,d)&=& \frac{e_n^\de(I, \mathcal{B}_{\mathcal{W}_p^r(D)}\cap C(D))}{\sup_{\mathcal{E}\in {\rm UEq}(\mathcal{B}_{\mathcal{W}_p^r(D)})} e_n^\q(I,\mathcal{E})}
\\
\gamma^\ran(n;I,r,p,d)&=& \frac{e_n^\ran(I, \mathcal{B}_{\mathcal{W}_p^r(D)})}{\sup_{\mathcal{E}\in {\rm UEq}(\mathcal{B}_{\mathcal{W}_p^r(D)})} e_n^\q(I,\mathcal{E})}.
\end{eqnarray*}
The principal exponents are defined for $\set\in\{\det,\ran\}$ as
\begin{eqnarray}
\theta^\set(J,r,p,q,d)&=& \lim_{n\to \infty}\frac{\log\gamma^\set(n;J,r,p,q,d)}{\log n}
\label{AM3}\\
\theta^\set(I,r,p,d)&=& \lim_{n\to \infty}\frac{\log\gamma^\set(n;I,r,p,d)}{\log n}.\label{AM4}
\end{eqnarray}
Since under the assumptions \eqref{AM5} all quantum rates are up to powers of $\log n$ of the form $n^{-\alpha}$ for some $\alpha>0$, it follows that the limit in \eqref{AM3} and \eqref{AM4} exist. We define the maximal speedup 
\begin{equation*}
\theta^\set_{\max}(J)=\sup\,\theta^\set(J,r,p,q,d), \quad \theta^\set_{\max}(I)=\sup\,\theta^\set(I,r,p,d), 
\end{equation*}
where the supremum is taken over the considered domain of parameters $r,p,q,d$, respectively $r,p,d$. Finally, we write $\theta^\de_{\max}(J)=a^-$, if $\sup\theta^\set(J,r,p,q,d)=a$, but the supremum is not attained (and similarly in the other cases).

We start with approximation.

Case $(i)\quad \frac{r}{d}\ge\frac{2}{p}-\frac{2}{q}\;\wedge\mbox{(EC)}$: Clearly,
\begin{equation*}
\theta^\set(J,r,p,q,d)=\left(\frac{1}{p}-\frac{1}{q}\right)_+, 
\end{equation*}
hence $\theta^\set_{\max}(J)=1$, attained for $p=1$, $q=\infty$, $r/d\ge 2$.

Case $(ii)\quad \frac{r}{d}\ge\frac{2}{p}-\frac{2}{q}\;\wedge\neg\mbox{(EC)}$:
The condition implies
\begin{eqnarray*}
2-\frac{2}{q}\le \frac{r}{d}
   <1\quad\mbox{if}\quad p=1,\qquad  
 \frac{2}{p}-\frac{2}{q}\le \frac{r}{d}
   \le \frac{1}{p}<1\quad\mbox{if}\quad p>1,  
\end{eqnarray*}
hence 
\begin{equation*}
\theta^\de(J,r,p,q,d)=\frac{r}{d}<1, \quad\theta^\ran(J,r,p,q,d)=\left(\frac{1}{p}-\frac{1}{q}\right)_+<\frac{1}{2}. 
\end{equation*}
Therefore choosing $p=1$, $q$ arbitrarily close to but smaller than 2, and $r/d=2-\frac{2}{q}$, we get $\theta^\de_{\max}(J)=1^{-}$, $\theta^\ran_{\max}(J)=(1/2)^{-}$.

Case $(iii)\quad \frac{1}{p}-\frac{1}{q}<\frac{r}{d}<\frac{2}{p}-\frac{2}{q}\;\wedge\mbox{(EC)}$: Then
\begin{equation*}
\theta^\set(J,r,p,q,d)=\frac{r}{d}-\frac{1}{p}+\frac{1}{q}<\frac{1}{p}-\frac{1}{q}<1. 
\end{equation*}
Letting $p=1$, $q=\infty$, and $r/d$ less than but arbitrarily close to 2, we obtain $\theta^\set_{\max}(J)=1^-$.

Case $(\romannumeral 4)\quad \frac{1}{p}-\frac{1}{q}<\frac{r}{d}<\frac{2}{p}-\frac{2}{q}\;\wedge\neg\mbox{(EC)}$. The condition implies $p<q$ and $\frac{r}{d}<1$. Hence 
\begin{equation*}
\theta^\ran(J,r,p,q,d)=\frac{r}{d}-\frac{1}{p}+\frac{1}{q}<\frac{r}{2d}<\frac{1}{2},\quad \theta^\de(J,r,p,q,d)=\frac{2r}{d}-\frac{2}{p}+\frac{2}{q}=2\theta^\ran(J,r,p,q,d)<1. 
\end{equation*}
Therefore with $p=1$, $q=2$ and $r/d$ less than but arbitrarily close to 1 we conclude $\theta^\ran_{\max}(J)=(1/2)^{-}$, and consequently $\theta^\de_{\max}(J)=1^{-}$.

Now we consider integration. 

Case $(\romannumeral 5)\quad \frac{r}{d}\ge\frac{2}{p}-1\;\wedge\mbox{(EC)}$: We have
\begin{equation*}
\theta^\de(I,r,p,d)=1, \quad\theta^\ran(I,r,p,d)=\frac{1}{\min(p,2)}, 
\end{equation*}
consequently $\theta^\de_{\max}(I)=1$, $\theta^\ran_{\max}(I)=1$, attained for $p=1$, $r/d\ge 1$.

Case $(\romannumeral 6)\quad \frac{r}{d}\ge\frac{2}{p}-1\;\wedge\neg\mbox{(EC)}$: The condition implies that $p=1$ cannot occur, and therefore $\frac{2}{p}-1\le \frac{r}{d}\le\frac{1}{p}$. Then
\begin{equation*}
\theta^\de(I,r,p,d)=\frac{r}{d}+1<2, \quad\theta^\ran(I,r,p,d)=\frac{1}{\min(p,2)}<1. 
\end{equation*}
With $p$ arbitrarily close to 1, we obtain
$\theta^\de_{\max}(I)=2^{-}$, $\theta^\ran_{\max}(I)=1^{-}$.

Case $(\romannumeral 7)\quad \frac{r}{d}<\frac{2}{p}-1$: Thus
\begin{equation*}
\theta^\ran(I,r,p,d)=\frac{r}{d}+1-\frac{1}{p}<\frac{1}{p}\le 1, \quad\theta^\de(I,r,p,d)=2\theta^\ran(I,r,p,d)<2. 
\end{equation*}
Setting $p=1$ and letting $r/d$ be smaller than but arbitrarily close to 1, we conclude $\theta^\ran_{\max}(I)=1^{-}$, $\theta^\de_{\max}(I)=2^{-}$.

In a table, this gives the following picture
\[\begin{array}{l|l|l}
\mbox{Approximation}\;\mathcal{W}_p^r(D)\to L_q(D)
& \ \theta^\de_{\max}(J)\  & \,  \theta^\ran_{\max}(J)\, \\ \hline 
&&\\
(i)\quad \frac{r}{d}\ge\frac{2}{p}-\frac{2}{q}\;\wedge\mbox{(EC)}& 1 & \,1  
\\
 &&\\
(ii)\quad \frac{r}{d}\ge\frac{2}{p}-\frac{2}{q}\;\wedge\neg\mbox{(EC)}&1^- & \, \left(\frac{1}{2}\right)^- 
\\
&&\\
(iii)\quad \frac{1}{p}-\frac{1}{q}<\frac{r}{d}<\frac{2}{p}-\frac{2}{q}\;\wedge\mbox{(EC)}& 1^- & \, 1^-
\\
 &&\\
(\romannumeral 4)\quad \frac{1}{p}-\frac{1}{q}<\frac{r}{d}<\frac{2}{p}-\frac{2}{q}\;\wedge\neg\mbox{(EC)}&1^- & \, \left(\frac{1}{2}\right)^-
\end{array}
\]

\medskip\medskip

\[
\hspace*{-.5cm}\begin{array}{l|l|l}
 \mbox{Integration} \;\mathcal{W}_p^r(D)& \ \theta^\de_{\max}(I)\  & \,  \theta^\ran_{\max}(I)\,\\ \hline 
&&\\
(\romannumeral 5)\quad \frac{r}{d}\ge\frac{2}{p}-1\;\wedge\mbox{(EC)} &1 \;& \,1
\\
&&\\
 (\romannumeral 6)\quad \frac{r}{d}\ge\frac{2}{p}-1\;\wedge\neg\mbox{(EC)} &2^- & \,1^-
\\
&&\\
(\romannumeral 7)\quad \frac{r}{d}<\frac{2}{p}-1\quad \big(\mbox{implies }\neg\mbox{(EC)}\big)   &2^- & \,1^-
\end{array}
\]

\medskip

Under the embedding condition we observe a possible improvement of $n^{-1}$ (up to log factors)
over the  randomized 
and also over the deterministic setting. This speedup was first found
in \cite{Hei01b} for integration 
and later for approximation \cite{Hei04b}. If the embedding condition does not hold, the speedup over the  randomized setting can be $n^{-1+\delta}$ for arbitrarily small $\delta>0$, while
it can be $n^{-2+\delta}$ over the deterministic setting.

\end{document}